\documentclass[3p,times,11pt]{elsarticle}
\usepackage[titletoc,title]{appendix}
\usepackage{float}
\usepackage{wrapfig,blindtext}
\usepackage{graphicx} % to include images
\usepackage{epstopdf} % to convert eps to pdf
\usepackage{pslatex} % to use postscript fonts
\usepackage{amsmath,amssymb}
\usepackage{caption}
\usepackage{subcaption}
\usepackage{amsfonts,amsthm} % Math packages
\usepackage[english]{babel} % English language/hyphenation
\usepackage[dvipsnames]{xcolor}
\usepackage{wasysym}
\usepackage{pdfpages}
\usepackage{float}
\usepackage{comment}

\usepackage{nomencl}%   nomenclature generation via makeindex

\usepackage[font=small,labelfont=bf,
   justification=justified,format=plain]{caption}

\usepackage{comment}

\newcommand{\MC}{\mathcal}
\usepackage{nomencl}%   nomenclature generation via makeindex
  \makenomenclature

\journal{Journal of Computational Physics, Nov. 2016}

\begin{document}

\begin{frontmatter}

\title{A Dynamic Subgrid Scale Model for Large Eddy Simulations Based on the Mori-Zwanzig Formalism}

\author{Eric J. Parish}\ead{parish@umich.edu}
\author{Karthik Duraisamy}\ead{kdur@umich.edu}
\address{Department of Aerospace Engineering, University of Michigan, Ann Arbor, MI 48109, USA}

%\fntext[fn1]{P.h.D. Candidate}
%\fntext[fn2]{Assistant Professor}

%\cortext[cor1]{Corresponding Author}

%\begin{keyword}
%data-driven modeling\sep
%machine learning\sep
%closure modeling
%\end{keyword}
%\input{abstract.tex}
\begin{abstract}
The development of reduced models for complex multiscale problems remains one of the principal challenges in computational physics. The optimal prediction framework of Chorin et al.~\cite{ChorinOptimalPrediction}, which is a reformulation of the Mori-Zwanzig (M-Z) formalism of non-equilibrium statistical mechanics, provides a  methodology for the development of mathematically-derived reduced models of dynamical systems. Several promising models have emerged from the optimal prediction community and have found application in molecular dynamics and turbulent flows. In this work, a new M-Z-based closure model that addresses some of the deficiencies of existing methods is developed. The model is constructed by exploiting similarities between two levels of coarse-graining via the Germano identity of fluid mechanics and by assuming that memory effects have a finite temporal support. The appeal of the proposed model, which will be referred to as the `dynamic-MZ-$\tau$' model, is that it is parameter-free and has a structural form imposed by the mathematics of the coarse-graining process (rather than the phenomenological assumptions made by the modeler, such as in classical subgrid scale models). To promote the applicability of M-Z models in general, two procedures are presented to compute the resulting model form, helping to bypass the tedious error-prone algebra that has proven to be a hindrance to the construction of M-Z-based models for complex dynamical systems. While the new formulation is applicable to the solution of general partial differential equations, demonstrations are presented in the context of Large Eddy Simulation closures for Burgers equation, decaying homogeneous turbulence, and turbulent channel flow. The performance of the model and validity of the underlying assumptions are investigated in detail. 
\end{abstract}

\end{frontmatter}

\section{Introduction}
%========================================================================
Direct computations of many complex problems, such as atomistic simulations and turbulence, are typically prohibitively expensive because of the presence of a wide range of spatial and temporal scales. Models of  reduced computational complexity are required to bridge the gap between reality and affordability. In the context of turbulent flows, the Large Eddy Simulation (LES) technique has proven to be a computationally feasible alternative to Direct Numerical Simulations (DNS). In LES, the governing equations of motion are coarse-grained to reduce the range of scales present in the system. Unresolved effects are accounted for through a so-called subgrid  model. For non-linear systems lacking scale separation, the coarse-graining procedure leads to non-Markovian effects~\cite{ChorinOptimalPrediction}. These non-local effects are challenging to quantify, understand, and model, and are typically ignored in  in LES models.
%, the majority of subgrid models are local effects by invoking scale separation arguments.

The optimal prediction framework developed by Chorin and co-workers~\cite{ChorinOptimalPrediction,GivonOrthogonal,HaldPredictFromData,StatisticalMechanics,ProblemReduction} provides a formal setting for both the development of closure models and the analysis of memory effects stemming from the coarse-graining process. This framework is a reformulation of the Mori-Zwanzig (M-Z) formalism of non-equilibrium statistical mechanics. The M-Z formalism \cite{MoriTransport,ZwanzigLangevin} provides a methodology to recast a high-dimensional Markovian dynamical system into a lower-dimensional, non-Markovian (non-local) system. In this lower-dimensional system, which is commonly referred to as the generalized Langevin equation (GLE), the effects of the unresolved scales on the resolved scales are non-local and appear as a convolution integral. The GLE is an exact statement of the original dynamics, but has the advantage that it depends only on the resolved variables.

It is possible, in theory, to use the GLE to determine the impact of the unresolved scales on the resolved scales using only the resolved variables. While such a process can  provide valuable insight into the non-locality of coarse-grained simulations and,  concurrently, model development, it does not result in a reduction of computational complexity compared to the original system. A variety of approximations have thus been developed. The $t$-model~\cite{bernsteinBurgers,stinisEuler,ChandyFrankelLES}, finite memory models~\cite{stinis_finitememory,stinisHighOrderEuler,ParishAIAA2016,ParishMZ1}, and renormalized M-Z models~\cite{RenormalizedMZ,Stinis-rMZ} are several such examples.

In the context of fluid dynamics, M-Z-based models have been successfully applied to a variety of problems, including Fourier-Galerkin solutions of the Navier-Stokes equations. The models, however, have several shortcomings. In the case of the finite memory models, the memory length of the M-Z memory integrand is required. While guidelines on how the memory length scales with grid resolution have been presented in ~\cite{ParishMZ1,Stinis-rMZ,GouasmiMZ1}, the a priori selection of the memory length is still challenging and somewhat unsatisfying. In the case of the Markovian t-model, it has been recognized~\cite{RenormalizedMZ} that the model form requires an additional scaling to maintain accuracy. Most prominently, Stinis~\cite{RenormalizedMZ,Stinis-rMZ} has developed a class of M-Z models that employ a renormalization procedure. The approach is shown to perform well for Burgers equation and the Euler equations. As in the $t$-model, however, the functional form of these models implicitly contains a time term. As such, it is not clear how the procedure can be applied to statistically steady problems.% Another drawback of the procedure is that it requires the solution of a linear system that becomes increasingly ill-conditioned as the resolution of the coarse-grained model grows. 

An additional limitation of M-Z-based models is that the simplifying assumptions  have been difficult to assess; thus the success of the models has remained somewhat a mystery. The challenge in assessing the simplifying assumptions arises from the fact that the evaluation of the non-local convolution integral requires the solution of the so-called orthogonal (or unresolved) dynamics equation, which itself is a high dimensional hyperbolic PDE. Instead of directly computing the orthogonal dynamics, a common approach has been to represent the convolution integral with an expressive basis~\cite{ChorinOptimalPrediction,bernsteinBurgers}. Unfortunately, this process is intractable for high-dimensional problems unless a (very) low-dimensional basis is used. The dimensionality of the basis functions severely limits the accuracy and insight that can be gained from such finite-rank projections. In Refs.~\cite{ParishAIAA2016,ParishMZ1,GouasmiMZ1}, Gouasmi and the present authors propose an approximate technique that can be used to obtain approximations to the memory. The methodology uses a set of auxiliary ODEs obtained from the unresolved equations to approximate solutions to the orthogonal dynamics. The convolution integral is then approximated by solving the auxiliary ODEs using the trajectories of the resolved variables as initial conditions. In Ref.~\cite{GouasmiMZ1}, the procedure was shown to provide accurate predictions for subgrid content in Burgers equation.  

In the present work, we develop a parameter-free closure model based on the M-Z formalism. The model combines ideas from the optimal prediction community with the Germano procedure from fluid mechanics. This work outlines the development of the model and assesses the performance and underlying assumptions of the model. Fourier-Galerkin solutions of Burgers equation and the incompressible Navier-Stokes equations will be considered. The outline of this paper is as follows: Section~\ref{sec:MZ} will provide a brief overview of the Mori-Zwanzig formalism. The orthogonal dynamics equation and an approximate methodology for evaluating the memory integral will be presented. Section~\ref{sec:dtau} will outline the formulation of the dynamic-$\tau$ model. In Section~\ref{sec:burg}, the model will be applied to Burgers equation. In Sections~\ref{sec:HIT} and~\ref{sec:Channel},
the model will be applied to decaying homogeneous turbulence and turbulent channel flow. Conclusions and perspectives will be provided in Section~\ref{sec:conclude}.

\section{The Mori-Zwanzig formalism}\label{sec:MZ}

The model reduction framework of Chorin et al.~\cite{ChorinOptimalPrediction}, which is a reformulation of the Mori-Zwanzig formalism, is now provided. We first consider a demonstrative example to introduce the unfamiliar reader to the Mori-Zwanzig formalism. Consider a two-state linear system given by
\begin{equation}\label{eq:linear1}
\frac{dx}{dt} = A_{11}x + A_{12} y
\end{equation}
\begin{equation}\label{eq:linear2}
\frac{dy}{dt} = A_{21}x + A_{22} y.
\end{equation}
Suppose that one wants to created a `reduced-order' model of the system given in Eqns.~\ref{eq:linear1} and~\ref{eq:linear2} by creating a reduced system that depends only on $x(t)$, i.e.
\begin{equation}\label{eq:linear3}
\frac{dx}{dt} = A_{11}x + F(x).
\end{equation}
The challenge for the  modeler is to construct the function $F(x)$ that accurately represents the effect of the unresolved variable $y$ on the resolved variable $x$. This is analogous to modeling subgrid scale effects in large eddy simulations. 

For this simple linear system, $F(x)$ can be exactly determined by solving Eq.~\ref{eq:linear2} for $y(t)$ in terms of a general $x(t)$. Through this process, the two-component Markovian system can be cast as a one-component non-Markovian system that has the form
\begin{equation}\label{eq:linear4}
\frac{dx}{dt} = A_{11}x + A_{12}A_{21}\int_0^t x(t-s)e^{A_{22}s}ds + A_{12}y(0)e^{A_{22}t}.
\end{equation}
Equation~\ref{eq:linear4} has no dependence on $y(t)$ and hence is closed. Note, however, that one had to formally integrate out the unresolved variable $y$ to obtain Eq.~\ref{eq:linear4}. The one-dimensional non-Markovian representation of the original system offers no decrease in computational complexity. Instead, it provides a starting point for the construction of closure models since the effect of the unresolved variable $y$ has been expressed purely in terms of the resolved variable, $x$. This is of importance since, in a reduced order simulation, one only has access to the resolved variable $x$. 

The reduction of a Markovian set of equations to a lower-dimensional, non-Markovian set of equations, for a general set of non-linear differential equations, is the essence of the Mori-Zwanzig formalism.

Before providing the mathematical foundations, we offer the following high level description of the Mori-Zwanzig formalism and how it can be used for model reduction:
\begin{enumerate}
\item Start with an $N$-component system of non-linear ordinary differential equations. In the case that the ODE system is a spatially discretized PDE (as is considered in this work), there is an underlying assumption that the ODE has some hierarchical structure. Spectral methods and high-order hierarchical finite element methods are two such examples.
\item Transform the non-linear set of ordinary differential equations into linear partial differential equations. The linear PDEs exist  in the space of the initial conditions and the transformation is exact. The purpose of this transformation is to perform the model reduction in the linear PDE space, and then transform the resulting equations back into ordinary differential equations. 
\item Split the right hand side dynamics of the PDEs into resolved and unresolved components through projection operators. 
\item Formally integrate out the unresolved dynamics through the Duhamel relation. This allows the effect of the unresolved scales on the resolved scales to be expressed purely as a memory integral (it will be shown that the noise term is inconsequential for this work). The arguments of the memory integral involve only the resolved variables.
\item Construct an efficient model to the memory integral. 
\item Transform the linear PDEs back into M-component non-linear ODEs (where $M<N$).
\end{enumerate}
The construction of an efficient model to the memory integral is the focus of this work.

\subsection{The Generalized Langevin Equations}
The mathematical basis of the Mori-Zwanzig formalism is now provided. Ref.~\citenum{ChorinOptimalPrediction} offers a more complete description.  Consider the semi-discrete system of ordinary differential equations
\begin{equation}\label{eq:baseODE}
\frac{d \mathbf{\phi} }{dt} = R(\mathbf{\phi}),
\end{equation}
%where $\mathbf{\phi} \in \mathbb{C}^N$.  The initial condition is $\mathbf{\phi}(0) = \mathbf{\phi}_0$ with $\phi_0 \in L^2$. Further, let
%$$\mathbf{\phi} = \hat{\mathbf{\phi}} + \tilde{\mathbf{\phi}},$$
%where the resolved modes are $\mathbf{\hat{\phi}} = \{\phi_1,\phi_2,\hdots, \phi_M, 0,0 , \hdots \}$ and the unresolved modes $\mathbf{\tilde{\phi}} = \{0,\hdots, 0,\phi_{M+1}, %\phi_{M+2} , \hdots, \phi_{N} \}$.
where $\mathbf{\phi} = \{\hat{\mathbf{\phi}},\tilde{\mathbf{\phi}}\}$,  $\hat{\mathbf{\phi}} \in {R}^M$ are the resolved modes, and  $\tilde{\mathbf{\phi}} \in {R}^{N-M}$ are the unresolved modes. The initial condition is $\mathbf{\phi}(0) = \mathbf{\phi}_0$ with $\phi_0 \in L^2$. 
The objective is to solve Eq.~\ref{eq:baseODE} for the resolved modes without explicitly computing the unresolved modes. 
The non-linear ODE can be posed as an N-dimensional linear partial differential equation by casting it in the Liouville form, 
\begin{equation}\label{eq:LiouvilleForm}
\frac{\partial}{\partial t} u(\phi_0,t) = \mathcal{L}u(\phi_0,t),
\end{equation}
with $u(\phi_0,0) = g(\phi(\phi_0,0))$ and 
$$\MC{L} = \sum_{k=1}^{N} R_k(\phi_0) \frac{\partial}{\partial \phi_{0k}},$$
where $\phi_{0k} = \phi_k(0)$. Note that Eq.~\ref{eq:LiouvilleForm} exists in the space of initial conditions, and has as many spatial dimensions as there are state variables. The link between Eq.~\ref{eq:baseODE} and Eq.~\ref{eq:LiouvilleForm} is intuitive by considering Eq.~\ref{eq:LiouvilleForm} as an N-dimensional advection equation where the characteristic paths are defined by the RHS of the original ODE.
It can be shown that the solution to Eq.~\ref{eq:LiouvilleForm} is given by
\begin{equation}\label{eq:LiouvilleSol}
u(\phi_0,t) = g(\phi(\phi_0,t)).
\end{equation}
The semigroup notation is now used, i.e. $u(\phi_0,t) =  e^{t\MC{L}}g(\phi_0$). By Eq.~\ref{eq:LiouvilleSol}, it is seen that the evolution operator $e^{t \MC{L}}$ is the Koopman operator~\cite{Koopman} for the original dynamic system.
By taking $g(\phi_0) = \phi_{0j}$, an equation for the trajectory of a resolved variable can be written as
\begin{equation}\label{eq:LiouvilleFormSG2}
\frac{\partial}{\partial t} e^{t \MC{L}}\phi_{0j}
= e^{t \MC{L}}\mathcal{L}\phi_{0j}.
\end{equation}
The right hand side can be partitioned into resolved and unresolved components through the use of projection operators. Let the space of the resolved variables be denoted by $\hat{L}^2$. Further, define $\mathcal{P}: L^2 \rightarrow \hat{L}^2$, as well as $\mathcal{Q} = I - \mathcal{P}$. 

In this work, a simple projection operator is used. For a function $f(\hat{\phi}_0,\tilde{\phi}_0)$, application of the projection operator yields $\mathcal{P}f(\hat{\phi}_0,\tilde{\phi}_0) = f(\hat{\phi}_0,0)$. More complex projections are possible (again see \cite{ChorinOptimalPrediction}). Using the identity $I = \MC{P + Q}$, the right hand side of Eq.~\ref{eq:LiouvilleFormSG2} can be split into resolved and unresolved components,
\begin{equation}\label{eq:LiouvilleFormSG3}
\frac{\partial}{\partial t} e^{t \MC{L}}\phi_{0j}
= e^{t \MC{L}}\mathcal{PL}\phi_{0j} + e^{t \MC{L}}\mathcal{QL}\phi_{0j}.
\end{equation}
At this point the Duhamel formula is utilized,
$$e^{t \mathcal{L}} = e^{t \mathcal{Q} \mathcal{L}} + \int_0^t e^{(t - s)\mathcal{L}} \mathcal{P}\mathcal{L} e^{s \mathcal{Q} \mathcal{L}} ds.$$
Inserting the Duhamel formula into Eq.~\ref{eq:LiouvilleFormSG3}, the generalized Langevin equation is obtained,
\begin{equation}\label{eq:M-Z_Identity}
\frac{\partial}{\partial t} e^{t\MC{L}}\phi_{0j} =   \underbrace{e^{t\MC{L}}\MC{PL}\phi_{0j}}_{\text{Markovian}} +   \underbrace{e^{t\MC{QL}}\MC{Q}L\phi_{0j}}_{\text{Noise}} + 
 \underbrace{ \int_0^t e^{{(t - s)}\mathcal{L}} \mathcal{P}\mathcal{L} e^{s \mathcal{Q} \mathcal{L}} \MC{QL}\phi_{0j}ds}_{\text{Memory}}.
\end{equation}
The system described in Eq.~\ref{eq:M-Z_Identity} is precisely equivalent to the original ODE system. The authors note that, although Langevin equations are traditionally thought of as stochastic differential equations, Eq.~\ref{eq:M-Z_Identity} is deterministic. Equation~\ref{eq:M-Z_Identity} demonstrates that coarse-graining leads to non-local (in time) effects, which are referred to as memory. For notational purposes, define
\begin{equation}\label{eq:orthoNotation}
F_j(\phi_0,t) = e^{t\MC{QL}}\MC{QL}\phi_{0j}, \qquad K_j(\phi_0,t) = \MC{PL}F_j(\phi_0,t).
\end{equation}
By definition, $F_j(\phi_0,t)$ satisfies the orthogonal dynamics equation
\begin{equation}\label{eq:orthogonalDynamicsEq}
\frac{\partial}{\partial t} F_j(\phi_0,t) = \MC{QL}F_j(\phi_0,t),
\end{equation}
where $F_j(\phi_0,0) = \MC{QL}\phi_{0j}$. It can be shown that solutions to the orthogonal dynamics equation are in the null space of $\MC{P}$, meaning $\MC{P}F_j(\phi_0,t) = 0$. Further simplifications can be achieved by applying
the projection $\MC{P}$ to Eq.~\ref{eq:M-Z_Identity} to eliminate the dependence on the noise term,
\begin{equation}\label{eq:M-Z_Identity3}
\frac{\partial}{\partial t} \MC{P}\phi_j(\phi_0,t) =  \MC{P} R_j(\hat{\phi}(\phi_0,t)) +  
 \MC{P} \int_0^t K_j \big(\hat{\phi}(\phi_0,t-s),s \big)ds.
\end{equation}
Equations~\ref{eq:orthogonalDynamicsEq} and~\ref{eq:M-Z_Identity3} provide a set of equations for the resolved modes $\mathbf{\hat{\phi}}$ that only depend on $\mathbf{\hat{\phi}}$. In this sense, Eq.~\ref{eq:M-Z_Identity3} is closed. Evaluation of the memory kernel is, however, not tractable as it involves the solution of a high dimensional partial differential equation (which itself depends on both resolved and unresolved modes). Instead, Eq.~\ref{eq:M-Z_Identity3} provides a starting point for the derivation of mathematically-based closure models. It is additionally worth noting that the derivation presented above relies on the projection operators $\MC{P}$ and $\MC{Q}$ being orthogonal to each other. In the context of traditional LES, this can be conceptualized  as a sharp cutoff filter. The above discussion does not apply to non-orthogonal spatial filters (such as a Gaussian filter). 

\subsection{The orthogonal dynamics equation and the memory kernel}\label{ssec:ortho}
For analysis purposes, it is desirable to directly evaluate the convolution integral in Eq.~\ref{eq:M-Z_Identity3}. The evaluation of the integral is made challenging by the orthogonal dynamics equation. The orthogonal dynamics equation is an $N$-dimensional hyperbolic partial differential equation that exists in free-space. The orthogonal dynamics can be thought of as a type of $N$-dimensional linear advection equation (with spatially varying wave speeds) projected onto the space of the unresolved variables. For high dimensional problems, a numerical solution to the orthogonal dynamics is not tractable. A procedure for approximating the orthogonal dynamics by the solution of a set of ODEs was introduced in ~\cite{ParishAIAA2016} and further developed in~\cite{GouasmiMZ1}. This procedure is briefly outlined below.

Recall that the generalized Langevin equation is an $N$-dimensional PDE, but its solution may be obtained by solving a set of ordinary differential equations. This is due to the fact that $e^{t\MC{L}}$ is a transfer (or Koopman~\cite{Koopman}) operator,
$$e^{t \MC{L}} g(\phi_0) = g(\phi(\phi_0,t)).$$
Next, recall that the orthogonal dynamics is given by 
$$F_j(\phi_0,t) = e^{t \MC{QL}} \MC{QL}\phi_{0j}.$$
For  general nonlinear systems, $e^{tQ\MC{L}}$ is not a Koopman operator. In the general case, solutions to the orthogonal dynamics equation can not be obtained by solving a set of ordinary differential equations~\footnote{While $\mathcal{L}$ is the generator of a Koopman operator, $\mathcal{QL}$ is not}. To approximate the orthogonal dynamics, we make the assumption that $e^{tQ\MC{L}}$ is a Koopman operator,
$$e^{t \MC{QL}}g( \phi_0^{\MC{Q}} )= g(\phi(\phi_0^{\MC{Q}},t)),$$
where $\phi^{\MC{Q}}$ obeys what we refer to as the orthogonal ODE,
\begin{equation}\label{eq:orthoODE}
\frac{d \phi_j^{\MC{Q}}}{dt} = R_j(\{\hat{\phi}^{\MC{Q}},\tilde{\phi}^{\MC{Q}}\}) - R_j(\{\hat{\phi}^{\MC{Q}},0\}).
\end{equation}
For the observable of interest, $e^{(t-s)} \MC{PL}e^{s \MC{QL}} \MC{QL}\phi_{0j}$, the approximation  is equivalent to linearizing the original dynamical system about $\phi(t)$ and solving the resulting orthogonal dynamics equation. The memory can be evaluated by solving the orthogonal ODE using the trajectories of the resolved simulation. Step by step details are provided in Section~\ref{ssec:tmodel_shortcomings}, where the method is applied to Burgers equation. For details on the methodology, the interested reader is referred to~\cite{GouasmiMZ1}.

\section{Formulation of the Dynamic-$\tau$ Model}\label{sec:dtau}
The construction of a closure model involves approximating the memory integral in Eq.~\ref{eq:M-Z_Identity3}. Various approximation strategies exist, the most explored of which is the $t$-model. In this section, we first outline the well studied $t$-model and expose its shortcomings and necessary modifications. The resulting insight is used as the basis for the dynamic-$\tau$ model. 

\subsection{Shortcomings of the $t$-model}\label{ssec:tmodel_shortcomings}
The $t$-model~\cite{ChorinOptimalPrediction} can be derived under the guise of various mathematical assumptions. For example, the $t$-model can be derived by expanding the memory integral in a Taylor series expansion about $s=0$ or by approximating the orthogonal dynamics evolution operator ($e^{t \MC{QL}}$) with the evolution operator of the full dynamics ($e^{t \MC{L}}$). The most straightforward and physically intuitive derivation is to use a left hand side quadrature rule to approximate the memory integral using its value at $s=0$,
\begin{equation}\label{eq:tModel}
\MC{P} \int_{0}^t e^{{(t-s)}\mathcal{L}} \mathcal{P}\mathcal{L} e^{s \mathcal{Q} \mathcal{L}} \MC{QL}\phi_{0j} ds \approx t e^{t \MC{L}}\MC{PLQL}\phi_{0j}.
\end{equation}
With this approximation one does not need to solve the troublesome orthogonal dynamics equation as the dependence on $e^{t\MC{QL}}$ is eliminated. The $t$-model has been applied with varying success to a number of problems, including Burgers equation and the incompressible Navier-Stokes equations. Work by Stinis~\cite{Stinis-rMZ} has argued that the $t$-model needs an additional scaling constant to maintain accuracy,
\begin{equation}\label{eq:tModelScale}
\MC{P} \int_{0}^t e^{{(t-s)}\mathcal{L}} \mathcal{P}\mathcal{L} e^{s \mathcal{Q} \mathcal{L}} \MC{QL}\phi_{0j} ds \approx C_t t e^{t \MC{L}}\MC{PLQL}\phi_{0j}.
\end{equation}
Stinis motivates the need for these ``renormalization" constants by the idea that the full-order model that the $t$-model is approximating is, itself, under-resolved. 

Assessing the validity of the underlying assumptions of the $t$-model (as well as a renormalized $t$-model) is challenging as it requires one to directly evaluate the memory kernel. This involves solving the $N$-dimensional orthogonal dynamics equation, which is thus far intractable. Here, we use the orthogonal ODE presented in Section~\ref{ssec:ortho} to \textit{approximate} solutions to the orthogonal dynamics in an attempt to directly assess the underlying assumptions of the $t$-model and Stinis' renormalized models. Burgers equation is considered as a numerical example as it has been studied extensively by various authors. Burgers equation in Fourier space is given by
\begin{equation}\label{eq:VBE_freq}
\frac{\partial u_k}{\partial t} + \frac{\imath  k}{2} \sum_{\substack{ p + q = k \\ p ,q \in F \cup G }}u_p u_{q} = -\nu k^2 u_k, \qquad k \in F \cup G.
\end{equation}
The Fourier modes have been separated into two sets, $F$ and $G$. Modes in $F$ are resolved and modes in $G$ are unresolved. The purpose of the following discussion is to use the orthogonal ODE to approximate the memory kernel so as to understand how the modes in $G$ impact the modes in $F$. Numerical details for the simulation considered are given in Table~\ref{table:ortho1}. The initial condition in physical space is $u(x) = \sin(x)$. Again, this initial condition is considered because it has been studied extensively in~\cite{bernsteinBurgers,RenormalizedMZ}. The reconstruction of the memory kernel through the orthogonal ODE is performed as follows:
\begin{enumerate}
\item Run the full order simulation to obtain $\mathbf{u}(t_j)$, where $j$ are the number of desired quadrature points for the eventual discrete reconstruction of the memory integral.
\item From the full order simulation data, compute $K(\mathbf{u}(t_j),0) = e^{t_j \MC{L}}\MC{PLQL}\mathbf{u}_{0}$.
\item For each $t_j$, evolve the orthogonal ODE using the initial condition $K(\mathbf{u}(t_j),0)$ to obtain $K(\mathbf{u}(t_j),s_i).$
\item For each $t_j$ discretely reconstruct the memory integrand, $\mathcal{M}(t) = \int_{0}^t K(\mathbf{u}(t-s),s) ds$, using any numerical quadrature rule.
\end{enumerate}
\begin{table}%Table of simulation parameters
\begin{center}\scriptsize
\vskip -0.1in
\begin{tabular}{| c | c | c | c | c | } \hline 
L &   $\nu$ & $N_f$ & $\Delta t$ & Quadrature $\Delta t$ \\ \hline 
$2\pi$ & $5 \times 10^{-3}$ & 1028 & $1 \times 10^{-3}$ & $5 \times 10^{-3}$ \\ \hline
\end{tabular}
\caption{Physical and numerical details for reconstruction of the kernel via the orthogonal ODE.}
\label{table:ortho1}
\end{center}
\end{table}

\begin{figure}
	\begin{center}
	%\begin{subfigure}[t]{0.45\textwidth}
	%\includegraphics[trim={0cm 0cm 0cm 0cm},clip,width=1.\textwidth]{burgmem1_tm.pdf}
	%\caption{Evaluation of the approximate (normalized) memory integrand for $k=7$ at $t=1$.}
	%\label{fig:burgOrthoA1}
	%\end{subfigure}
	%\begin{subfigure}[t]{0.45\textwidth}
	%\includegraphics[trim={0cm 0cm 0cm 0cm},clip,width=1.\textwidth]{burgmem2_tm.pdf}
	%\caption{As in Figure~\ref{fig:burgOrthoA}, but at $t=3.0$.}
	%\label{fig:burgOrthoB1}
	%\end{subfigure}
	\begin{subfigure}[t]{0.45\textwidth}
	\includegraphics[trim={0cm 0cm 0cm 0cm},clip,width=1.\textwidth]{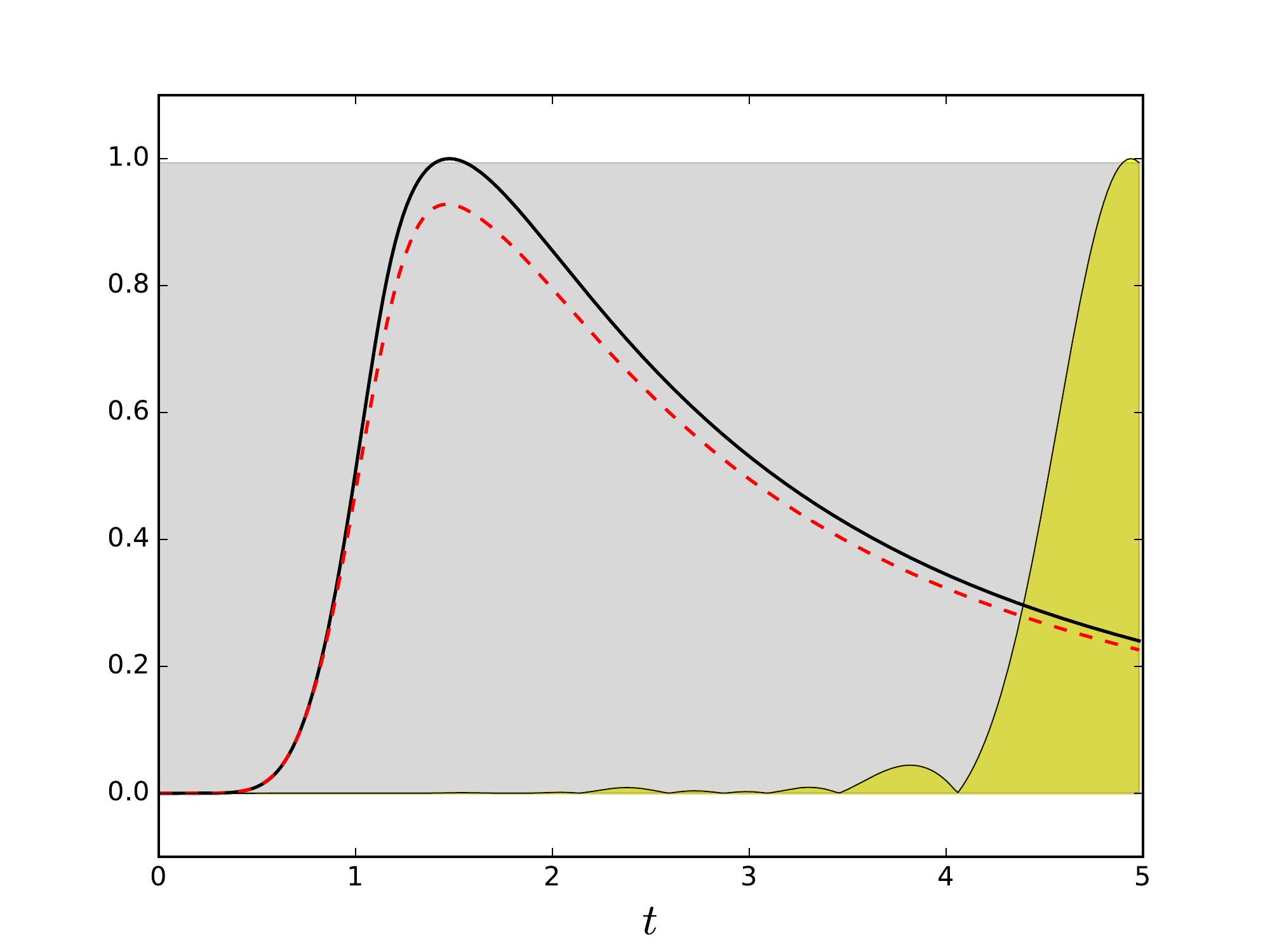}
	\caption{As in Figure~\ref{fig:burgOrthoA}, but at $t=5.0$.}
	\label{fig:burgOrthoC1}
	\end{subfigure}
	\begin{subfigure}[t]{0.45\textwidth}
	\includegraphics[width=1.\textwidth]{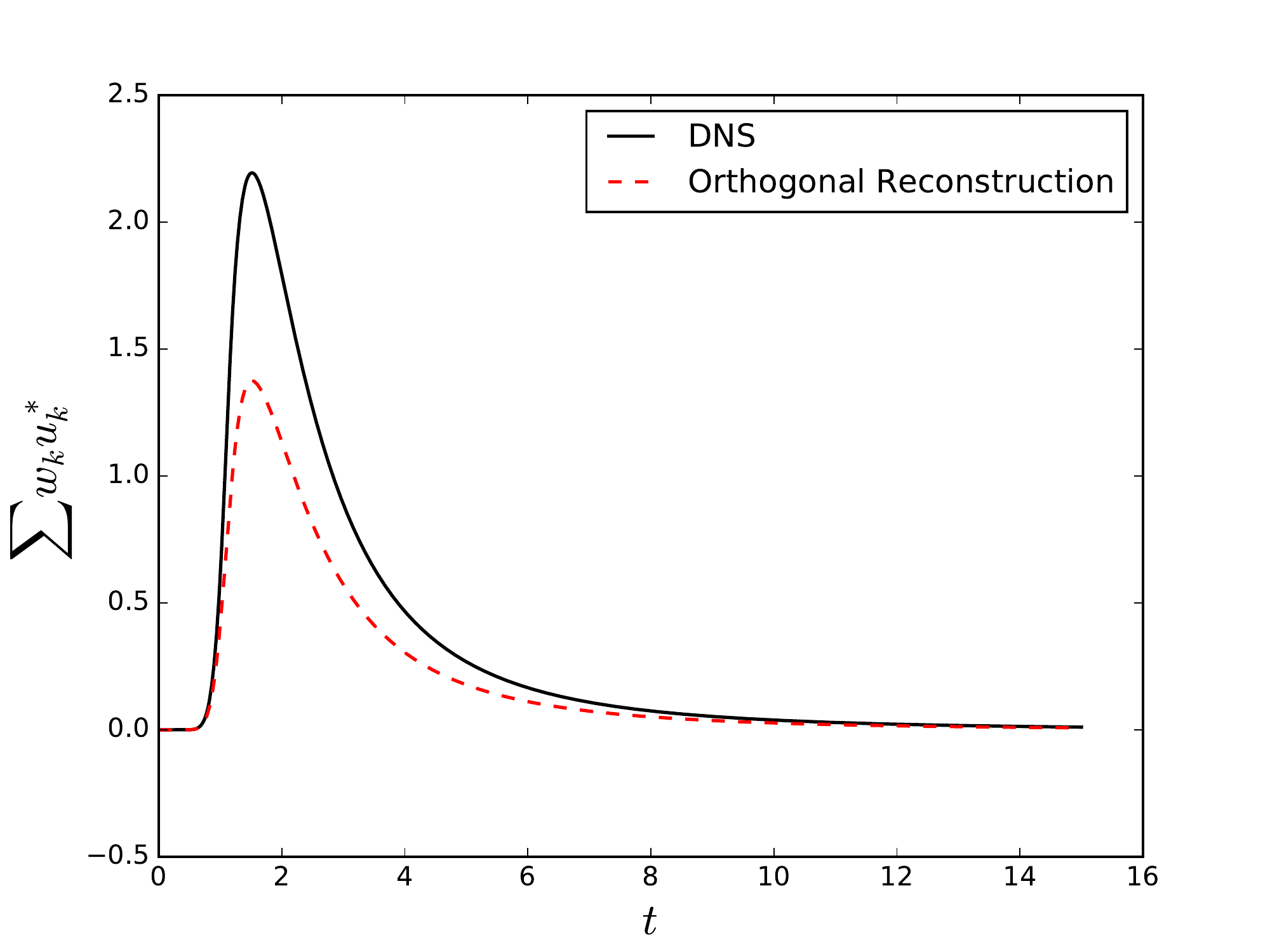}
	\caption{The total subgrid energy transfer.}
	\label{fig:burgOrthoD1}
	\end{subfigure}
	\end{center}
	\caption{Reconstruction of the memory kernel via the orthogonal ODE. }
	\label{fig:burgOrtho1}
\end{figure}

\begin{figure}
	\begin{center}
	\begin{subfigure}[t]{0.45\textwidth}
	\includegraphics[trim={0cm 0cm 0cm 0cm},clip,width=1.\textwidth]{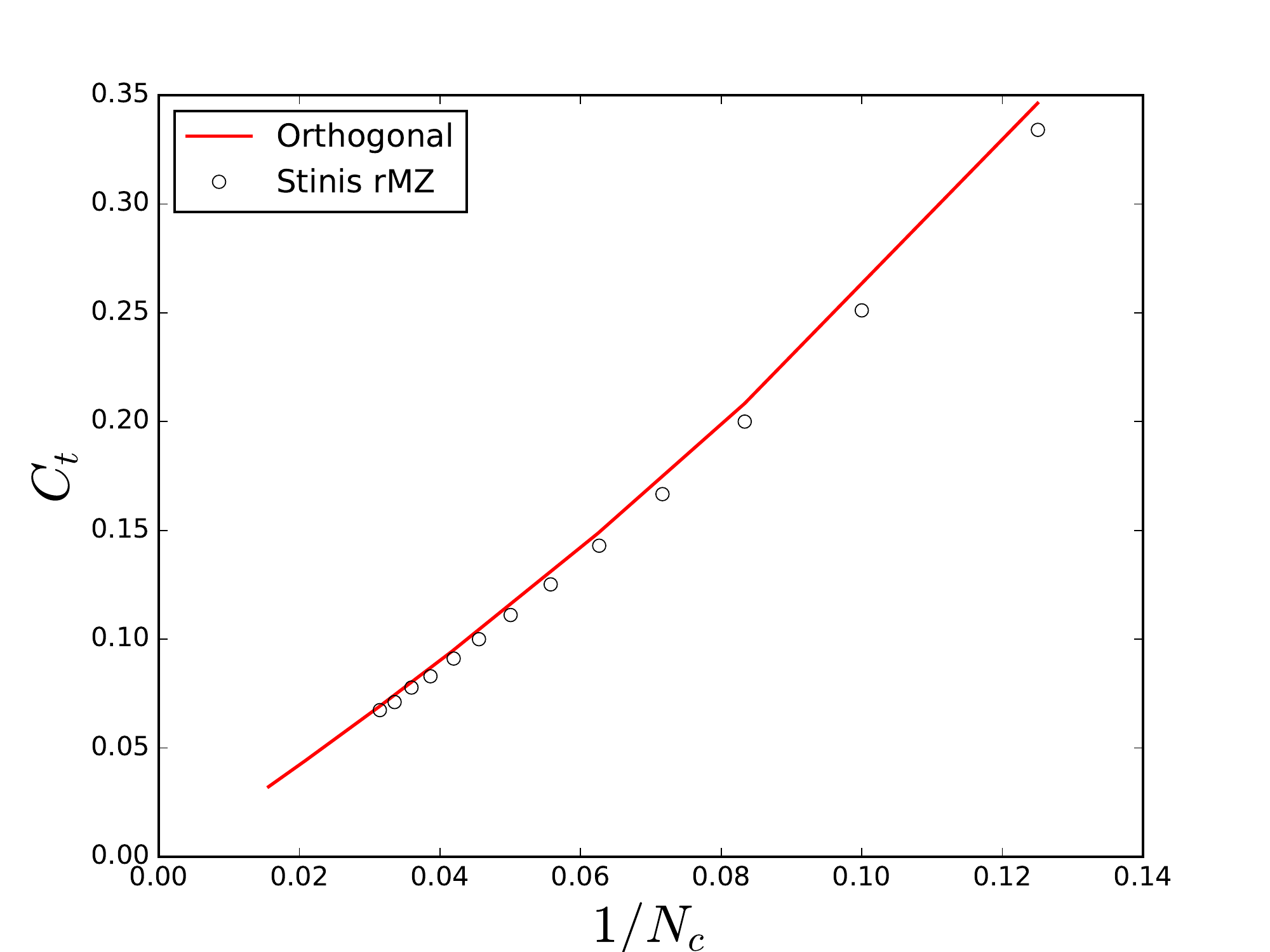}
	\caption{Ratio of the orthogonal dynamics integral to the assumed $t$-model integral for various $N_c$.}
	\label{fig:burgScalingA}
	\end{subfigure}
	\begin{subfigure}[t]{0.45\textwidth}
	\includegraphics[trim={0cm 0cm 0cm 0cm},clip,width=1.\textwidth]{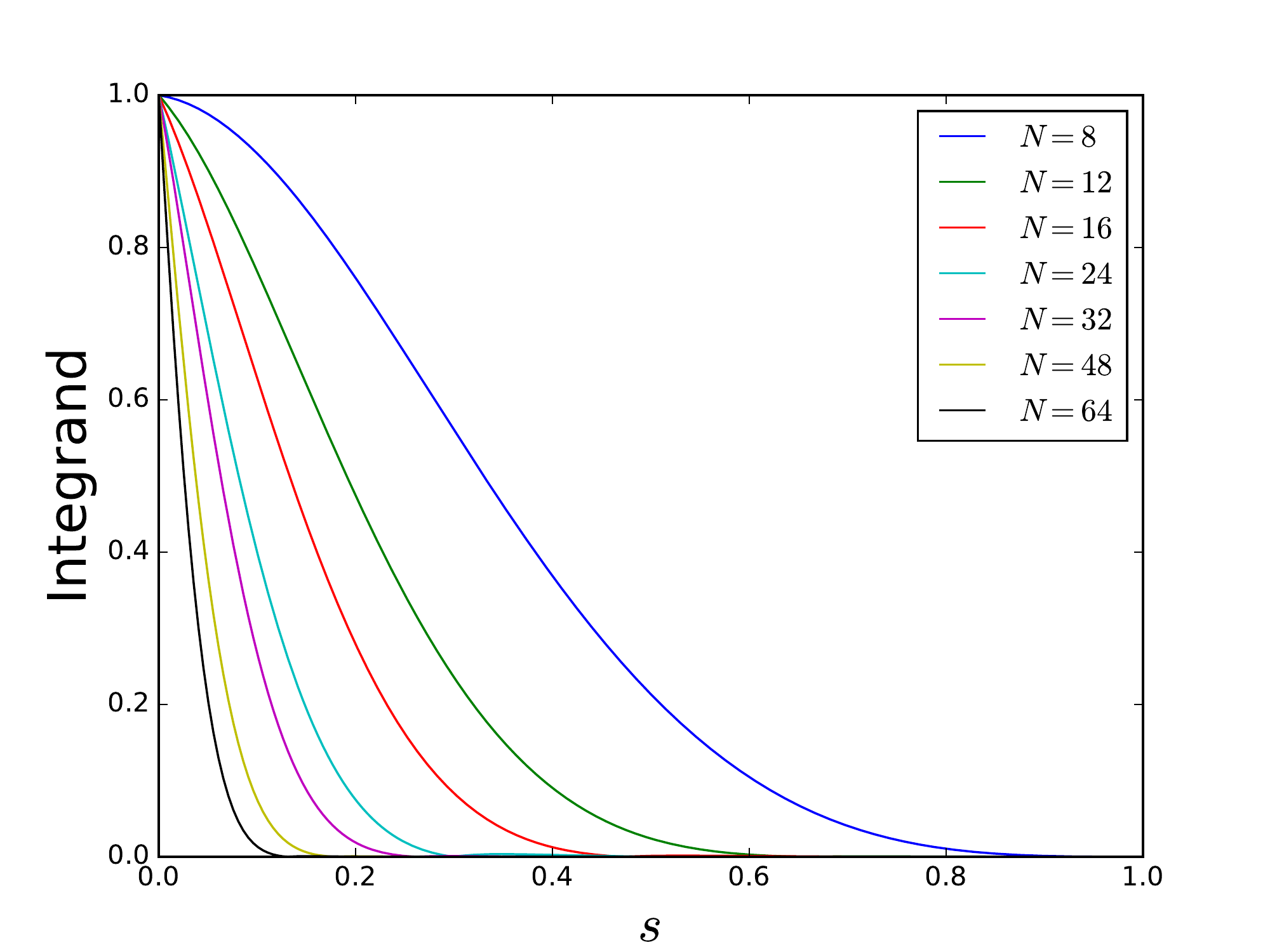}
	\caption{Normalized integrands for various $N_c$.}
	\label{fig:burgScalingB}
	\end{subfigure}
	\end{center}
	\caption{Comparison of the ratio of the true memory integral to the t-model memory integral (i.e. the ratio of the area in the shaded yellow region in Fig.~\ref{fig:burgOrthoC1} to the area in the shaded gray region) versus the renormalization coefficients obtained by Stinis.}
	\label{fig:burgScaling}
\end{figure}

Figure~\ref{fig:burgOrtho1} shows the reconstructed memory integrand at $t=5$, as well as the total subgrid energy transfer. The reconstructed terms are compared to those extracted from DNS data. The integrand assumed by the $t$-model is also shown for reference. The subgrid term given by M-Z is the area of the shaded regions. The yellow shaded region is that predicted by the orthogonal ODE procedure, while the gray shaded region is that assumed by the $t$-model. It is seen that the comparison of the memory term approximated through the orthogonal ODE to the DNS is reasonable, although some error is notably present for the total subgrid energy transfer. Nonetheless, the important observation to make from Figure~\ref{fig:burgOrtho1} is the presence of a decaying memory kernel. It is quite clear that the $t$-model will grossly overestimate the value of the integrand and that an additional scaling is required. The required scaling is the ratio of the true integral (area of the shaded yellow region) to the assumed $t$-model integral (area of the shaded gray region). Figure~\ref{fig:burgScaling} shows these ratios computed for various coarse-grained model sizes (i.e. varying the cut-off wave number $k_c$) at $t=1$. The   ratios are compared to the renormalized $t$-model coefficients as computed by Stinis. The agreement is excellent. 

This numerical evidence shows that, as suggested by Stinis, the $t$-model requires an additional scaling coefficient. 
%The need for the scaling coefficient is due to inaccuracies in the $t$-model. 
The simple left hand side quadrature rule assumed by the $t$-model is inappropriate and unjustified. To accurately model the memory, one needs to  account for the decaying kernel.

\subsection{Derivation of the dynamic-$\tau$ model}
The numerical evidence in the previous section suggests that the memory kernel has a finite support,
\begin{equation}\label{eq:FM_approx}
\MC{P} \int_{0}^t e^{{(t-s)}\mathcal{L}} \mathcal{P}\mathcal{L} e^{s\mathcal{Q} \mathcal{L}} \MC{QL}\phi_{0j} ds \approx \MC{P} \int_{t - \tau_{\MC{P}}}^t e^{{(t-s)}\mathcal{L}} \mathcal{P}\mathcal{L} e^{s \mathcal{Q} \mathcal{L}} \MC{QL}\phi_{0j} ds.
\end{equation}
In addition to numerical evidence, there is also a theoretical argument for the presence of a finite memory. For problems that reach a statistically steady state (such as a turbulent channel), the memory kernel must also reach a statistically steady state. This directly implies that any model for the memory term must not explicitly contain a time term. 

In this work, we present a Markovian model that represents non-local effects through a dynamic approximation to the memory length. The most straightforward derivation of the model is to use a single-point quadrature rule starting from $s=0$ to approximate the memory integrand and assume a temporal support of $\tau_{\MC{P}}$. The finite memory assumption implies that $K(\hat{\phi}(t -\tau),\tau) = 0$. In this case, the memory kernel can be simplified as  
\begin{equation}\label{eq:tauModel}
\MC{P} \int_{t - \tau_{\MC{P}}}^t e^{{(t-s)}\mathcal{L}} \mathcal{P}\mathcal{L} e^{s \mathcal{Q} \mathcal{L}} \MC{QL}\phi_{0j} ds \approx C_q \tau_{\MC{P}} (t) e^{t \MC{L}}\MC{PLQL}\phi_{0j},
\end{equation}
where $C_q$ is the constant associated with the quadrature rule. For example, a trapezoidal quadrature would assume $C_q=0.5$. One could alternatively assume the memory decays in an exponential fashion, sinusoidal fashion, etc. 
Eq.~\ref{eq:tauModel} will be referred to as the $\tau$-model. As in the $t$-model, the key simplification in Eq.~\ref{eq:tauModel} is that the dependence on the orthogonal dynamics has been removed. Additionally, note that the $\tau$-model can recover the $t$-model if $C_q\tau_{\MC{P}} = t$.
%Note that, while we have derived the $\tau$-model based on a trapezoidal quadrature with a finite memory, this physical interpretation of the model should not be taken too literally. The accuracy of the $\tau$-model relies on the integral in Eq.~\ref{eq:tauModel} being well correlated to its value at $s=0$. This will be further commented on at the end of this section. 
%For completeness, the t-model is also described. If one assumes that the orthogonal component of the flow $e^{t \MC{QL}} %\approx e^{t\MC{L}}$, then the memory integral in Eq.~\ref{eq:M-Z_Identity} simplifies to
%\begin{equation}\label{eq:tModel}
%\MC{P} \int_{0}^t e^{{s}\mathcal{L}} \mathcal{P}\mathcal{L} e^{(t-s) \mathcal{Q} \mathcal{L}} \MC{QL}\phi_{0j} ds \approx t e^{t %\MC{L}}\MC{PLQL}\phi_{0j}.
%\end{equation}
%%At first glance, the t-model appears flawed. 
%In the limiting case when the resolution of the coarse-grained model goes to zero one must have $e^{t \MC{QL}} \rightarrow %e^{t \MC{L}}$ and the assumptions of the t-model are justified.  The $\tau$-model can be considered to be a generalization of %the $t$-model if $\tau_\MC{P} = t$, although it is derived in a different setting.

The model described in Eq.~\ref{eq:tauModel} requires the selection of the memory length $\tau_{\MC{P}}$.  Here a dynamic procedure using the Germano identity~\cite{Germano} is derived for the $\tau$-model. To proceed, decompose the resolved variable $\hat{\phi}$ into two sets such that
$$\hat{\phi} = \{\overline{\phi}, \phi'\}, \qquad \phi = \{ \overline{\phi}, \phi', \tilde{\phi} \}.$$ 
Further, define the sharp cutoff filters $\hat{\MC{G}}$ and $\overline{\MC{G}}$ that satisfy
$$\hat{\MC{G}}\phi = \{ \hat{\phi} , 0 \} = \{ \overline{\phi}, \phi' , 0 \}  \qquad \overline{\MC{G}}\phi = \{ \overline{\phi} , 0 , 0\}.$$
Note that these filters are not Zwanzig projection operators (i.e. $\hat{\MC{G}}f(\phi) \ne \overline{f}(\hat{\MC{G}}{\phi})$) and they do not commute with non-linear functions; instead they act as a traditional sharp spectral cutoff filter. To derive an expression for $\tau_{\MC{P}}$, note the following identity for the specified projection operators with fully resolved initial conditions, 
%\begin{equation}\label{eq:memResidEquiv}
% \int_0^t K_j(\hat{\mathbf{\phi}}(s),t-s)ds = e^{t \MC{L}} \big( \MC{QL} \phi_{0j} - e^{-t \MC{L}} e^{t \MC{QL}} \MC{QL} \phi_{0k} \big).
%\end{equation}
%For the zero-variance projection $e^{t \MC{QL}} \MC{QL} \phi_{0k} = 0$ and the typical unclosed equations of LES are obtained,
\begin{equation}\label{eq:memResidEquiv}
\MC{P} \int_0^t K_j(\hat{\mathbf{\phi}}(s),t-s)ds =  \hat{\MC{G}}R_j(\mathbf{\phi}) - R_j(\hat{\MC{G}}\mathbf{\phi}).
\end{equation}
Applying the "test" filter $\overline{\MC{G}}$ to Eq.~\ref{eq:memResidEquiv} and adding and subtracting $R_j(\MC{\hat{G}}\mathbf{\phi})$ yields
%\begin{comment}
%\begin{equation}\label{eq:MZGermano}
%\MC{G'}\MC{P}\int_0^t K_j(\hat{\mathbf{\phi}}(s),t-s)ds =   \big[ \MC{\overline{G}\hat{G}}R_j(\mathbf{\phi}) - R_j(\MC{\overline{G}\hat{G}}\mathbf{\phi}) \big] + %\big[ R_j(\MC{\overline{G}\hat{G}}\mathbf{\phi}) - \MC{\overline{G}}R_j(\MC{\hat{G}}\mathbf{\phi}) \big].
%\end{equation}
%Using the identities $\MC{\overline{G}\hat{G}} f(\phi) = \MC{\overline{G}} f(\phi)$ and $\MC{\overline{G}}\phi = \overline{\phi}$,
%\end{comment}
\begin{equation}\label{eq:MZGermano2}
\MC{\overline{G}}\MC{P} \int_0^t K_j(\hat{\mathbf{\phi}}(s),t-s)ds =   \big[ \MC{\overline{G}}R_j(\mathbf{\phi}) - R_j(\overline{G}{\mathbf{\phi}}) \big] + \big[ R_j(\overline{G}{\mathbf{\phi}}) - \MC{\overline{G}}R_j(\hat{G}{\mathbf{\phi}}) \big].
\end{equation}
Equation~\ref{eq:MZGermano2} is a statement of the Germano identity. The first term on the RHS is simply the unclosed term that arises from coarse-graining at a level $\MC{\overline{G}}$. The second bracketed term on the RHS can be computed from the under-resolved simulation. The memory length $\tau_{\MC{P}}$ can be approximated by using the $\tau$-model to model the LHS and the first term on the RHS,
\begin{equation}\label{eq:MZGermanoTau}
\MC{\overline{G}} e^{t \MC{L}} C_q \tau_{\MC{P}} \MC{PLQL} \phi_{0j}=  e^{t \MC{L}}C_q \tau_{\MC{\overline{P}}} \MC{\overline{P}L\overline{Q}L} \phi_{0j} + \big[ R_j(\overline{\MC{G}}{\mathbf{\phi}}) - \MC{\overline{G}}R_j(\hat{\MC{G}}{\mathbf{\phi}}) \big],
\end{equation}
where $\MC{\overline{P}}$ and $\MC{\overline{Q}}$ are the corresponding Zwanzig projection/remainder operators for coarse-graining at the level $\MC{\overline{G}}$ (i.e. $\MC{\overline{P}}f(\phi) = f(\overline{\phi})$), $\tau_{\MC{P}}$ is the time-scale for coarse-graining at the level $\MC{P}$, and $\tau_{\MC{\overline{P}}}$ is the time-scale for coarse-graining at the level $\MC{\overline{P}}$. 

\subsection{Scaling Laws for the Memory Length}\label{sec:burgScaling}
\begin{figure}
	\begin{center}
	\includegraphics[trim={0cm 0cm 0cm 0cm},clip,width=0.6\textwidth]{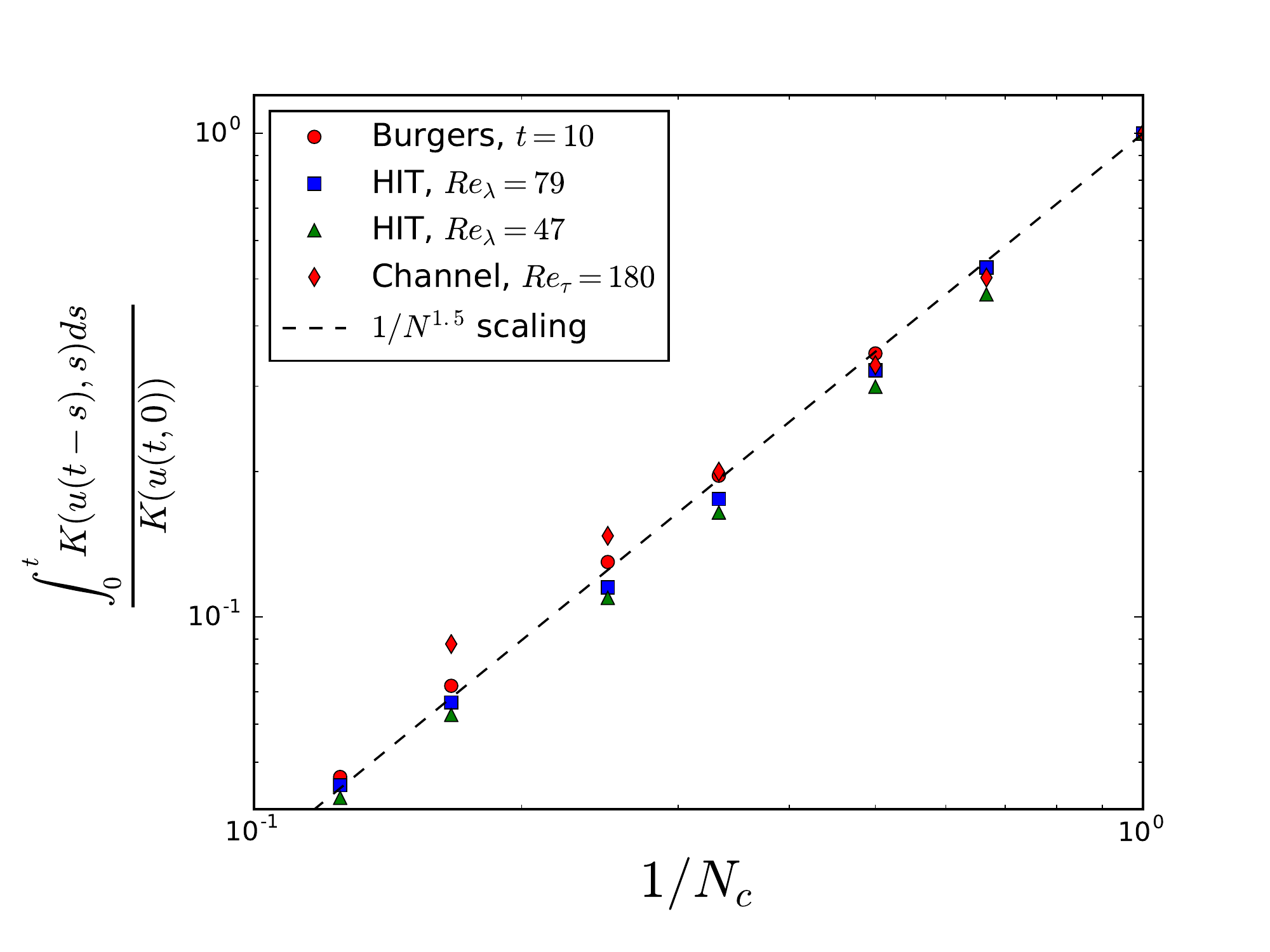}
	\caption{Ratio of the memory kernel integral (as extracted from the DNS) to the value of the kernel at $s=0$ for Burgers equation, homogeneous turbulence, and channel flow. All terms are normalized. Note that the integrated kernel can be computed from the full-order simulation as $e^{t \MC{L}}\MC{QL}u_{0j}$.}
	\label{fig:1.5Scaling}
	\end{center}
\end{figure}

To close Eq.~\ref{eq:MZGermanoTau}, a constitutive relation between the time-scales $\tau_{\MC{P}}$ and $\tau_{\MC{\overline{P}}}$ needs to be established. It is seen in Figure~\ref{fig:burgScaling} that decreasing the size of the coarse-grained model increases the memory length. This increase is seen to obey a power-law scaling. Figure~\ref{fig:1.5Scaling} shows the ratio of the integrated kernel to that of the memory kernel at $s=0$ for Burgers equation, homogeneous turbulence at several Reynolds numbers, and channel flow. Remarkably, in all cases, the memory-length is seen to approximately obey a $1/N^{1.5}$ scaling.  
As such, the relationship between $\tau_{\MC{P}}$ and $\tau_{\MC{\overline{P}}}$ is straightforward,
\begin{equation}\label{eq:tauRelation}
\tau_{\MC{\overline{P}}} = \bigg(\frac{\Delta_{\MC{P}}}{\Delta_{\MC{\overline{P}}}} \bigg)^{1.5}  \tau_{\MC{P}},
\end{equation}
where $\Delta$ is the corresponding filter scale. The resulting equation for the memory length, $\tau_{\MC{P}}$, is 
\begin{equation}\label{eq:MZGermanoTau2}
\tau_{\MC{P}}= \frac{1}{C_q} \frac{ R_j(\overline{\MC{G}}{\mathbf{\phi}}) - \MC{\overline{G}}R_j(\hat{\MC{G}}{\mathbf{\phi}})}{\MC{\overline{G}} e^{t \MC{L}} \MC{PLQL} \phi_{0j} - \big( \frac{\Delta_{\MC{P}}}{\Delta_{\MC{\overline{P}}}} \big)^{1.5} e^{t \MC{L}}   \MC{\overline{P}L\overline{Q}L} \phi_{0j}}.
\end{equation}
It is noted that Eq.~\ref{eq:MZGermanoTau2} still contains the unknown quadrature constant, $C_q$. This term, however, vanishes when Eq.~\ref{eq:MZGermanoTau2} is injected into Eq.~\ref{eq:tauModel}.

\subsection{Energy Transfer Constraint}
Equation~\ref{eq:MZGermanoTau2} is an over-determined system for $\tau_\MC{P}$. It is unlikely that one scalar $\tau_{\MC{P}}$ will exist that can satisfy Eq.~\ref{eq:MZGermanoTau2} for each mode. As in the Dynamic Smagorinsky model, Eq.~\ref{eq:MZGermanoTau2} is understood to be valid in an average sense only. To constrain the equation for a single scalar $\tau_{\MC{P}}$, the coarse-graining time-scale is estimated by equating the energy transferred by the subgrid model out of the modes $\overline{\phi}$,
\begin{equation}\label{eq:tau_equation}
\tau_{\MC{P}} = \frac{1}{C_q}\frac{ \mathbb{R} \bigg( \sum \overline{\phi}_j^*  \big[ R_j(\overline{\mathbf{\phi}}) - \MC{\overline{G}}R_j(\hat{\mathbf{\phi}}) \big] \bigg)} { \mathbb{R} \bigg( \sum  \overline{\phi}_j^* e^{t \MC{L}}\MC{PLQL}\phi_{0j} -  \big( \frac{\Delta_{\MC{P}}}{\Delta_{\MC{\overline{P}}}} \big)^{1.5} \sum  \overline{\phi}_j^*e^{t \MC{L}} \MC{\overline{P}L\overline{Q}L}\phi_{0j} \bigg)}.
\end{equation}
With Eqns~\ref{eq:tauModel} and \ref{eq:tau_equation}, the dynamic model for the memory integral is closed. The final expression for the closure model is
\begin{equation}\label{eq:dtauModelF}
\MC{P} \int_{0}^t e^{{s}\mathcal{L}} \mathcal{P}\mathcal{L} e^{(t-s) \mathcal{Q} \mathcal{L}} \MC{QL}\phi_{0j} ds \approx \frac{ \mathbb{R} \bigg( \sum \overline{\phi}_j^*  \big[ R_j(\overline{\mathbf{\phi}}) - \MC{\overline{G}}R_j(\hat{\mathbf{\phi}}) \big] \bigg)} { \mathbb{R} \bigg( \sum  \overline{\phi}_j^* e^{t \MC{L}}\MC{PLQL}\phi_{0j} -  \big( \frac{\Delta_{\MC{P}}}{\Delta_{\MC{\overline{P}}}} \big)^{1.5} \sum  \overline{\phi}_j^*e^{t \MC{L}} \MC{\overline{P}L\overline{Q}L}\phi_{0j} \bigg)} e^{t \MC{L}} \MC{PLQL}\phi_{0j}.\end{equation}
Note that $C_q$ is canceled out in the dynamic procedure. The key assumptions in the dynamic-model are that the memory integral is well correlated to its value at $s=0$ (Eq.~\ref{eq:tauModel}) and that the scaling law (Eq.~\ref{eq:tauRelation}) holds. The stronger of these two assumptions is that the memory integral is well correlated to its value at $s=0$. For increasingly under-resolved simulations, the orthogonal dynamics is expected to grow in complexity and the Markovian approximation to the integral may not be sufficient.
\subsection{Evaluation of $e^{t \MC{L}}\MC{PLQL}\phi_{0j}$}\label{sec:numericalEvaluation}
The dynamic-$\tau$ model (as well as many other M-Z-based models) requires the evaluation of $e^{t \MC{L}}\MC{PLQL}\phi_{0j}$. This term can be computed by analytically evaluating the Liouville and projection operators (note that $\MC{PLQL}$ is applied to the initial conditions and hence the evaluation is simply an exercise in algebra; examples can be found in~\cite{StatisticalMechanics}). The naive analytic evaluation of such terms, however, is extremely tedious for complex systems of equations. The complexity of the derivations can be greatly simplified by recognizing that application of the Liouville operator to a vector $\mathbf{v}(\mathbf{\phi}_0)$ can be written as a matrix vector product,
\begin{equation}\label{eq:liouville_matvec}
\mathcal{L}\mathbf{v} = \frac{\partial  \mathbf{v}}{\partial \mathbf{\phi_0} } \mathbf{R}(\mathbf{\phi_0}  ).
\end{equation}
Evaluating the Jacobian of $\mathbf{v}$ and matrix vector product in Eq.~\ref{eq:liouville_matvec} can be costly and tedious. Fortunately, the only quantity of interest is the final matrix vector product, which is simply the Fr\'echet derivative of $\mathbf{v}$ in the direction of $\mathbf{R}$,
\begin{equation}\label{eq:LFdirec}
\MC{L}\mathbf{v} = \nabla_{R(\mathbf{\phi_0})} \mathbf{v} =\lim_{\epsilon \rightarrow 0} \frac{\mathbf{v} \big(\mathbf{\phi_0} + \epsilon \mathbf{R}(\mathbf{\phi_0}) \big) - \mathbf{v}(\mathbf{\phi_0}) }{\epsilon}.
\end{equation}

Two techniques are considered to compute the Fr\'echet derivative. The most straightforward approach is to approximate the derivative with a first order finite difference,
$$\mathcal{L}\mathbf{v}(\mathbf{\phi}_0) \approx   \frac{\mathbf{v} \big(\mathbf{\phi_0} + \epsilon \mathbf{R}(\mathbf{\phi_0}) \big) - \mathbf{v}(\mathbf{\phi_0}) }{\epsilon}.$$
%The first order finite difference approximation of the Fr\'echet derivative is computationally simple as it only involves two evaluations of the vector $\mathbf{v}$. 
%
%Consider a vector $F(\mathbf{\phi_0})$ that exists in the space of the initial conditions. Application of the Liouville operator to $F$ requires derivatives with respect to each initial condition,
%\begin{equation}\label{eq:LFdef}
%\MC{L}F(\mathbf{\phi_0},0) = \sum_{j=1}^N R_j(\mathbf{\phi_0}) \frac{\partial}{\partial \phi_{0j}} F.
%\end{equation}
%The naive approach to numerically evaluating the Liouville operator is to numerically differentiate $F$ with respect to each degree of freedom in the initial conditions, $\phi_{0j}$. One can realize, however, that application of the Liouville operator to $F$ is simply the directional derivative of $F$ in the direction of $R(\mathbf{\phi}_0)$,
%
%where $\vec{R}(\mathbf{\phi_0})$ is a unit vector. 
%Application of the Liouville operator to the vector $F(\mathbf{\phi_0})$ can be numerically evaluated with one sensitivity evaluation. 

To evaluate $\MC{PLQL}\phi_{0j}$ using the first order finite difference approximation for the specified projection operators, define the vector field $\mathbf{v}$ to be
\begin{equation}\label{eq:QL_init}
\mathbf{v}(\mathbf{\phi}_0) = \mathcal{QL} \mathbf{\phi}_0 = \mathbf{R}(\mathbf{\phi}_0) -  \mathbf{R}(\hat{\MC{G}}\mathbf{{\phi}}_0).
\end{equation}
Applying the projection operator and the finite difference approximation of the Liouville operator to Eq.~\ref{eq:QL_init} yields
\begin{equation}\label{eq:PLQL_directional}
\MC{PLQL}\phi_{0j} = \bigg[ \frac{R_j \big(\hat{\MC{G}}\mathbf{{\phi}_0} + \mathbf{R}(\mathbf{\hat{\MC{G}}{\phi}_0}) \big) - R_j \big( \hat{G} \big[ \mathbf{\hat{\MC{G}}{\phi}_0} + \epsilon \mathbf{R}(\mathbf{\hat{\MC{G}}{\phi}_0}) \big] \big) }{\epsilon} - \frac{R_j \big(\mathbf{\hat{\MC{G}}{\phi}_0}\big) - R_j \big( { \mathbf{\hat{\MC{G}}{\phi}_0} } \big) }{\epsilon} \bigg].
\end{equation}
Noting that the last term on the right hand side is zero, one obtains
\begin{equation}\label{eq:PLQL_directional2}
e^{t \MC{L}}\MC{PLQL}\phi_{0j} = \frac{R_j \big(\hat{\MC{G}}\mathbf{{\phi}}(t) + \epsilon \mathbf{R}(\hat{\MC{G}}\mathbf{{\phi}}(t)) \big) - R_j \big( \hat{G} \big[ \hat{\MC{G}}\mathbf{{\phi}}(t) + \epsilon \mathbf{R}(\mathbf{\hat{\MC{G}}{\phi}}(t)) \big] \big) }{\epsilon} .
\end{equation}
The practical evaluation of Eq.~\ref{eq:PLQL_directional2} is as follows:
\begin{enumerate}
\item Evaluate the right hand side residual, $ \mathbf{R}(\hat{\MC{G}}\mathbf{{\phi}})$. This evaluation has to be performed in an enriched space.
\item Compute the filtered quantity $\hat{G} \big[  \hat{\MC{G}}{\mathbf{\phi}} +  \mathbf{R}(\hat{\MC{G}}\mathbf{{\phi}}) \big]$
\item Evaluate the right hand side residual with arguments $\hat{\MC{G}}{\mathbf{\phi}} +  \mathbf{R}(\hat{\MC{G}}\mathbf{{\phi}}) $  This has to be performed in an enriched space, although only the first $M$ residual components are sought.
\item Evaluate the right hand side residual in the full order space with the filtered arguments $\hat{G} \big[  \hat{\MC{G}}{\mathbf{\phi}} +  \mathbf{R}(\hat{\MC{G}}\mathbf{{\phi}}) \big]$. This evaluation can be performed in the reduced order space and, again, only the first $M$ components of the residual are sought.
\item Evaluate $e^{t \MC{L}}\MC{PLQL}\phi_{0j}$ via Eq.~\ref{eq:PLQL_directional2}
\end{enumerate}
The finite difference approximation to $\MC{PLQL}\phi_{0j}$ is attractive due to its simplicity. Further, the procedure can be used to evaluate higher order terms (i.e. $\MC{PL}(\MC{QL})^n \phi_{0j}$) through recursive differentiation of functions, which provides a potential pathway to the generation of automated M-Z-based closures. Several issues arise, however. First, the selection of the finite difference step size $\epsilon$ can be problematic. While the authors have not found the finite difference approximation to be overly sensitive to $\epsilon$ for the problems considered, it is still a point of concern (especially for the evaluation of higher order terms). The second issue with the above approach is the cost. Evaluation of $e^{t \MC{L}}\MC{PLQL}\phi_{0j}$ requires three residual evaluations, several of which are in an enriched space. We do note that this cost can be reduced for certain problems due to symmetries (this will be seen later). Nonetheless, this approach increases the computational cost of a reduced model significantly. 

The second approach considered to evaluate $\MC{PLQL}\phi_{0j}$ is to compute the exact Fr\'echet derivative. The exact derivative can be computed by linearizing the functional form produced by $\mathcal{QL}\mathbf{\phi_0}$, and then evaluating the resulting function at $R(\hat{\MC{G}}\mathbf{{\phi}})$. An example is provided for Burgers equation in Appendix A. While more tedious, this approach is attractive in that the exact functional form of the model is obtained and it is computationally more affordable. The model evaluation requires $N-M$ components of an enriched space evaluation as well as an additional linearized right hand side evaluation that is comparable in cost to step 2 in the finite difference approach. In addition to being more affordable than the finite difference approach, many terms in the linearized right hand side evaluation become zero such that the cost of the linearized right hand side evaluation can be decreased substantially.

\section{Application to Burgers equation}\label{sec:burg}
In this section the dynamic-$\tau$ model is applied to Burgers equation. The memory length predicted by the dynamic model is compared to the memory length computed by directly evaluating the memory kernel via solutions of the auxiliary orthogonal ODE system.

The viscous Burgers equation in Fourier space is given by
\begin{equation}\label{eq:VBE_freqA}
\frac{\partial u_k}{\partial t} + \frac{\imath  k}{2} \sum_{\substack{ p + q = k \\ p ,q \in F \cup G }}u_p u_{q} = -\nu k^2 u_k, \qquad k \in F \cup G
\end{equation}
with $u_k(0) = u_{0k}$. The Fourier modes $u = \{\hat{u},\tilde{u} \}$ are contained within the union of two sets, $F$ and $G$. In the construction of the reduced order model, the resolved modes are $\hat{u} \in F$ and the unresolved modes are $\tilde{u} \in G$. Partitioning Eq.~\ref{eq:VBE_freqA} into the resolved and unresolved sets, the evolution equation for the resolved variables is written as
\begin{equation}\label{eq:LESVBE_freqA}
\frac{\partial {u}_k}{\partial t} + \frac{\imath k}{2} \sum_{\substack{ p + q = k \\ p \in F ,q \in F }}{u}_{p}{u}_{q} = -\nu k^2 {u}_k -  \frac{\imath k}{2}\bigg( \sum_{\substack{ p + q = k \\ p \in G ,q \in G }}{u}_{p} {u}_{q} +   \sum_{\substack{ p + q = k \\ p \in F ,q \in G }}{u}_{p} {u}_{q}  +\sum_{\substack{ p + q = k \\ p \in G ,q \in F }}{u}_{p} {u}_{q}  \bigg) \qquad k \in F . 
\end{equation}
Eq.~\ref{eq:LESVBE_freqA} is equivalent to the LES form of the Burgers equation with a sharp spectral cutoff filter. The last term on the RHS of Eq.~\ref{eq:LESVBE_freqA} contains the effect of the unresolved scales on the resolved scales and must be modeled. The dynamic-$\tau$-model takes the form
\begin{equation}\label{eq:tau_burg}
-\frac{\imath k}{2}\bigg( \sum_{\substack{ p + q = k \\ p \in G ,q \in G }}{u}_{p} {u}_{q} +   \sum_{\substack{ p + q = k \\ p \in F ,q \in G }}{u}_{p} {u}_{q}  +\sum_{\substack{ p + q = k \\ p \in G ,q \in F }}{u}_{p} {u}_{q}  \bigg) 
\approx
 -0.5 \tau_{\MC{P}} \imath k \sum_{\substack{ p + q = k \\ p \in F ,q \in G }}u_{p}  \bigg[-\frac{\imath q}{2}\sum_{\substack{ r + s = q \\ r,s \in F  }}u_{r} u_{s} \bigg] 
 \qquad k \in F
 \end{equation}
where $\tau_{\MC{P}}$ is determined through the dynamic procedure described in Section~\ref{sec:dtau}. Note that the RHS of Eq.~\ref{eq:tau_burg} was derived by analytically evaluating $\MC{PLQL}u_{0j}$ (see Appendix A), but it can also be evaluated numerically.

The initial condition considered in the numerical example is $u(x,0) = \sin(x)$ with $x \in [0,2\pi)$. The viscosity is taken to be $\nu = 5e-3$, which leads to a standing shock slightly after $t=1$. The low viscosity allows for the solution to be mostly inviscid but keeps the PDE from being singular. 
The reduced model is taken to be of size $N = 16$ (corresponding to a cutoff frequency at $k=8$), while the full order model assumes $N = 1024$. Note that, due to symmetries in Burgers equation, one only needs to take the unresolved wave numbers $q \in G$ to have  $2N$ support when evaluating Eq.~\ref{eq:tau_burg}. The subgrid model must be capable of removing the energy transferred to the high frequency modes resulting from the shock. The simulation is evolved until $t=15$. 

\begin{figure}
	\begin{center}
	\begin{subfigure}[t]{0.45\textwidth}
	\includegraphics[width=1.\textwidth]{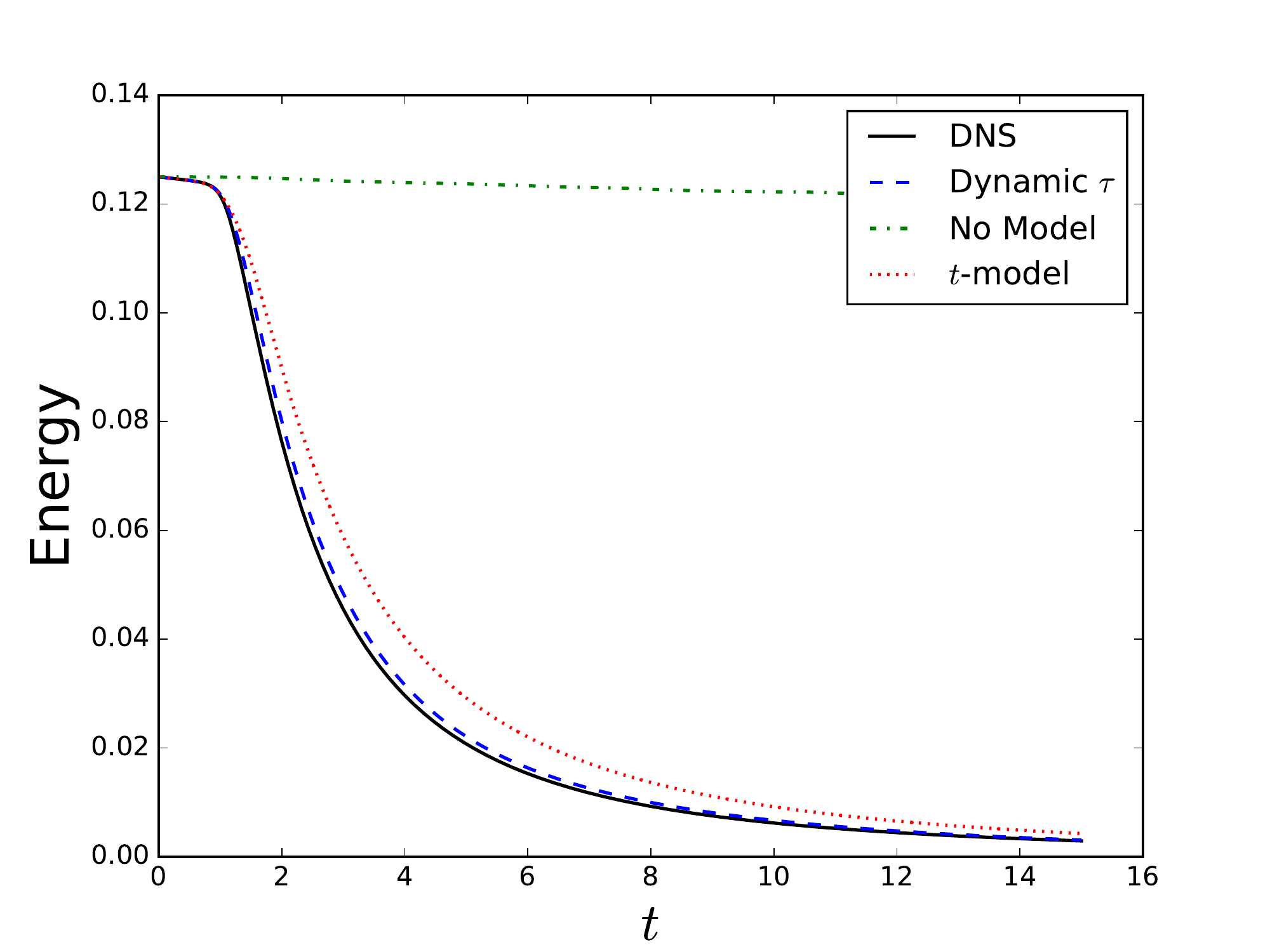}
	\caption{Total kinetic energy in the resolved modes.}
	\end{subfigure}
	\begin{subfigure}[t]{0.45\textwidth}
	\includegraphics[width=1.\textwidth]{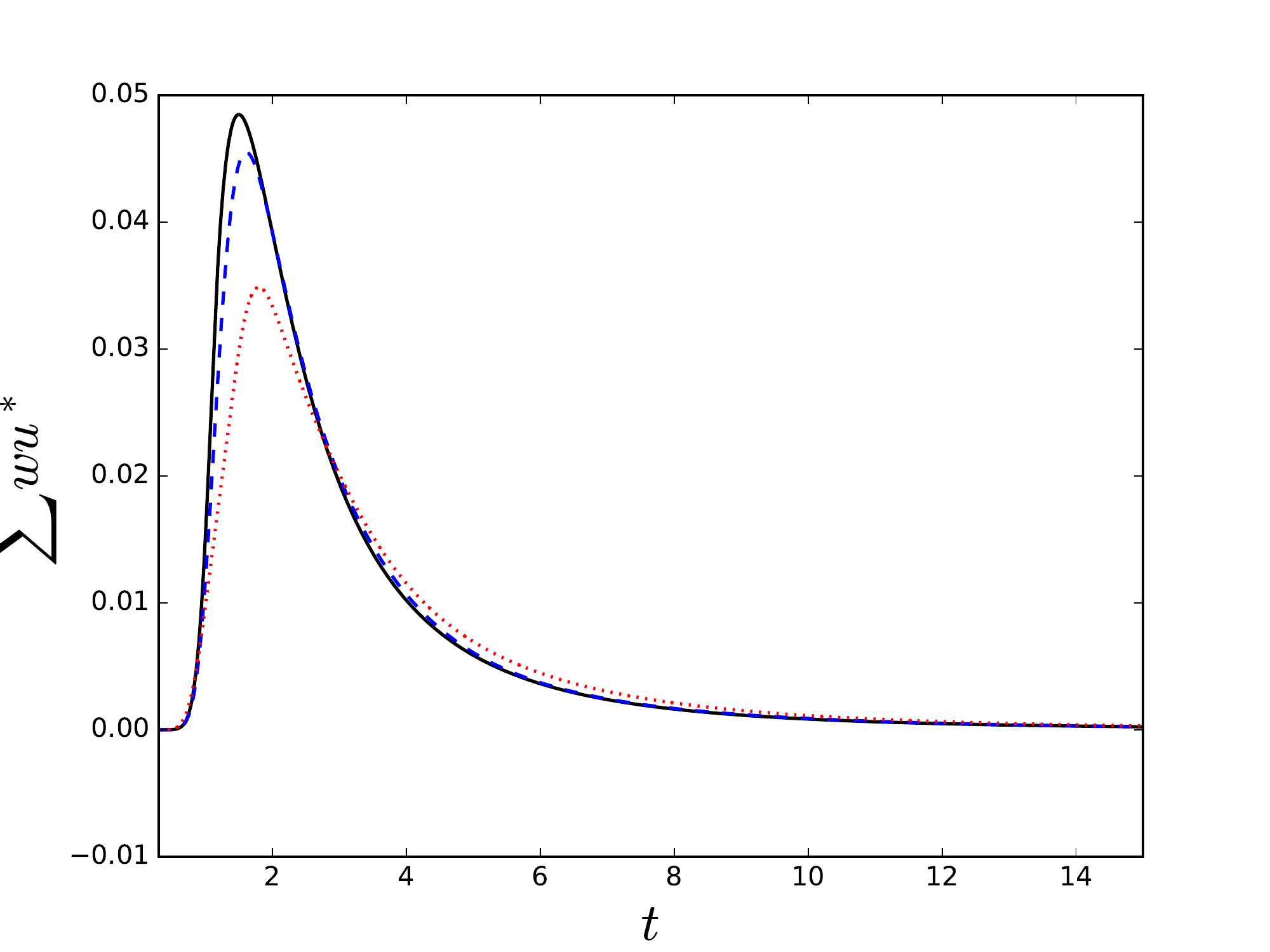}
	\caption{Mean magnitude of the subgrid content.}
	\end{subfigure}
	\caption{ Evolution of integral quanities for Burgers equation. }
	\label{fig:Burgers_energy}
	\end{center}
\end{figure}

\begin{figure}
	\begin{center}
	\begin{subfigure}[t]{0.45\textwidth}
	\includegraphics[width=1.\textwidth]{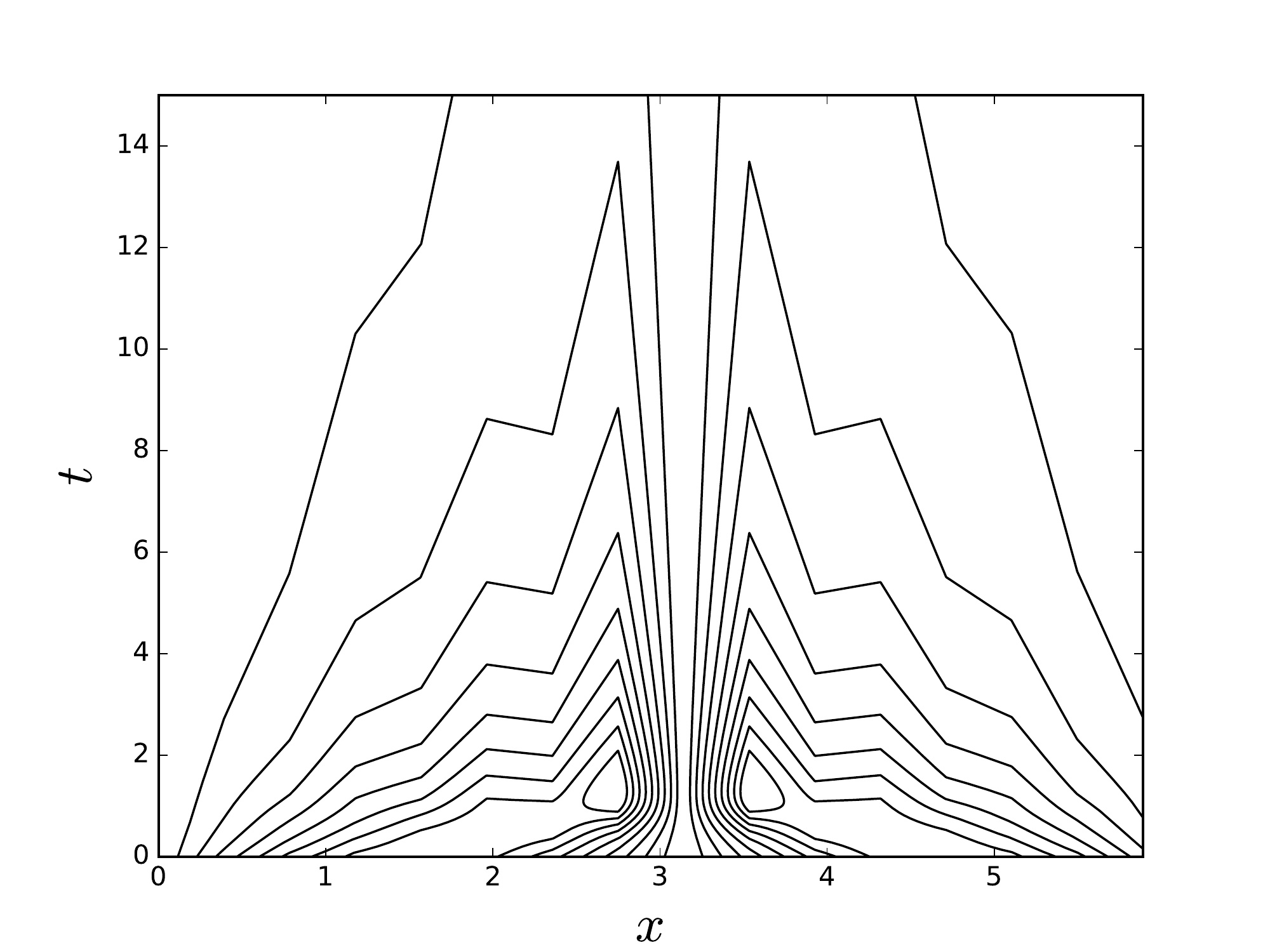}
	\caption{Filtered DNS}
	\end{subfigure}
	\begin{subfigure}[t]{0.45\textwidth}
	\includegraphics[width=1.\textwidth]{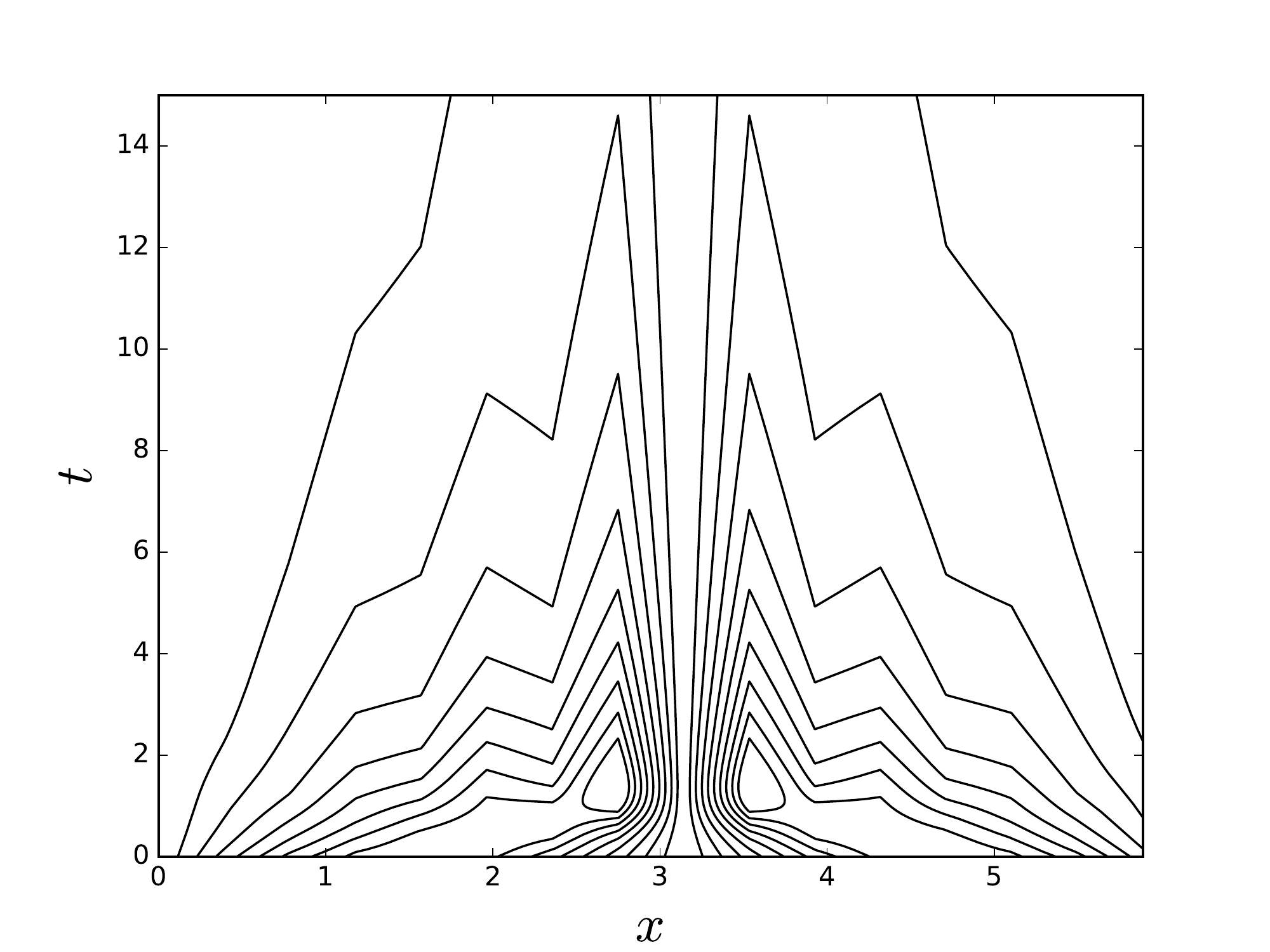}
	\caption{LES with dynamic-$\tau$}
	\end{subfigure}
	\begin{subfigure}[t]{0.45\textwidth}
	\includegraphics[width=1.\textwidth]{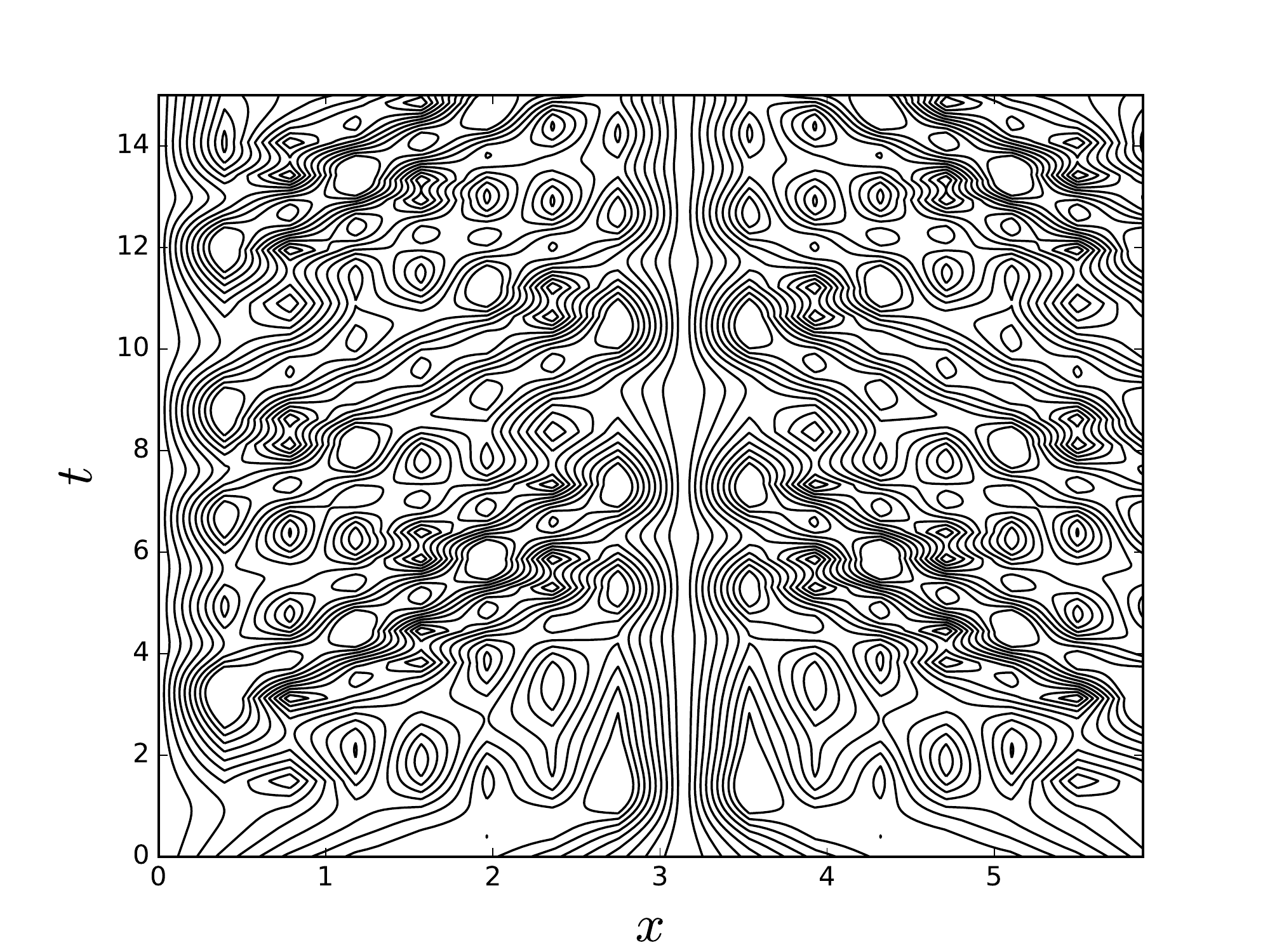}
	\caption{LES with no model}
	\end{subfigure}
	\begin{subfigure}[t]{0.45\textwidth}
	\includegraphics[width=1.\textwidth]{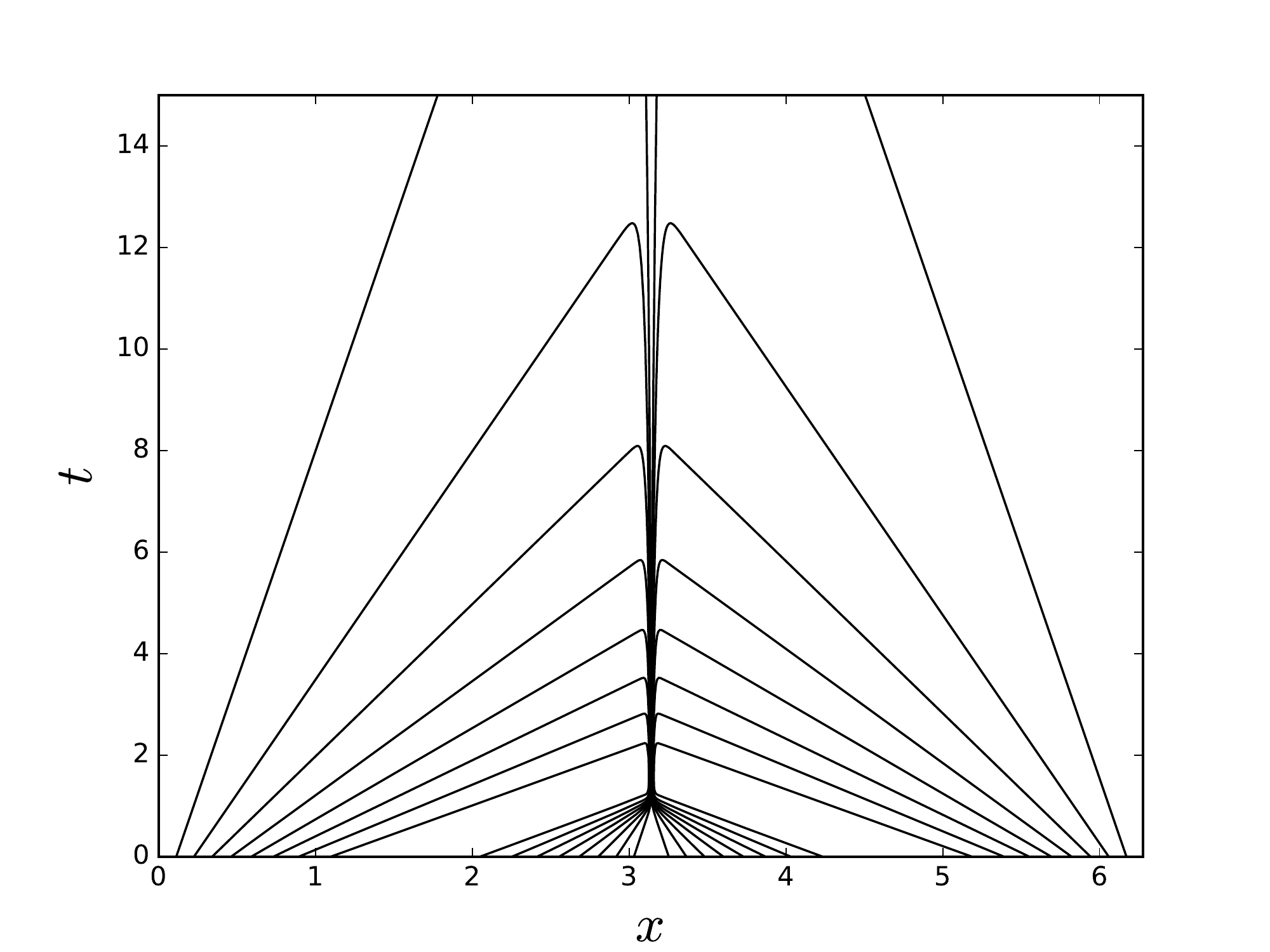}
	\caption{Unfiltered DNS}
	\end{subfigure}
	\caption{ $x-t$ diagrams for the filtered velocity field in physical space. The unfiltered field shows that the DNS solution is fully resolved. }
	\label{fig:Burgers_xt}
	\end{center}
\end{figure}

The numerical results for total resolved kinetic energy and the mean magnitude of the subgrid content are shown in Figure~\ref{fig:Burgers_energy}. The results are compared to the t-model as well as a coarse-grained simulation using no subgrid model. The dynamic model is seen to accurately predict the decay of kinetic energy and the subgrid content. Figure~\ref{fig:Burgers_xt} shows the $x-t$ diagrams of the numerical solutions in physical space. The trajectories of the coarse-grained simulation using the dynamic model compare well with the DNS data.

To assess the underlying assumptions of the model, we return to approximating the convolution integral using the orthogonal ODE, as in Section~\ref{sec:burgScaling}.
Similar to Figure~\ref{fig:burgOrthoC1}, Figure~\ref{fig:burgOrthoA} shows the evaluation of the imaginary component of the memory integrand for $k=7$ at $t=3$. The area of the shaded yellow region yields the subgrid content (dashed red line). The support of the integrand is seen to decay in time, indicating that only a finite time history of the resolved variables is needed to reconstruct the subgrid content. Figure~\ref{fig:burgOrthoE} shows the coarse-graining time-scale $\tau_{\MC{P}}$ as predicted by the dynamic-$\tau$ model and compares it to the time-scale of the orthogonal ODE integrand. The memory length for the orthogonal ODE is defined as being the support in time of the integrand that is required for the area under the curve to be $95\%$ of the total value. Note that, although the final model form is independent of the quadrature constant $C_q$, the equation for $\tau_{\MC{P}}$ is not. We take $C_q = 0.5$, corresponding to a trapezoidal rule quadrature. It is emphasized that this selection is only for the qualitative comparison of the memory length predicted by the model to that extracted from the orthogonal ODE, it is inconsequential in the model performance. The memory length is additionally compared to the ratio
$$ \tau_r = \frac{1}{C_q} \frac{e^{t \MC{L}}\MC{QL}\phi_{0j}}{e^{t \MC{L}}\MC{PLQL}\phi_{0j}},$$ as computed by the DNS data The dynamic-$\tau$ well predicts this memory length and almost perfectly recovers the DNS ratio. Figure~\ref{fig:burgOrthoF} shows non-dimensional integrands for various wave numbers and time instances, where it is seen that the memory integrand is somewhat self-similar. This implies that the memory integral for the entire field can be well correlated to its value at $s=0$. Finally, the slope of the memory length $\tau_{\MC{P}}$ for $0 \le t \le 1$ for varying grid resolutions is compared to the results obtained by Stinis' dynamic procedure~\cite{Stinis-rMZ} in Figure~\ref{fig:burgOrthoTau}. We compare the slopes for early time because Stinis obtains his renormalized coefficients while the model is still fully resolved.

We make several important observations. 

\begin{itemize}
\item
The support of the memory integrand is seen to grow approximately linearly with respect to time. Further, the shape of the integrand is somewhat self-similar. As such, the memory for this example is well described by
\begin{equation}\label{eq:ApproxMemory}
\int_{0}^t e^{(t-s)\MC{L}}\MC{PL}e^{s \MC{QL}} \MC{QL} \phi_{0j} ds \approx  C t e^{t \MC{L}}\MC{PL}\MC{QL} \phi_{0j} ds,
\end{equation}
where $C$ is a scalar constant that is related to the self-similar area under the curve. 

\item Eq.~\ref{eq:ApproxMemory} shows that the  impact of the unresolved scales on the resolved scales is well-described by a Markovian functional of the instantaneous field. The authors would like to emphasize the significance of this point and its implications for model development. The reader will note that Eq.~\ref{eq:ApproxMemory} is nothing but the renormalized (scaled) t-model and the reason for its surprising accuracy becomes clear. 

\item It is observed that the "memory length" computed by the dynamic model compares well with the true memory (as approximated by the orthogonal ODE procedure) in that it grows mostly linearly with time. This growth is consistent with the DNS data. This predicted memory is quantitatively consistent with the results predicted by Stinis' renormalized models. We reiterate that Stinis' renormalized models are derived under different assumptions and implicitly assume a linear dependence on time (the renormalization procedure determines the slope), while the dynamic-$\tau$ model predicts it.

\begin{figure}
	\begin{center}
	\begin{subfigure}[t]{0.42\textwidth}
	\captionsetup{justification=justified,singlelinecheck=false,format=hang}
	\includegraphics[trim={0cm 0cm 0cm 0cm},clip,width=1.\textwidth]{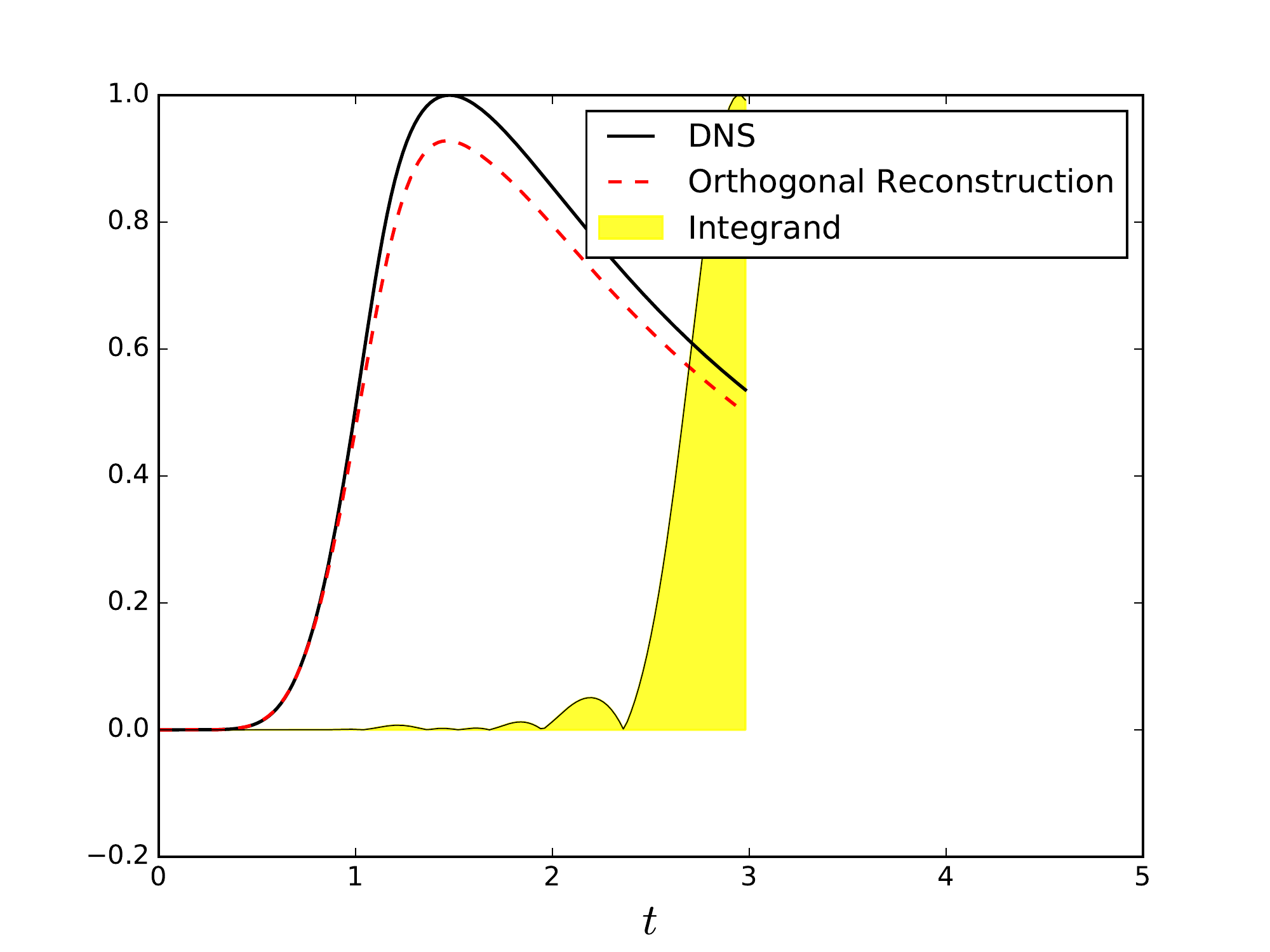}
	\caption{Evaluation of the approximate (normalized) memory integrand for $k=7$ at $t=3$.}
	\label{fig:burgOrthoA}
	\end{subfigure}
	%\begin{subfigure}[t]{0.42\textwidth}
	%\captionsetup{justification=justified,singlelinecheck=false,format=hang}
	%\includegraphics[trim={0cm 0cm 0cm 0cm},clip,width=1.\textwidth]{burgmem3.pdf}
	%\caption{As in Figure~\ref{fig:burgOrthoA}, but at $t=5.0$.}
	%\label{fig:burgOrthoB}
	%\end{subfigure}
	%\begin{subfigure}[t]{0.45\textwidth}
	%\includegraphics[trim={0cm 0cm 0cm 0cm},clip,width=1.\textwidth]{burgmem3.pdf}
	%\caption{As in Figure~\ref{fig:burgOrthoA}, but at $t=5.0$.}
	%\label{fig:burgOrthoC}
	%\end{subfigure}
	%\begin{subfigure}[t]{0.42\textwidth}
	%\captionsetup{justification=justified,singlelinecheck=false,format=hang}
	%\includegraphics[width=1.\textwidth]{orthogonalReconstruction.pdf}
	%\caption{The total subgrid energy transfer as computed by approximating the orthogonal dynamics with the residual equations.}
	%\label{fig:burgOrthoD}
	%\end{subfigure}
	\begin{subfigure}[t]{0.42\textwidth}
	\captionsetup{justification=justified,singlelinecheck=false,format=hang}
	\includegraphics[trim={0cm 0cm 0cm 0cm},clip,width=1.\textwidth]{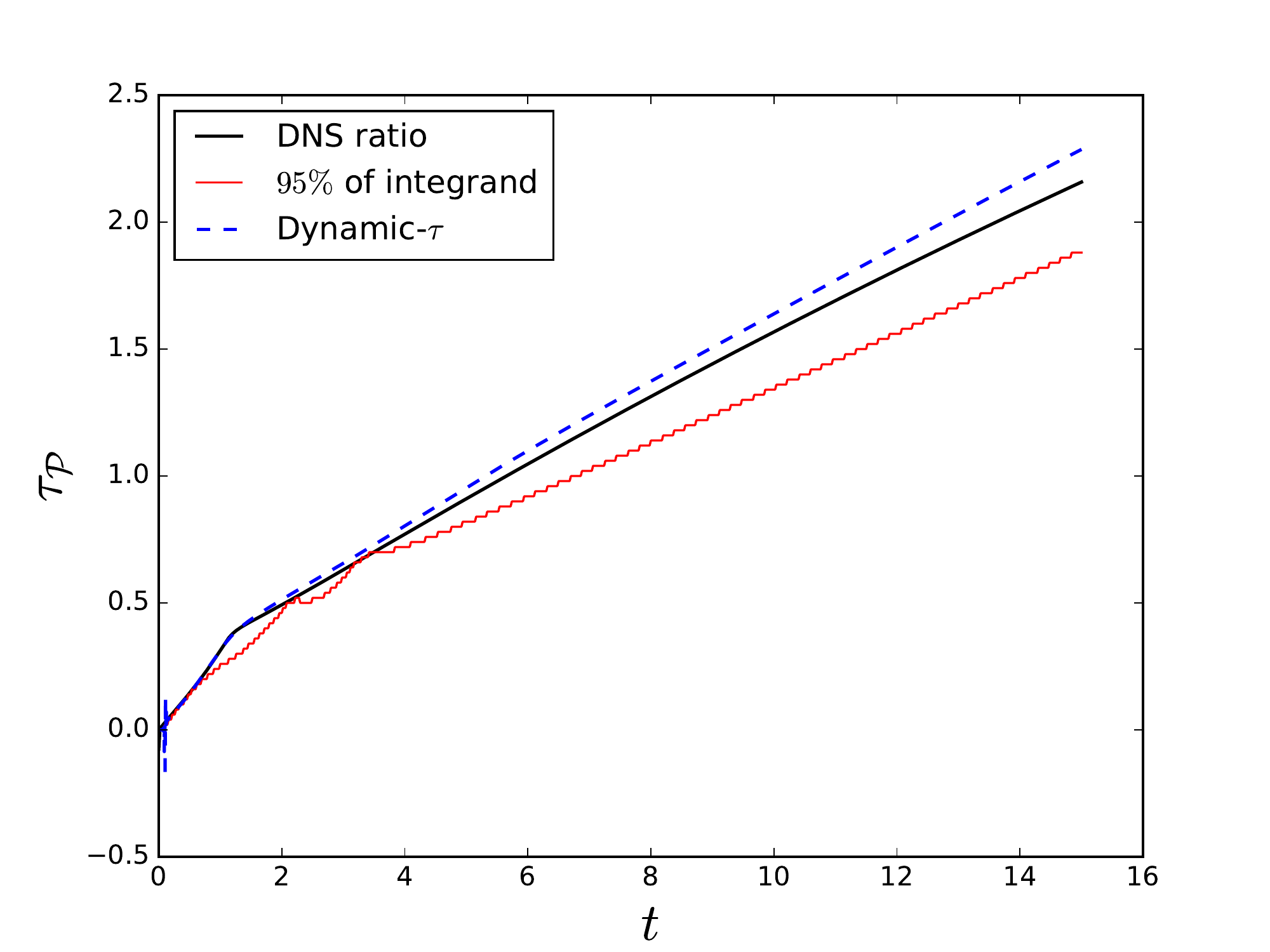}
	\caption{Growth of the computed memory length with time.}
	\label{fig:burgOrthoE}
	\end{subfigure}
	\begin{subfigure}[t]{0.42\textwidth}
	\captionsetup{justification=justified,singlelinecheck=false,format=hang}
	\includegraphics[trim={0cm 0cm 0cm 0cm},clip,width=1.\textwidth]{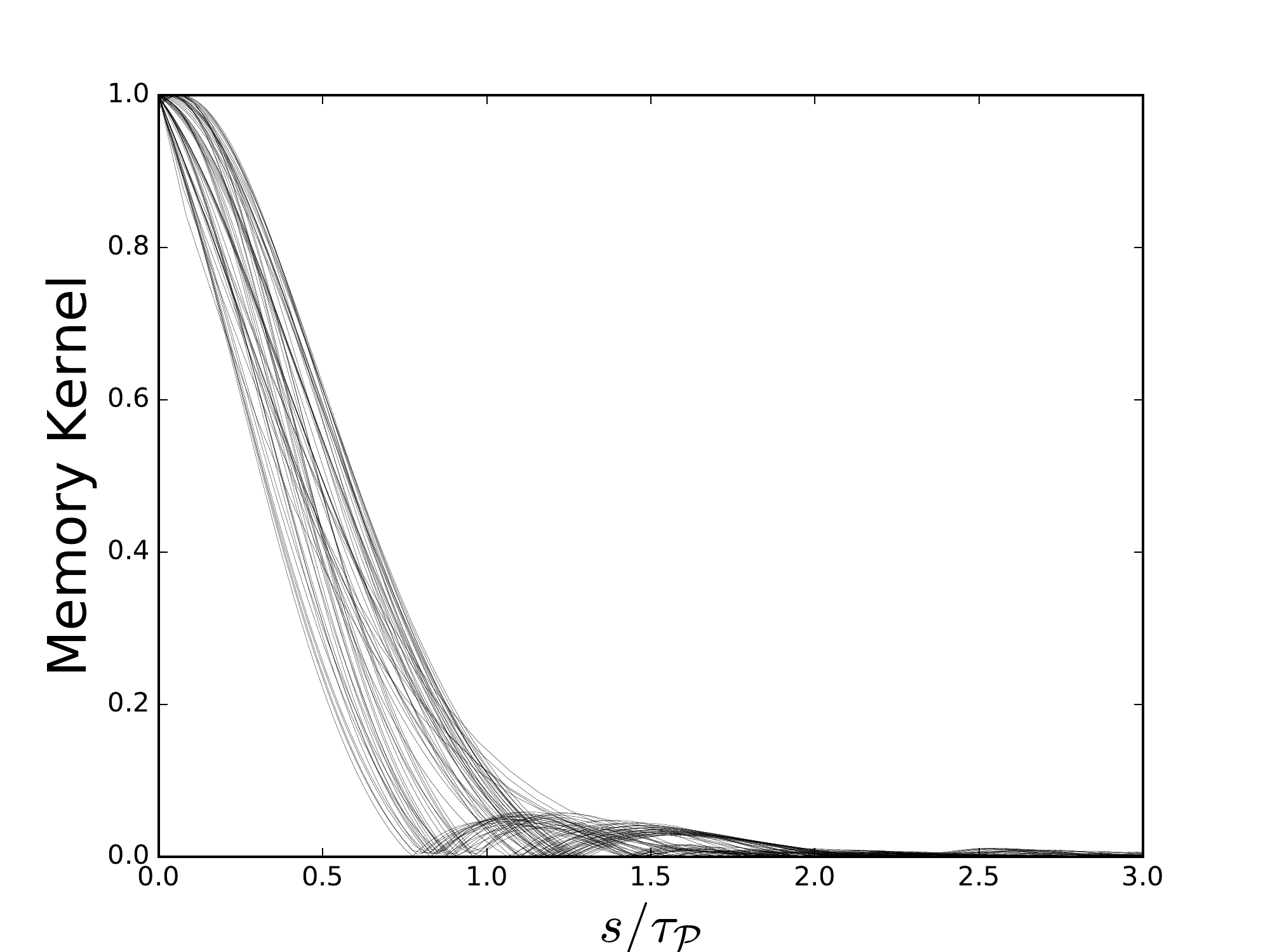}
	\caption{Non-dimensional integrand profiles.}
	\label{fig:burgOrthoF}
	\end{subfigure}
	\begin{subfigure}[t]{0.42\textwidth}
	\captionsetup{justification=justified,singlelinecheck=false,format=hang}
         \includegraphics[trim={0cm 0cm 0cm 0cm},clip,width=1.\textwidth]{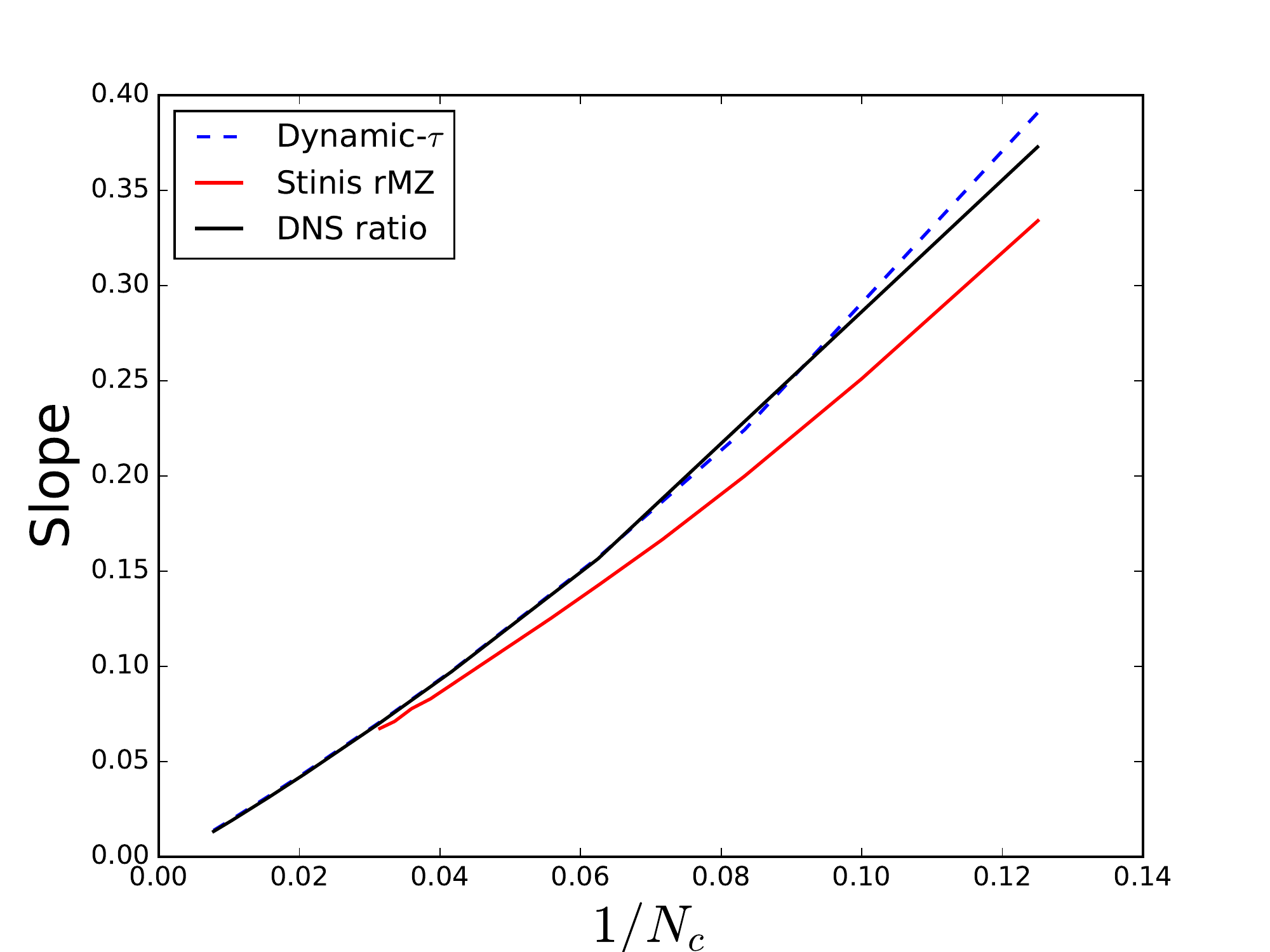}
	\caption{Comparison of the slope of the memory length (with respect to time) for varying grid resolutions to the renormalized coefficients obtained by Stinis.}
	\label{fig:burgOrthoTau}
	\end{subfigure}
	\caption{Summary of the reconstruction of the subgrid terms via the orthogonal ODE and a comparison of the predicted memory lengths. The support of the integrand and hence the memory length grows with time. The non-dimensional integrand profiles are seen to be somewhat self-similar, meaning that the area under the curve can be well correlated to the integrand at $s=0$.}
	\label{fig:Burgers_kernel}
	\end{center}
\end{figure}

\item For this specific case, it was shown that the M-Z-based dynamic-$\tau$ model is able to  predict the decay of kinetic energy, the mean magnitude of the subgrid content, and the trajectories of the resolved field in physical space. 

\end{itemize}

While it is not prudent to assume that more complex problems will have a self-similar integrand (for increasingly complex problems this would be very surprising), the results presented in this section provide strong evidence for the relative accuracy of the t-model and show that the dynamic-$\tau$ model provides accurate results for the right reasons. 

\section{Application to the Triply Periodic Navier-Stokes Equations}\label{sec:HIT}
Large Eddy Simulations of Fourier-Galerkin approximations to the Navier-Stokes equations are now considered. The incompressible Navier-Stokes equations are given by
\begin{equation}\label{eq:cont}
\frac{\partial u_i}{\partial x_i} = 0
\end{equation}
\begin{equation}\label{eq:momentum}
\frac{\partial u_i}{\partial t} + \frac{\partial u_i u_j}{\partial x_j} = -\frac{1}{\rho} \frac{\partial p}{\partial x_i} + \frac{\partial}{\partial x_j} \bigg(\nu \frac{\partial u_i}{\partial x_j} \bigg).
\end{equation}
For triply periodic problems, Eqns.~\ref{eq:cont} and~\ref{eq:momentum} can be Fourier-transformed in all directions. The pressure Poisson equation can be directly solved in Fourier space, allowing the governing equations to be written compactly as
\begin{equation}\label{eq:NSspec}
\bigg(\frac{\partial}{\partial t} + \nu k^2 \bigg){\hat{u}}_{i} (\mathbf{k},t) + \bigg(\delta_{im} - \frac{k_i k_m}{k^2} \bigg) \imath k_j \sum_{\substack{ \mathbf{p} + \mathbf{q} = \mathbf{k} \\ \mathbf{p,q} \in F \cup G } } {\hat{u}}_j (\mathbf{p},t)  {\hat{u}}_m (\mathbf{q},t)  = 0  \qquad \mathbf{k} \in F \cup G,
\end{equation}
where the Fourier modes have been written to belong to the union of two sets (with $F$ being the resolved set and $G$ being the unresolved set). Separating the modes into the resolved and unresolved sets yields the reduced system
\begin{equation}\label{eq:NSspec_filtered}
\bigg(\frac{\partial}{\partial t} + \nu k^2 \bigg){\hat{u}}_{i} (\mathbf{k},t) + \bigg(\delta_{im} - \frac{k_i k_m}{k^2} \bigg) \imath k_j \sum_{\substack{ \mathbf{p} + \mathbf{q} = \mathbf{k} \\ \mathbf{p,q} \in F } } {\hat{u}}_j (\mathbf{p},t)  {\hat{u}}_m (\mathbf{q},t) = - \bigg(\delta_{im} - \frac{k_i k_m}{k^2} \bigg) \imath k_j  \hat{\tau}_{jm}(\mathbf{k},t) \qquad k \in F,
\end{equation}
where
$$\hat{\tau}_{jm}(\mathbf{k},t) = \sum_{\substack{ \mathbf{p} + \mathbf{q} = \mathbf{k} \\ \mathbf{p,q} \in G } } {\hat{u}}_j (\mathbf{p},t)  {\hat{u}}_m (\mathbf{q},t) + 
 \sum_{\substack{ \mathbf{p} + \mathbf{q} = \mathbf{k} \\ \mathbf{p} \in G, \mathbf{q} \in F } } {\hat{u}}_j (\mathbf{p},t)  {\hat{u}}_m (\mathbf{q},t) + 
  \sum_{\substack{ \mathbf{p} + \mathbf{q} = \mathbf{k} \\ \mathbf{p} \in F, \mathbf{q} \in G } } \hat{u}_j (\mathbf{p},t)  \hat{u}_m (\mathbf{q},t).$$
Note that, in Fourier space, the pressure term appears as a projection. This projection leads to additional non-linear interactions between the resolved and unresolved scales. Through the use of the Mori-Zwanzig formalism, the RHS of Eq.~\ref{eq:NSspec_filtered} can be alternatively written as a convolution integral
\begin{multline}\label{eq:NS_M-Z}
\bigg(\frac{\partial}{\partial t} + \nu k^2 \bigg)\hat{u}_{i} (\mathbf{k},t) + \bigg(\delta_{im} - \frac{k_i k_m}{k^2} \bigg) i k_j \sum_{\substack{ \mathbf{p} + \mathbf{q} = \mathbf{k} \\ \mathbf{p,q} \in F } } \hat{u}_j (\mathbf{p},t)  \hat{u}_m (\mathbf{q},t) = \MC{P}\int_0^t K_i(\mathbf{\hat{u}}(t-s),s)ds  \qquad k \in F.
\end{multline}
For the spectral Navier-Stokes equations in their primitive form (Eq.~\ref{eq:NSspec}), the analytic derivation of the M-Z models is tractable. Details regarding the derivations can be found in~\cite{stinisHighOrderEuler}. Here only the final form of the dynamic-$\tau$ model is given, which can be shown to be,
\begin{multline}\label{eq:NS_PLQLu}
\MC{P}\int_0^t K_i(\mathbf{\hat{u}}(t),t-s)ds \approx  0.5 \tau_{\MC{P}} \bigg[ -\bigg(\delta_{im} +  \frac{k_i k_m}{k^2} \bigg)\imath k_j   \sum_{\substack{ \mathbf{p} + \mathbf{q} = \mathbf{k} \\ \mathbf{p} \in F, \mathbf{q} \in  G }} \hat{u}_{j}(\mathbf{p},t)  e^{t \MC{L}}\MC{PL}\hat{u}_{m}(\mathbf{q},0) -  \\
 \bigg(\delta_{im} + \frac{k_i k_m}{k^2} \bigg) \imath k_j\sum_{\substack{ \mathbf{p} + \mathbf{q} = \mathbf{k} \\ \mathbf{p} \in F, \mathbf{q} \in  G }} u_{m}(\mathbf{p},t) e^{tL} \MC{PL}\hat{u}_{j}(\mathbf{q},0) \bigg],
\end{multline}
where $e^{t \MC{L}} \MC{PL}\hat{u}_j$ is the coarse-grained right-hand side. 

The triply periodic spectral Navier-Stokes equations are solved in their primitive form using a Galerkin pseudo-spectral method. The main solvers are written in Python and use the fast Fourier transform to evaluate the convolutions. The LES solver uses 2x padding for FFTs. The DNS solver utilizes 2/3 dealiasing. A low storage RK4 time integration scheme is used for all calculations. The relevant details for the simulations are given in Table~\ref{table:1}.

\begin{table}%Table of simulation parameters
\begin{center}\scriptsize
\vskip -0.1in
\begin{tabular}{l |l l l l   l l  c c  c c c c} \hline 
Case &  & L & $Re_{\lambda}$ & $\hat{Re_{\lambda}}$ & $\nu$ &  $\qquad$ & N (DNS) & $\Delta t$ (DNS) &  $\qquad$ & N (LES) & $\Delta t$ (LES) & $k_c$\\ \hline
Low $Re$  & & $2 \pi$ & 65  & 82  & 0.001 &  & $512^3$ & $0.0025$ &  & $64^3$ & $0.025$ & 32\\ \hline
Moderate $Re$  & & $2 \pi$ & 75  & 130 & 0.0005 &  & $512^3$ & $0.0025$ &  & $64^3$ & $0.025$ & 32 \\ \hline
High $Re$ 1  & & $2 \pi$ & 164 & 482 & 0.00015 &  & $1024^3$ & $0.00125$ &  & $64^3$ & $0.025$ & 32 \\ \hline
High $Re$  2 & & $2 \pi$ & 164  & 324 & 0.00015 &  & $1024^3$ & $0.00125$ &  & $128^3$ & $0.02$ & 64 \\ \hline
\end{tabular}
\caption{Physical and numerical details for homogeneous turbulence simulations. The Taylor microscale-based Reynolds number of the filtered field ($\hat{Re}_{\lambda} = \hat{E} \sqrt{ \frac{20}{3\nu \hat{\epsilon}} }$) is higher than that of the unfiltered field due to the filtering of high wave number content.} 
\label{table:1}
\end{center}
\end{table}

\begin{comment}
$$e^{t \MC{L}}\MC{PL}\hat{u}_{i}(\mathbf{k},0) = - \bigg(\delta_{im} - \frac{k_i k_m}{k^2} \bigg) \imath k_j \sum_{\substack{ \mathbf{p} + \mathbf{q} = \mathbf{k} \\ \mathbf{p,q} \in F } } \hat{u}_j (\mathbf{p},t)  \hat{u}_m (\mathbf{q},t) \qquad k \in G.$$
\end{comment}

\subsection{Homogeneous Turbulence at a Low Reynolds Number}
\begin{figure}
	\begin{center}
	\begin{subfigure}[t]{0.45\textwidth}
	\includegraphics[width=1.\textwidth]{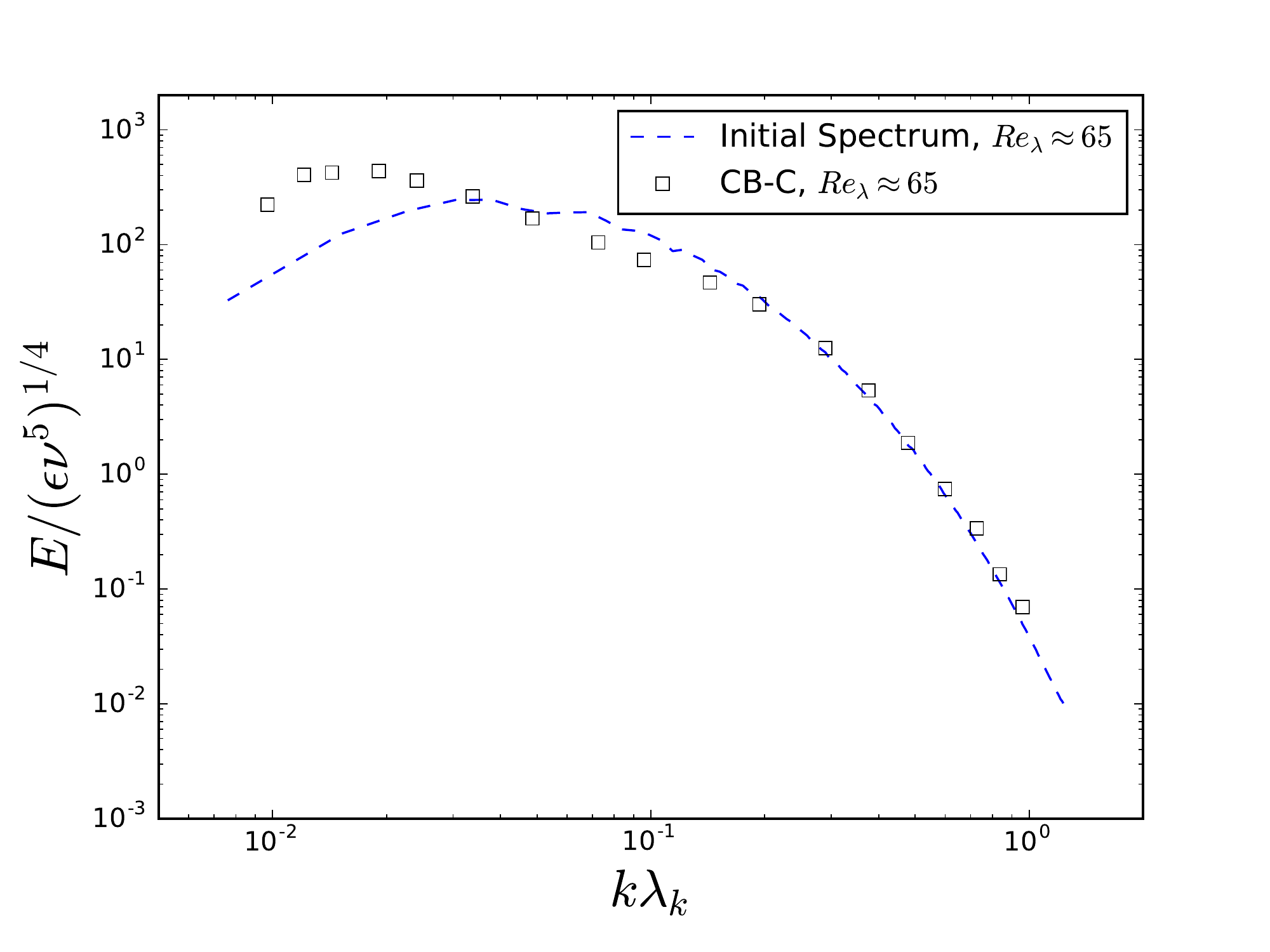}
	\end{subfigure}
	\begin{subfigure}[t]{0.45\textwidth}
	\includegraphics[trim={0 0 25cm 0cm},width=1.\textwidth]{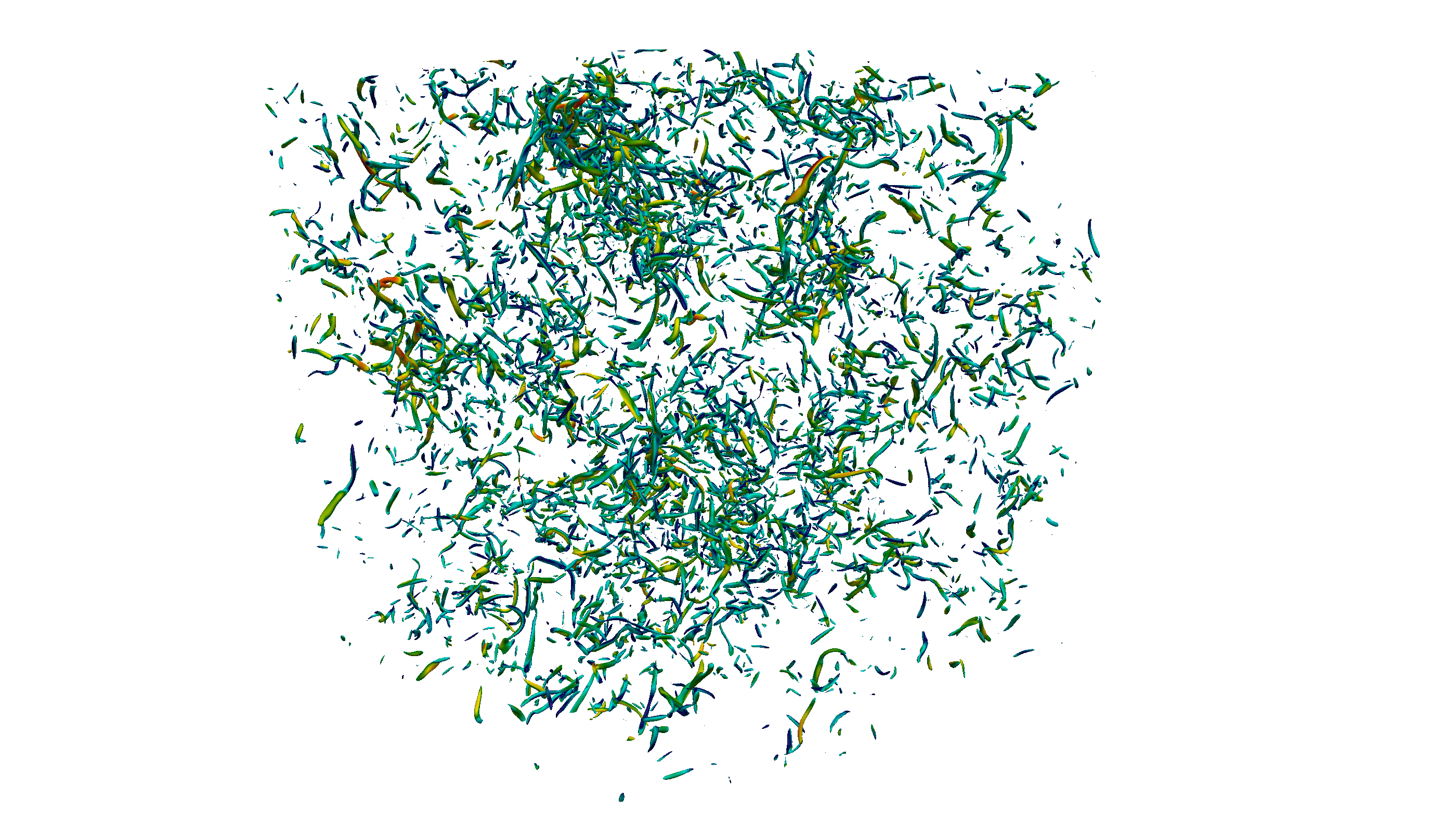}
	\end{subfigure}
	\caption{Energy spectra (left) and q-criterion (right) of the unfiltered DNS field used for the initialization of the low Reynolds number Large Eddy Simulations. The initial Reynolds number is $Re_{\lambda} \approx 65$.}
	\label{fig:LowReHIT_ICs}
	\end{center}
\end{figure}

Homogeneous isotropic turbulence with an initial Reynolds number of $Re_{\lambda} \approx 65$ is first considered. The Taylor microscale-based Reynolds number is defined by $Re_{\lambda} = 
E \sqrt{ \frac{20}{3\nu\epsilon}} $, where $\epsilon$ is the dissipation rate.
The DNS  velocity field is initialized using the Rogallo~\cite{Rogallo} procedure. The initial spectrum is taken to be
$$E(k,0) = \frac{q^2}{2A} \frac{1}{k_p^{\sigma+1}} k^{\sigma} \exp \bigg(-\frac{\sigma}{2} \big(\frac{k}{k_p}\big)^2 \bigg),$$
where $k_p$ is the wave number at which the energy spectrum is maximum, $\sigma$ is a parameter set to 4, and $A = \int_0^{\infty}k^{\sigma} \exp(-\sigma k^2/2)dk$. The parameters used are $\sigma = 4, k_p = 5,$ and $q^2 = 3$. The DNS simulation is evolved on a $512^3$ mesh until realistic homogeneous turbulence is present. The filtered velocity field is then used for initial conditions for the Large Eddy Simulations. Figure~\ref{fig:LowReHIT_ICs} shows the resulting spectrum and q-criterion. For reference the initial spectrum is compared to the experiment of Comte Bellot and Corrsin (CB-C)~\cite{CBC}. The initial Reynolds number of the simulations is $Re_{\lambda} \approx 65$. The DNS simulation using $512^3$ degrees of freedom resolves up to $k \lambda_k \approx 1$, where $\lambda_k = \big(\frac{\nu^3}{\epsilon} \big)^{\frac{1}{4}}$ is the Kolmogorov lengthscale. 

\begin{figure}
	\begin{center}
	\begin{subfigure}[t]{0.32\textwidth}
	\includegraphics[clip,width=1.\textwidth]{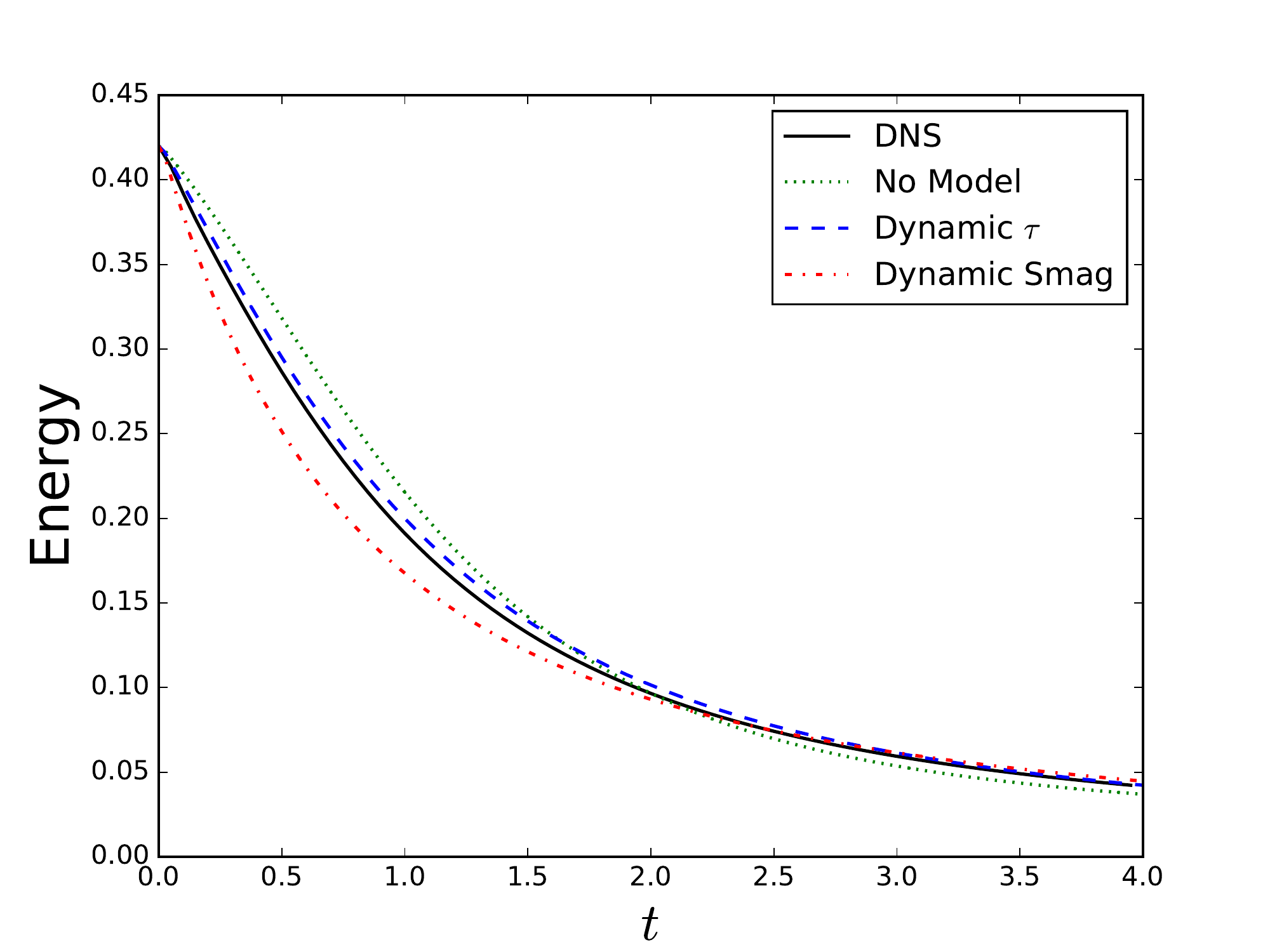}
	\caption{Evolution of the resolved kinetic energy.}
	\end{subfigure}
	\begin{subfigure}[t]{0.32\textwidth}
	\includegraphics[clip,width=1.\textwidth]{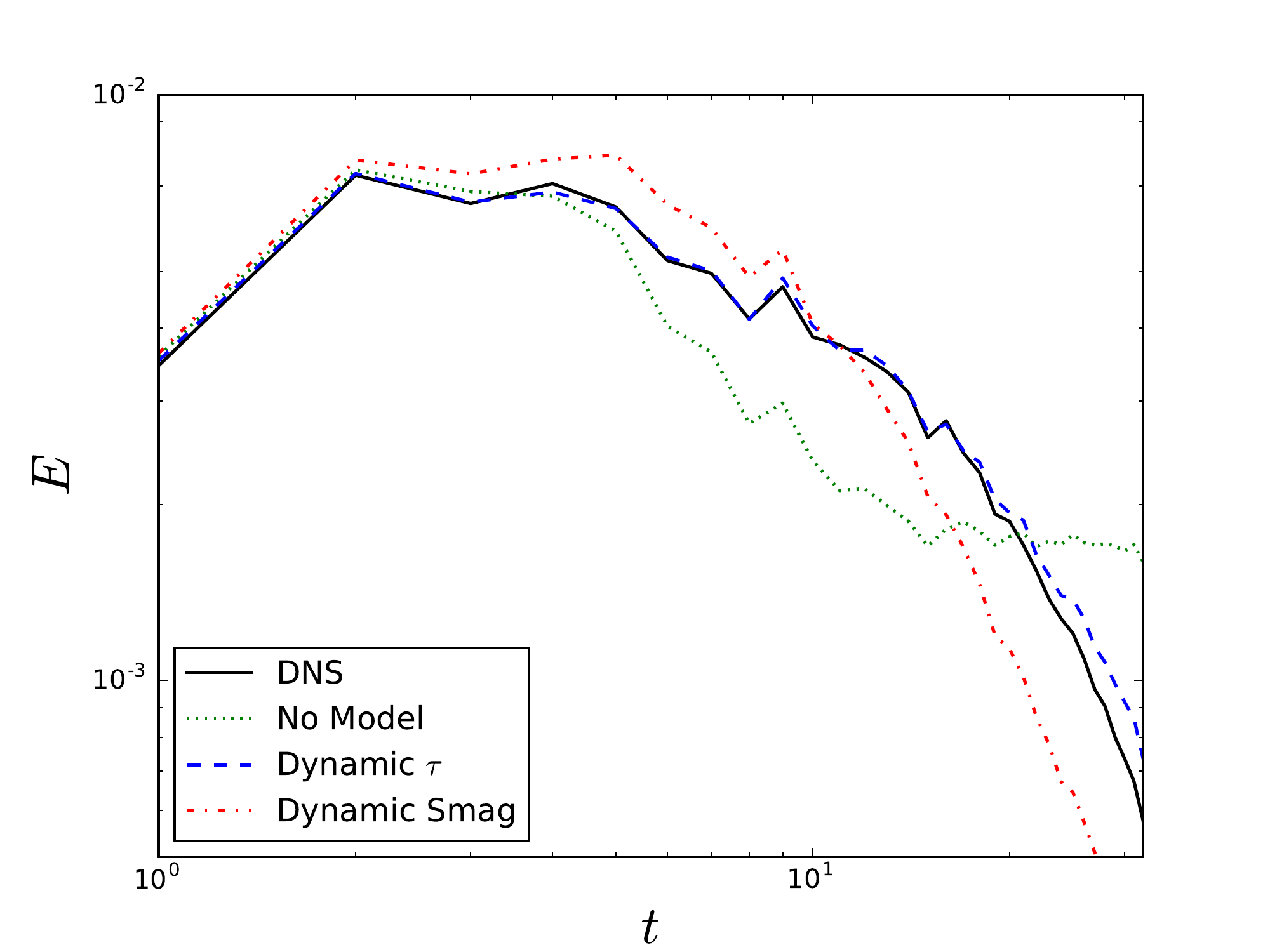}
	\caption{Energy spectrum at $t = 2.0$.}
	\label{fig:LowReHIT_energyb}
	\end{subfigure}
	%\begin{subfigure}[t]{0.32\textwidth}
	%\includegraphics[clip,width=1.\textwidth]{HIT/LowRe/spec3.pdf}
	%\caption{Filtered energy spectrum at $t = 3.0$.}
	%\end{subfigure}	
	\begin{subfigure}[t]{0.32\textwidth}
	\includegraphics[clip,width=1.\textwidth]{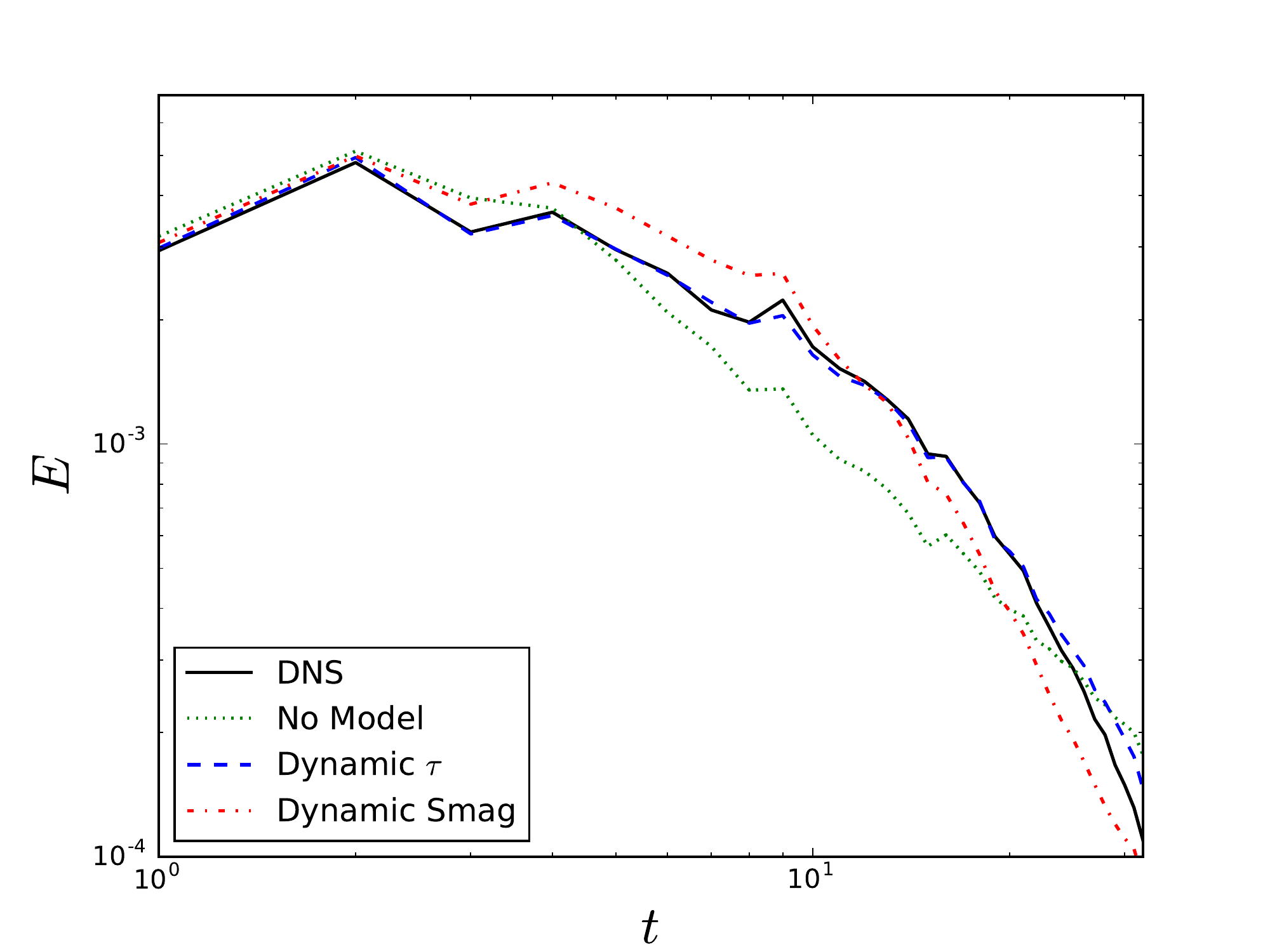}
	\caption{Energy spectrum at $t = 4.0$.}
        \label{fig:LowReHIT_energyc}
	\end{subfigure}		
	\caption{Evolution of the resolved kinetic energy and resolved spectrum.}
	\label{fig:LowReHIT_energy}
	\end{center}
\end{figure}

\begin{figure}
	\begin{center}
	\begin{subfigure}[t]{0.32\textwidth}
	\includegraphics[clip,width=1.\textwidth]{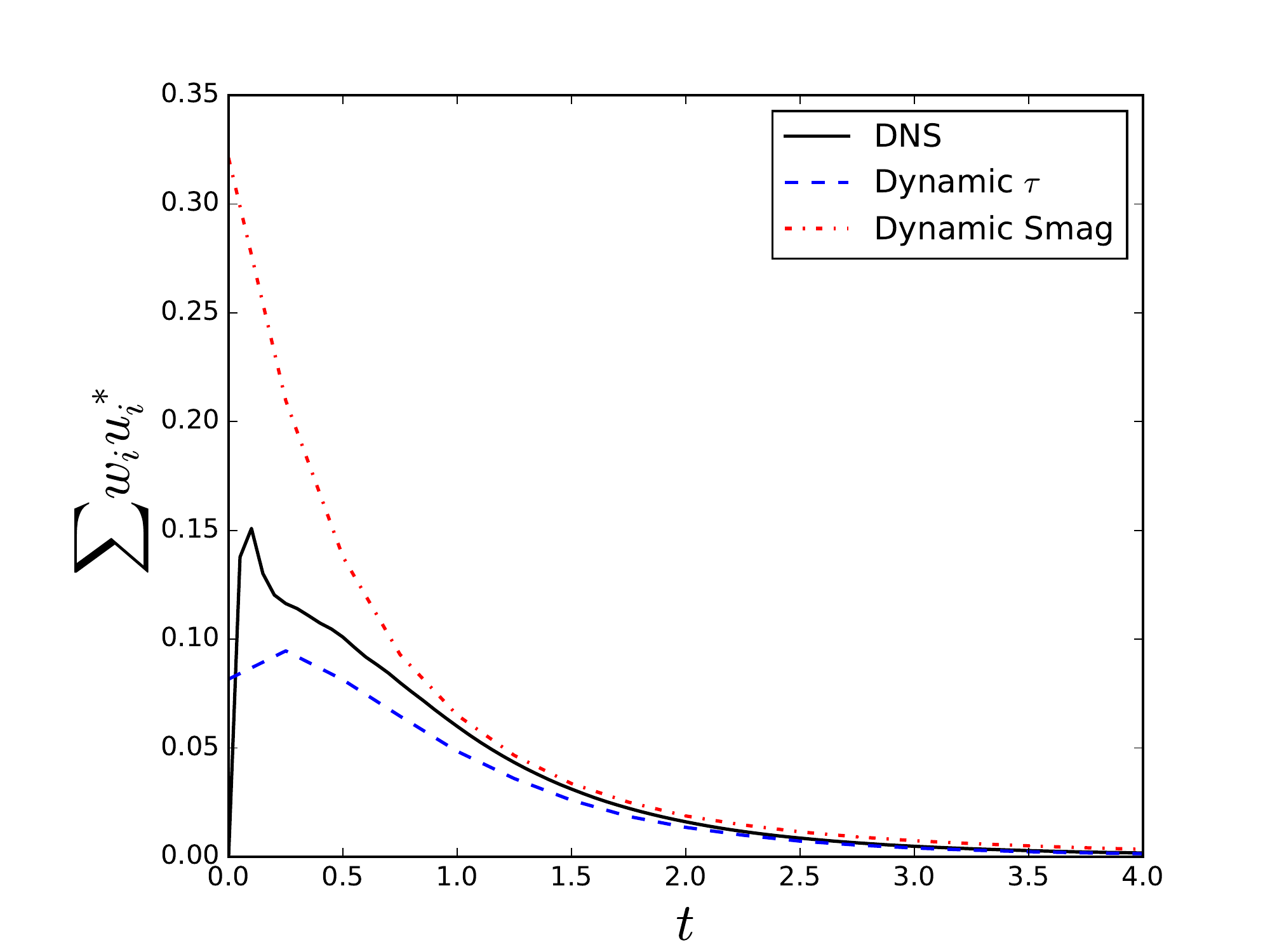}
	\caption{Evolution S-G energy transfer.}
	\end{subfigure}
	\begin{subfigure}[t]{0.32\textwidth}
	\includegraphics[clip,width=1.\textwidth]{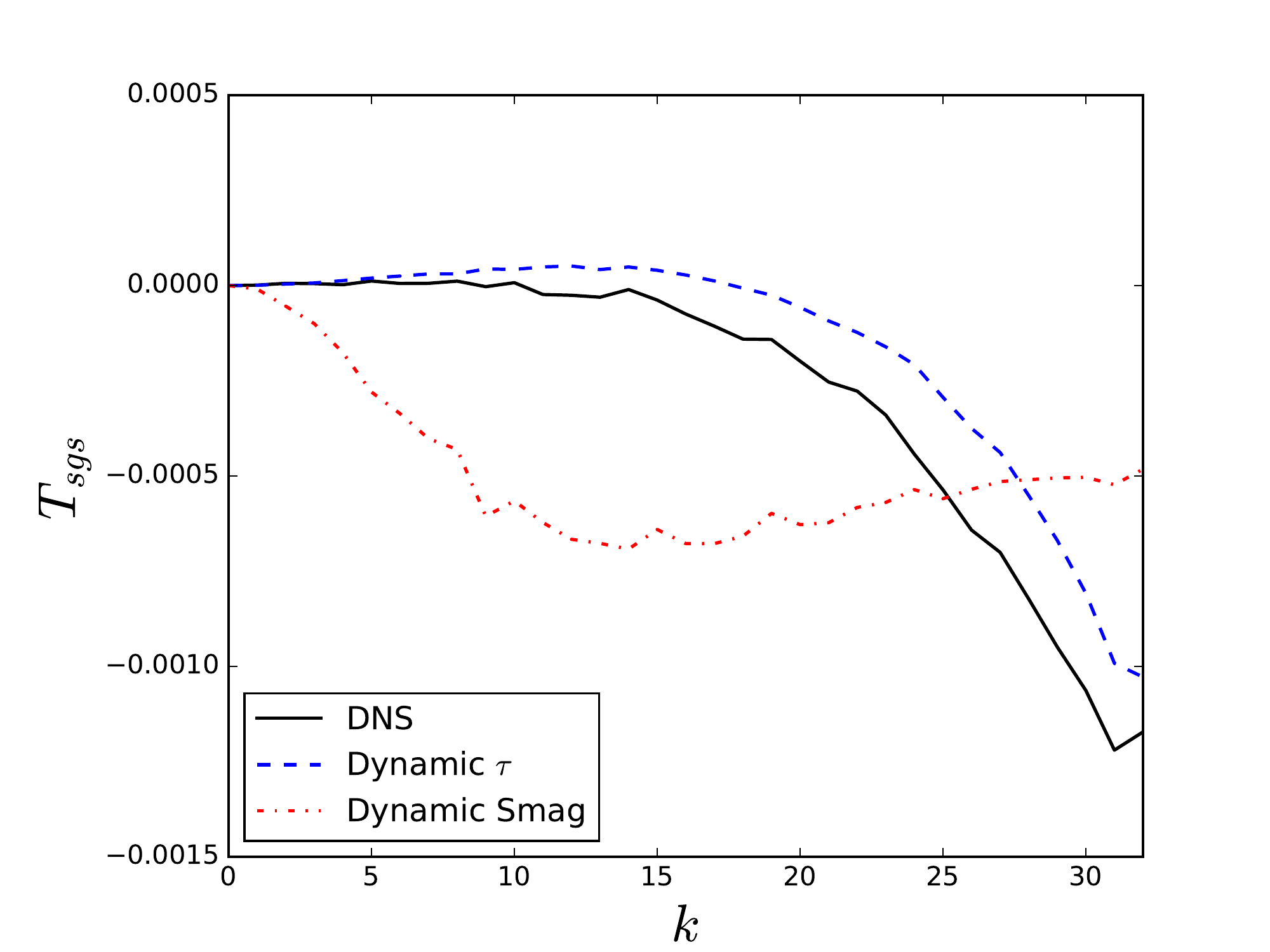}
	\caption{S-G energy transfer at $t = 2.0$.}
	\end{subfigure}
	%\begin{subfigure}[t]{0.32\textwidth}
	%\includegraphics[clip,width=1.\textwidth]{HIT/LowRe/tsgs3.pdf}
	%\caption{Subgrid energy transfer at $t = 3.0$.}
	%\end{subfigure}	
	\begin{subfigure}[t]{0.32\textwidth}
	\includegraphics[clip,width=1.\textwidth]{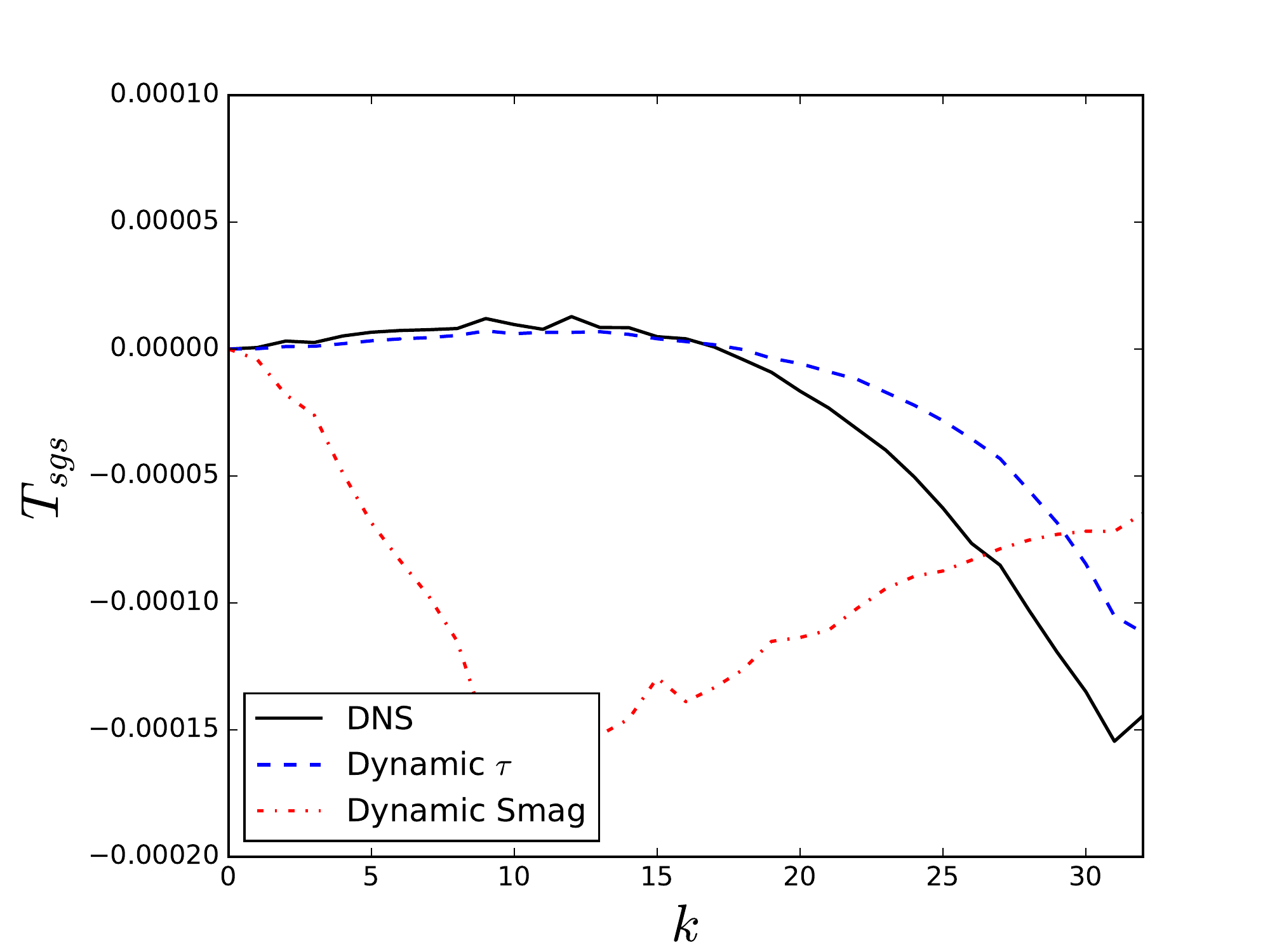}
	\caption{S-G energy transfer at $t = 4.0$.}
	\end{subfigure}		
	\caption{Evolution of the subgrid (S-G) energy transfer for the low Reynolds number case. The subgrid energy transfer is computed by $T_{sgs_k} = {w_k}{u_k}^*$.}
	\label{fig:LowReHIT_transfer}
	\end{center}
\end{figure}

Figure~\ref{fig:LowReHIT_energy} shows the evolution of the resolved kinetic energy and the energy spectra at various time instances. The dynamic-$\tau$ model is compared to the dynamic Smagorinsky model as well as a simulation without using any subgrid model. The simulation with no subgrid model predicts the decay of energy reasonably, but the energy spectra are seen to exhibit significant error. The dynamic Smagorinsky model slightly over predicts the decay of energy, but offers improved predictions for the energy spectra. The dynamic-$\tau$ model is seen to provide excellent predictions for both the energy decay and spectra. Figure~\ref{fig:LowReHIT_transfer} shows the evolution of the energy transferred by the subgrid model. The dynamic Smagorinsky model is seen to remove too much energy from the low wave numbers and too little from the high wave numbers. The dynamic-$\tau$ model is seen to compare well to the subgrid energy transfer extracted from the DNS data.  

An interesting feature to note about the dynamic-$\tau$ model is that it deposits a small amount of energy into the low wave numbers $(0 \le k \le 16)$. This phenomena indicates that the M-Z-based models are capable of predicting backscatter. Indeed, at $t=4.0$ (Figure~\ref{fig:LowReHIT_energyc}) it is clear that a small amount of backscatter is present in the DNS simulation at low wave numbers. The dynamic model is able to capture this phenomena. Note, however, that the model predicts backscatter at $t=2.0$ (Figure~\ref{fig:LowReHIT_energyb}) while none is present in the DNS.

\subsection{Homogeneous Turbulence at a Moderate Reynolds Number}
\begin{figure}
    \begin{center}
	\begin{subfigure}[t]{0.45\textwidth}
	\includegraphics[width=1.\textwidth]{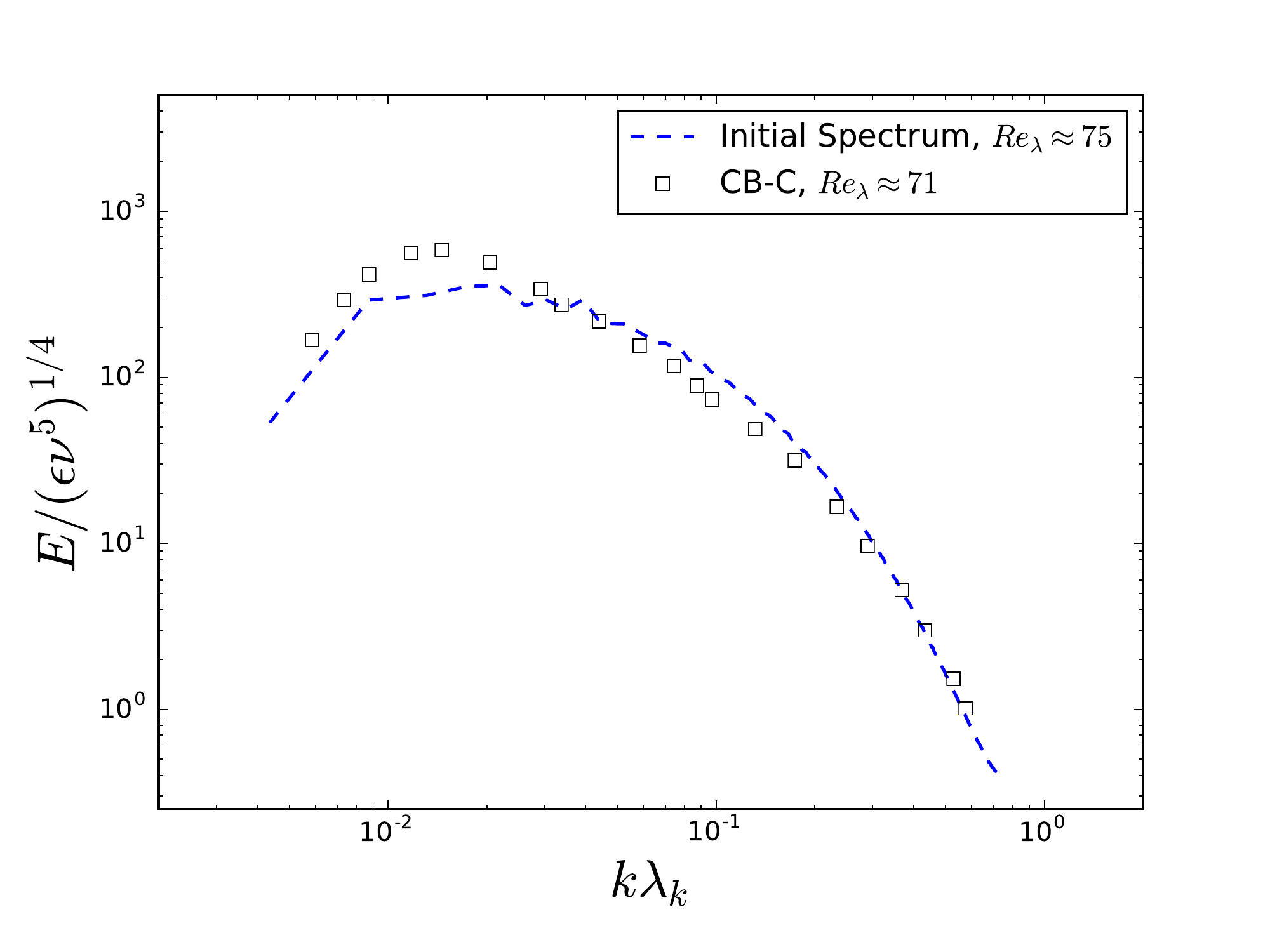}
	\end{subfigure}
	\begin{subfigure}[t]{0.45\textwidth}
	\includegraphics[trim={0 0 25cm 0cm},width=1.\textwidth]{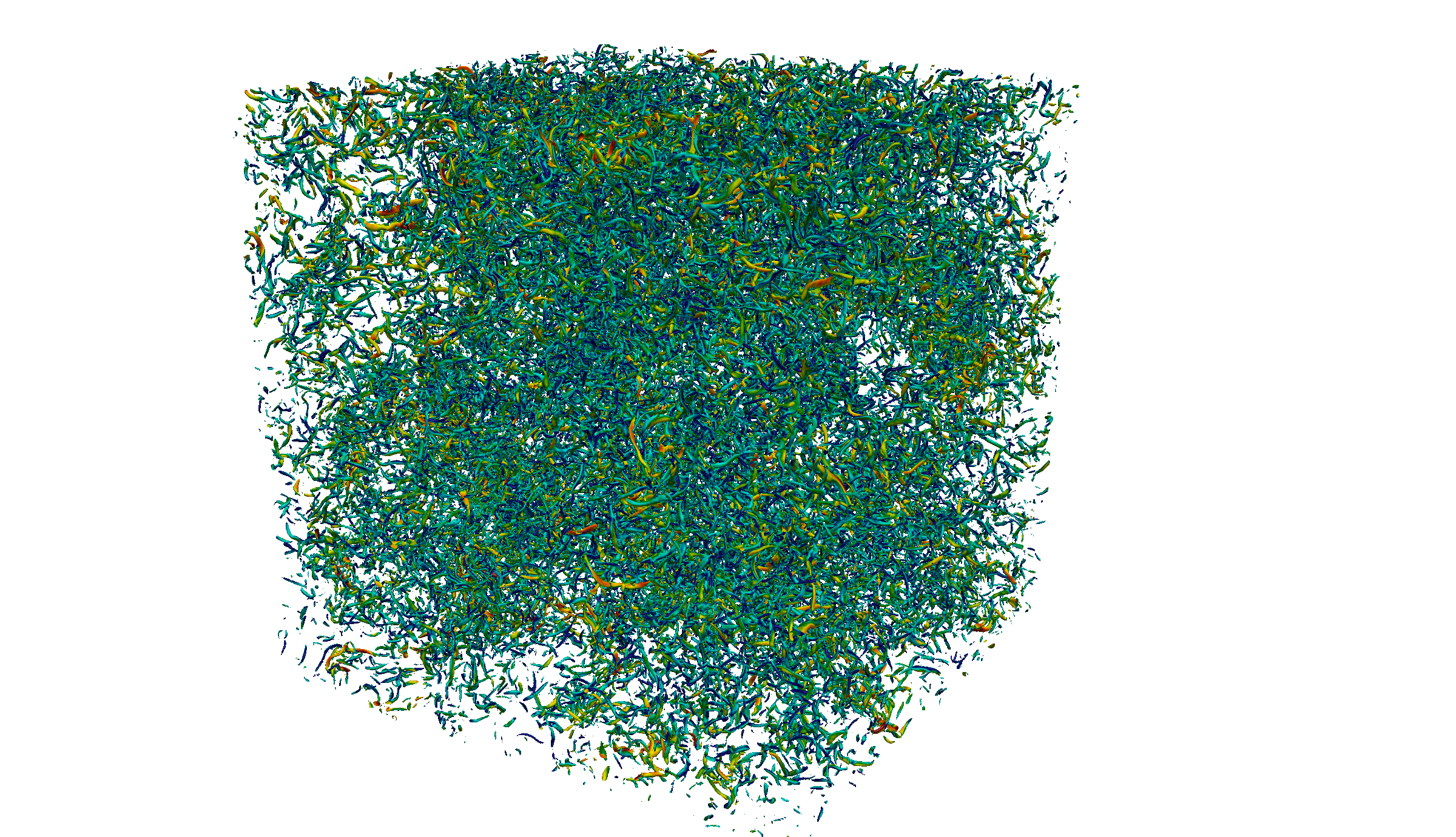}
	\end{subfigure}
	\caption{Energy spectra (left) and q-criterion (right) of the unfiltered DNS field used for the initialization of the moderate Reynolds number Large Eddy Simulations. The q-criterion is set at the same magnitude as in Fig.~\ref{fig:LowReHIT_ICs}. The initial Reynolds numbers is $Re_{\lambda} \approx 75$. }
	\label{fig:HIT_ICs}
	\end{center}
\end{figure}

The simulation of homogeneous turbulence at a moderate Reynolds number ($Re_{\lambda} \approx 75$) is now considered. The procedure described in the previous section is again used to initialize the flow field. Figure~\ref{fig:HIT_ICs} shows the resulting initial spectrum and q-criterion. The initial condition is seen to contain more small-scale vortical structures than in the previous case. At this Reynolds number, the DNS simulation using $512^3$ degrees of freedom resolves up to $k \lambda_k \approx 0.5$.

\begin{figure}
	\begin{center}
	\begin{subfigure}[t]{0.32\textwidth}
	\includegraphics[clip,width=1.\textwidth]{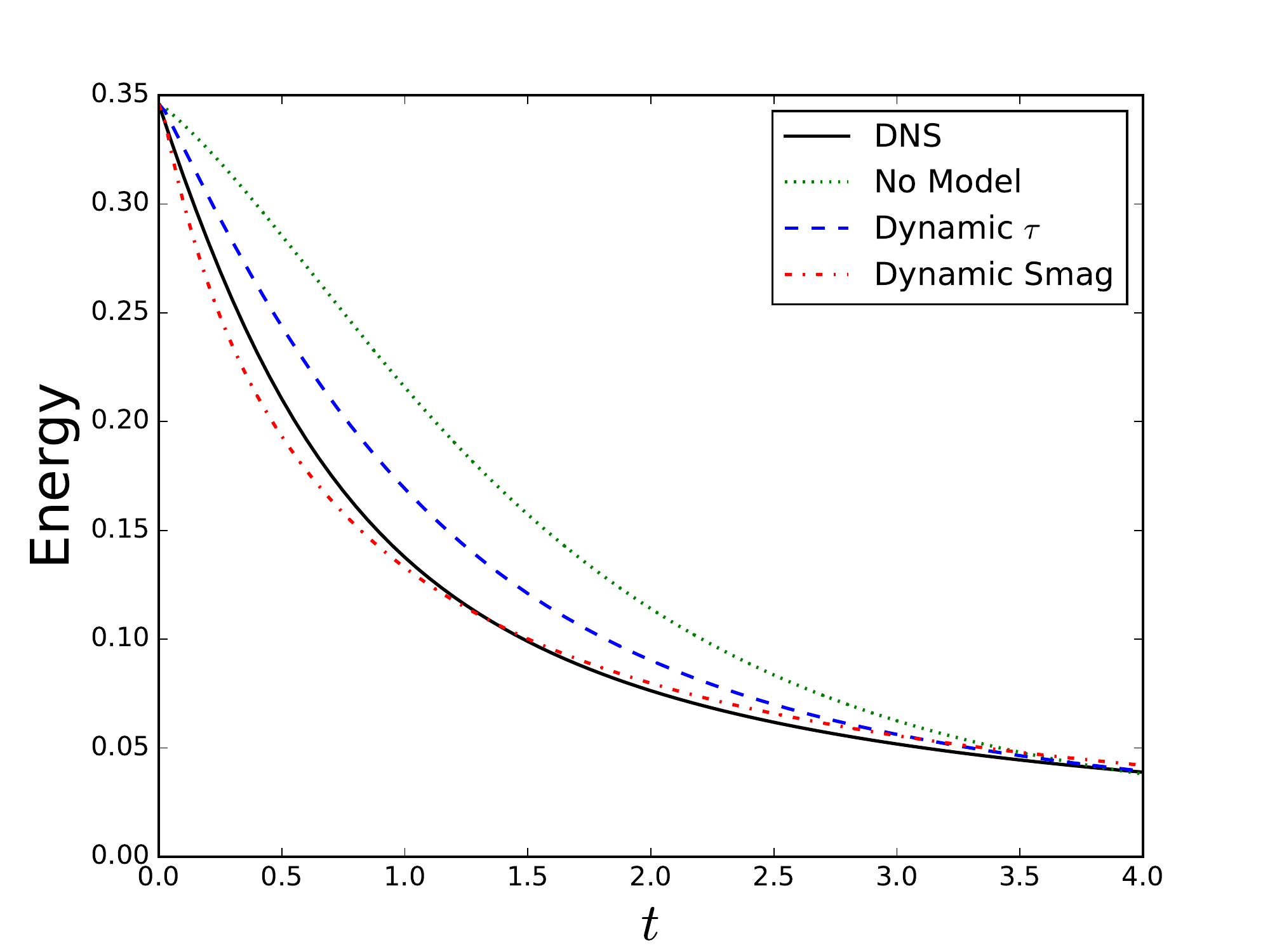}
	\caption{Evolution of the resolved kinetic energy.}
	\end{subfigure}
	\begin{subfigure}[t]{0.32\textwidth}
	\includegraphics[clip,width=1.\textwidth]{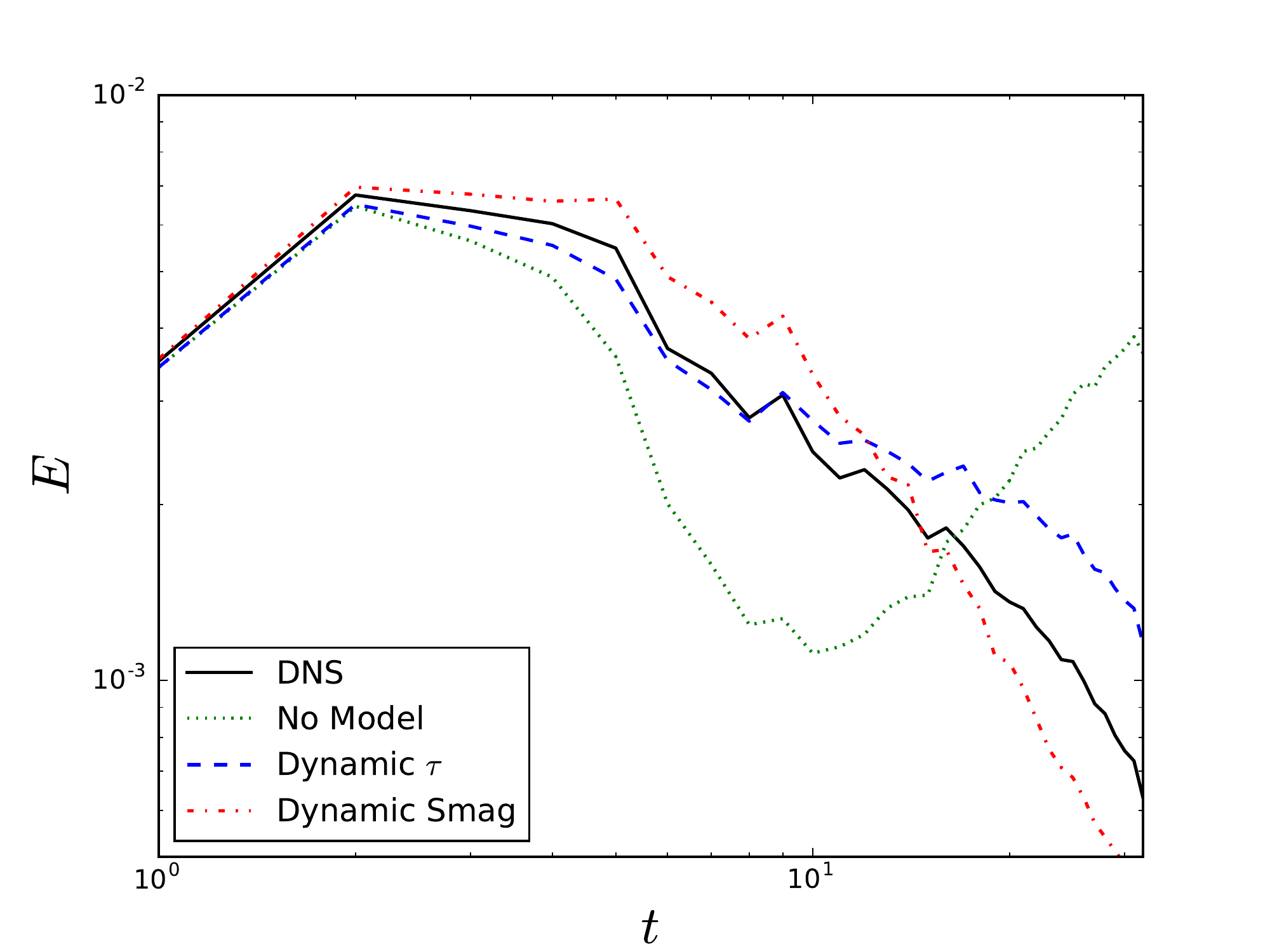}
	\caption{Energy spectrum at $t = 2.0$.}
	\end{subfigure}
	%\begin{subfigure}[t]{0.45\textwidth}
	%\includegraphics[clip,width=1.\textwidth]{HIT/spec3.pdf}
	%\caption{Filtered energy spectrum at $t = 3.0$.}
	%\end{subfigure}	
	\begin{subfigure}[t]{0.32\textwidth}
	\includegraphics[clip,width=1.\textwidth]{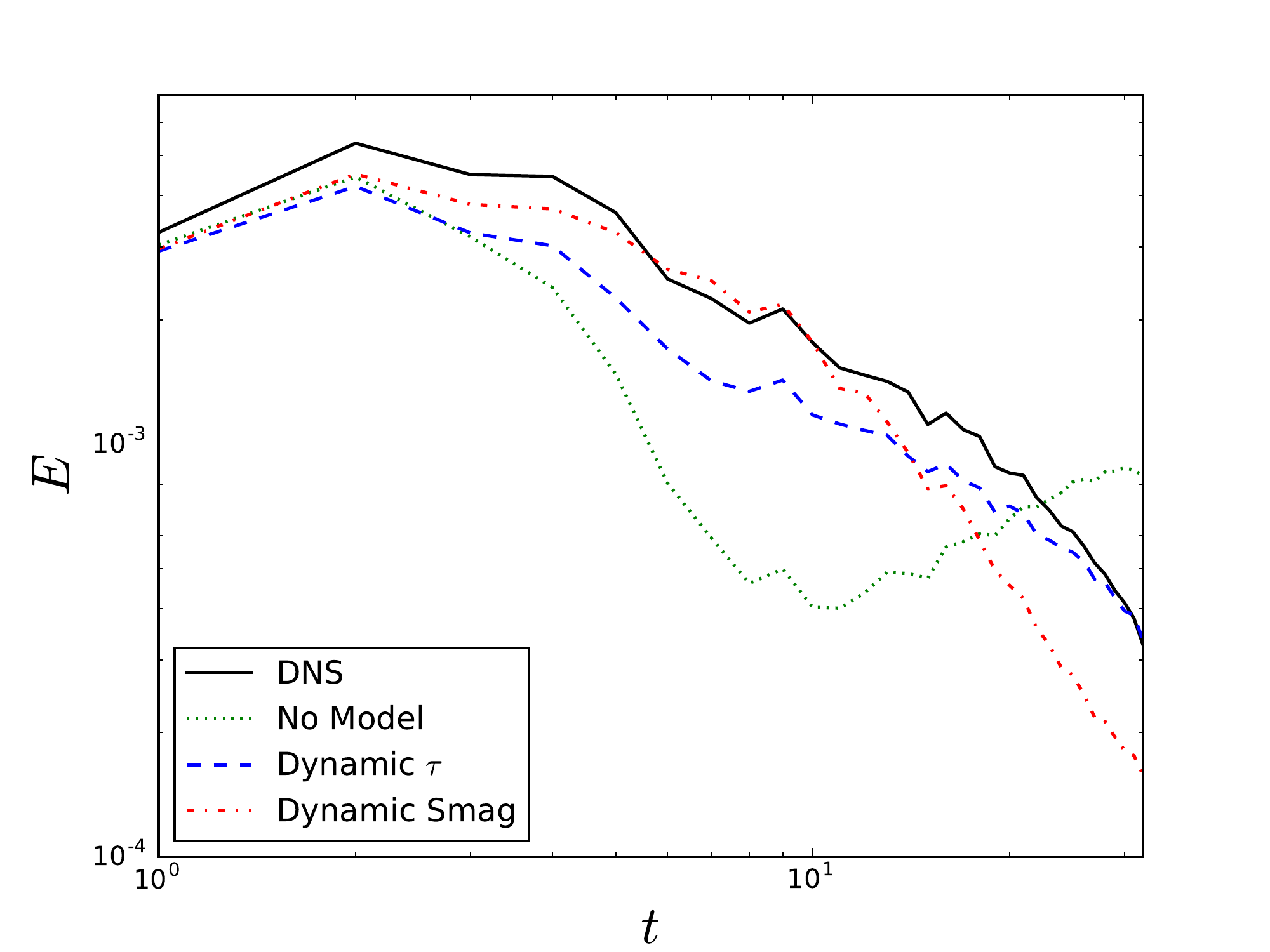}
	\caption{Energy spectrum at $t = 4.0$.}
	\end{subfigure}		
	\caption{Evolution of the resolved kinetic energy for the moderate Reynolds number case.}
	\label{fig:HIT_energy}
	\end{center}
\end{figure}

\begin{figure}
	\begin{center}
	\begin{subfigure}[t]{0.32\textwidth}
	\includegraphics[clip,width=1.\textwidth]{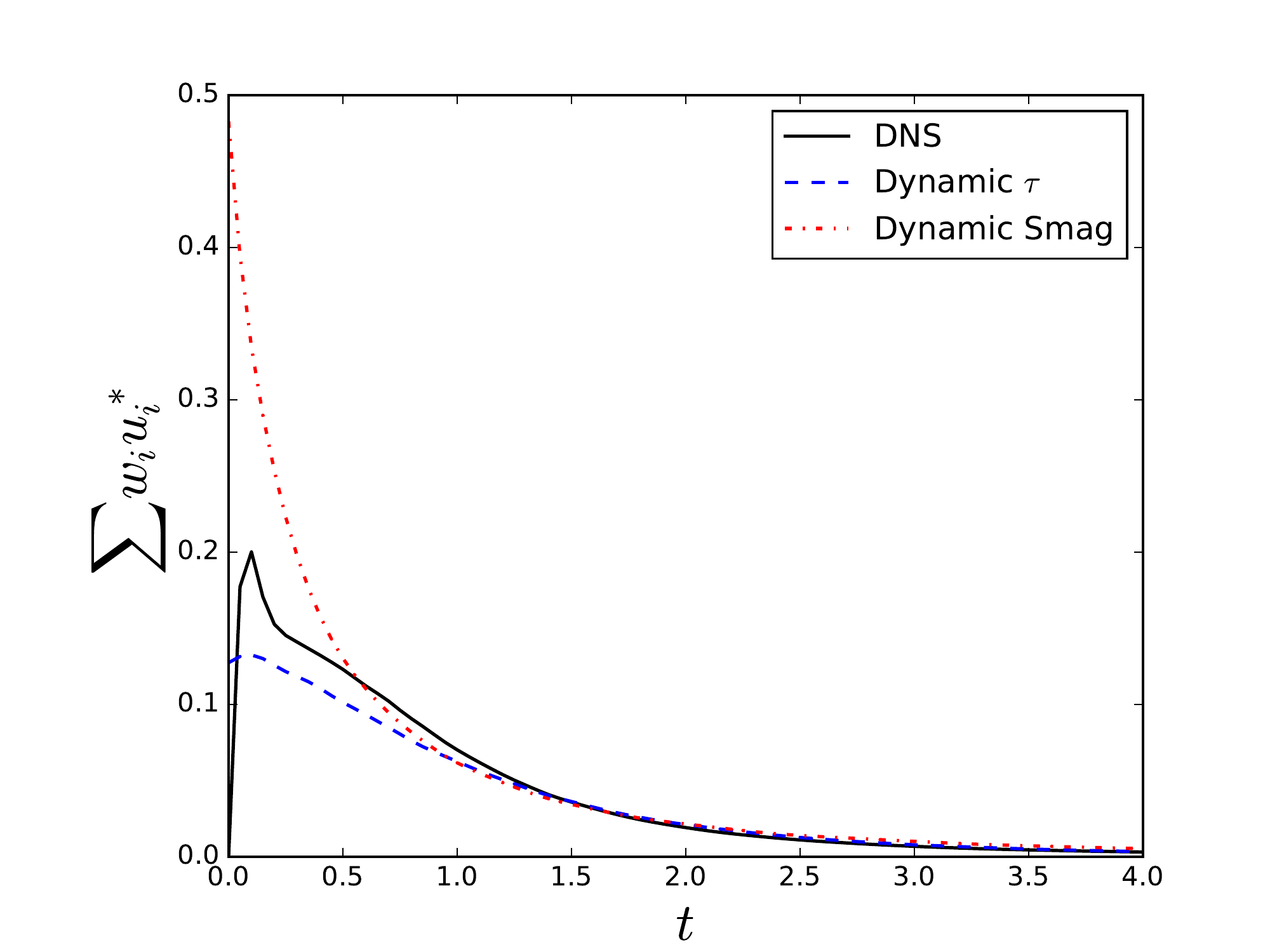}
	\caption{Evolution of S-G energy transfer.}
	\end{subfigure}
	\begin{subfigure}[t]{0.32\textwidth}
	\includegraphics[clip,width=1.\textwidth]{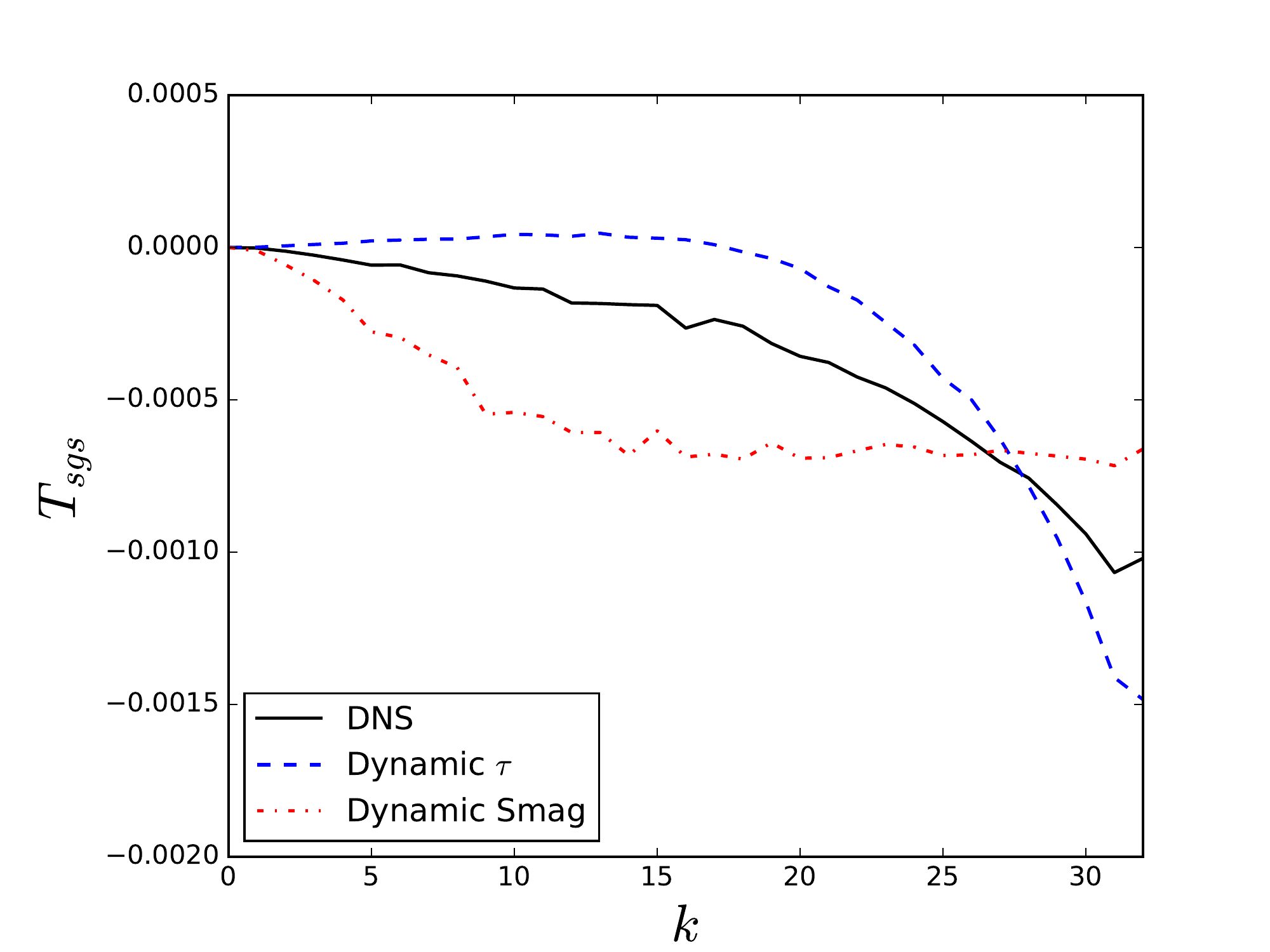}
	\caption{S-G energy transfer at $t = 2.0$.}
	\end{subfigure}
	%\begin{subfigure}[t]{0.32\textwidth}
	%\includegraphics[clip,width=1.\textwidth]{HIT/tsgs3.pdf}
	%\caption{Subgrid energy transfer at $t = 3.0$.}
	%\end{subfigure}	
	\begin{subfigure}[t]{0.32\textwidth}
	\includegraphics[clip,width=1.\textwidth]{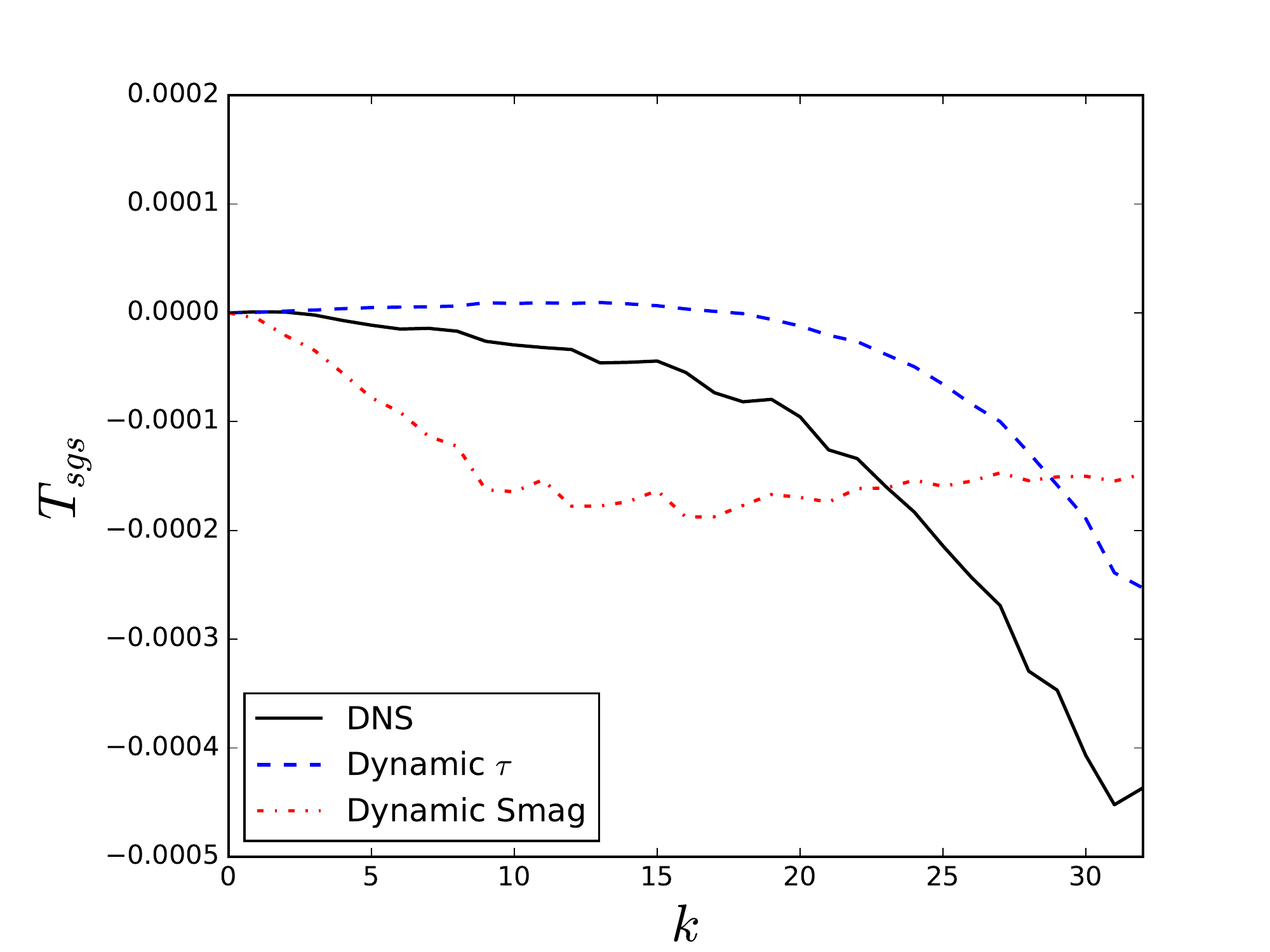}
	\caption{S-G energy transfer at $t = 4.0$.}
	\end{subfigure}		
	\caption{Evolution of the subgrid (S-G) energy transfer for the moderate Reynolds number case.}
	\label{fig:HIT_transfer}
	\end{center}
\end{figure}

The dynamic-$\tau$ model does not perform as well at the increased Reynolds number. Figure~\ref{fig:HIT_energy} shows the evolution of the resolved kinetic energy and the energy spectra for multiple time instances. The dynamic model is seen to under predict the initial dissipation of energy. A slight build up of energy is seen for the mid wave numbers at early times. As the simulation evolves, the spectrum predicted by the dynamic model improves and approaches the DNS simulation. Figure~\ref{fig:HIT_transfer} shows the evolution of the subgrid energy transfer and the subgrid energy transfer spectra at various time instances. The dynamic-$\tau$ model is seen to under-predict the energy transfer for low to mid wave numbers, while it over predicts the energy transfer for higher wave numbers. This leads to an under-prediction of the dissipation rate. 

\subsection{Homogeneous Turbulence at a High Reynolds Number}
\begin{figure}
	\begin{center}
	\begin{subfigure}[t]{0.32\textwidth}
	\includegraphics[clip,width=1.\textwidth]{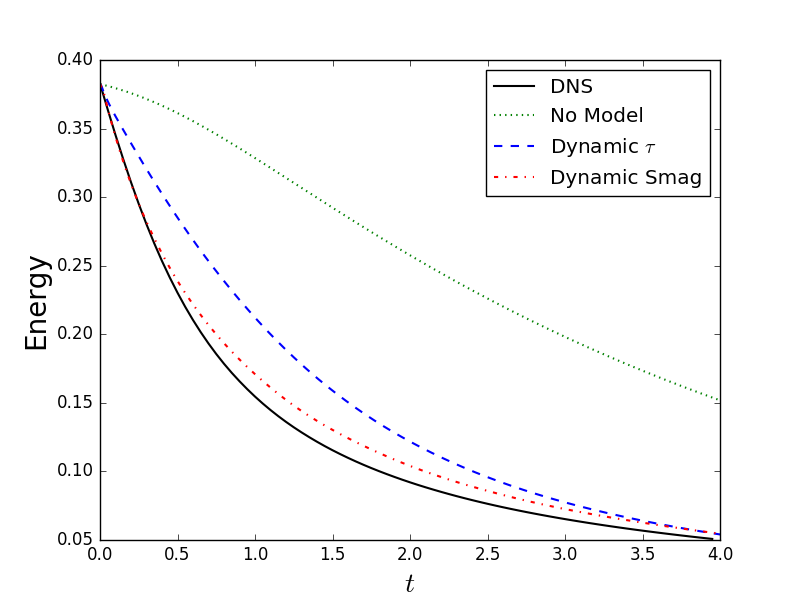}
	\caption{Evolution of the resolved kinetic energy.}
	\end{subfigure}
	\begin{subfigure}[t]{0.32\textwidth}
	\includegraphics[clip,width=1.\textwidth]{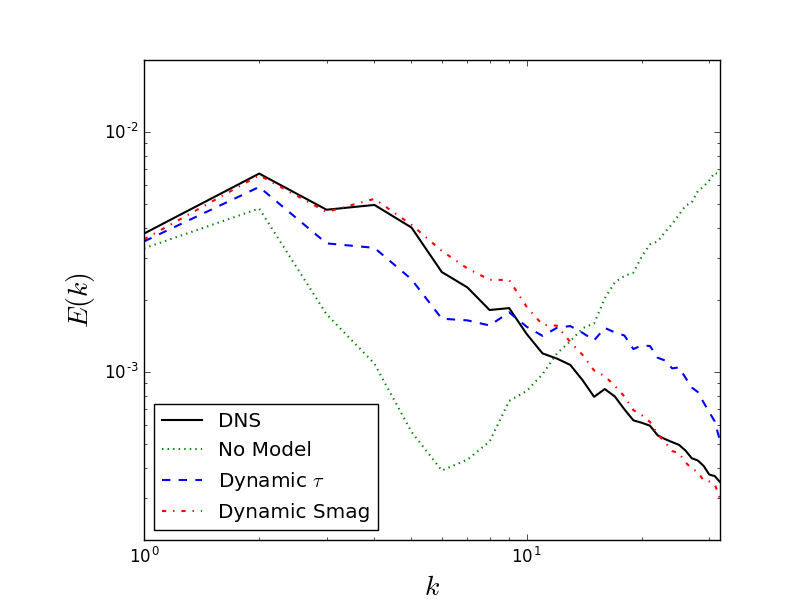}
	\caption{Energy spectrum at $t = 4.0$.}
	\end{subfigure}
	%\begin{subfigure}[t]{0.45\textwidth}
	%\includegraphics[clip,width=1.\textwidth]{HIT/spec3.pdf}
	%\caption{Filtered energy spectrum at $t = 3.0$.}
	%\end{subfigure}	
	\begin{subfigure}[t]{0.32\textwidth}
	\includegraphics[clip,width=1.\textwidth]{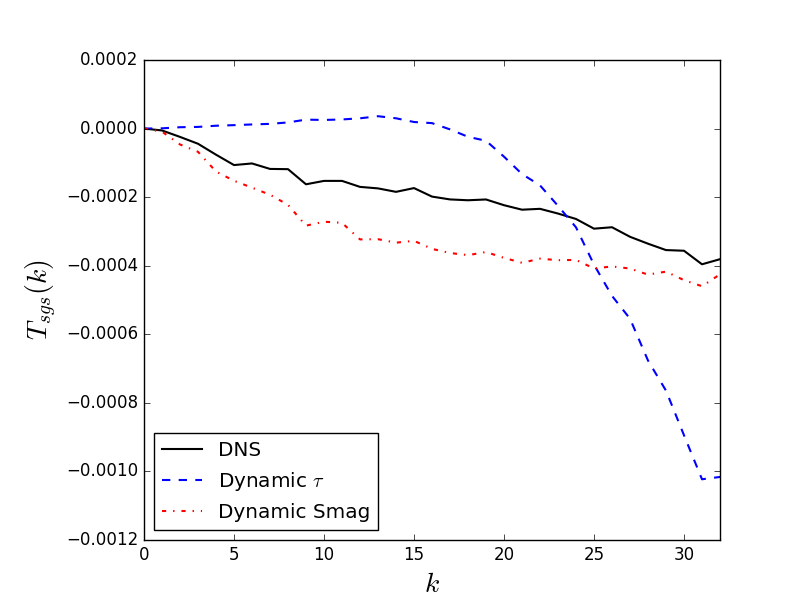}
	\caption{S-G energy transfer at $t = 4.0$.}
	\label{fig:HIT_highRe1c}
	\end{subfigure}			
	\caption{Summary of simulations of the high Reynolds number case using $64^3$ DOFs.}
	\label{fig:HIT_highRe1}
	\end{center}
\end{figure}

\begin{figure}
	\begin{center}
	\begin{subfigure}[t]{0.32\textwidth}
	\includegraphics[clip,width=1.\textwidth]{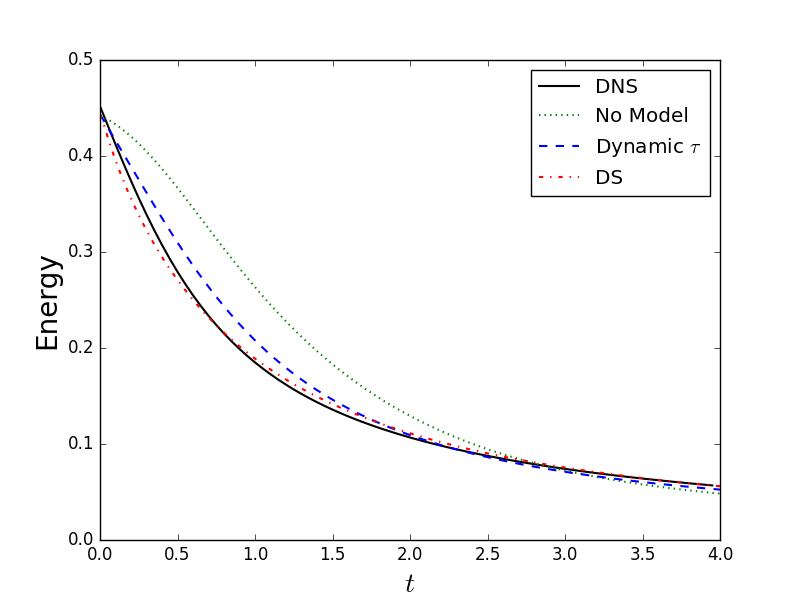}
	\caption{Evolution of the resolved kinetic energy.}
	\end{subfigure}
	\begin{subfigure}[t]{0.32\textwidth}
	\includegraphics[clip,width=1.\textwidth]{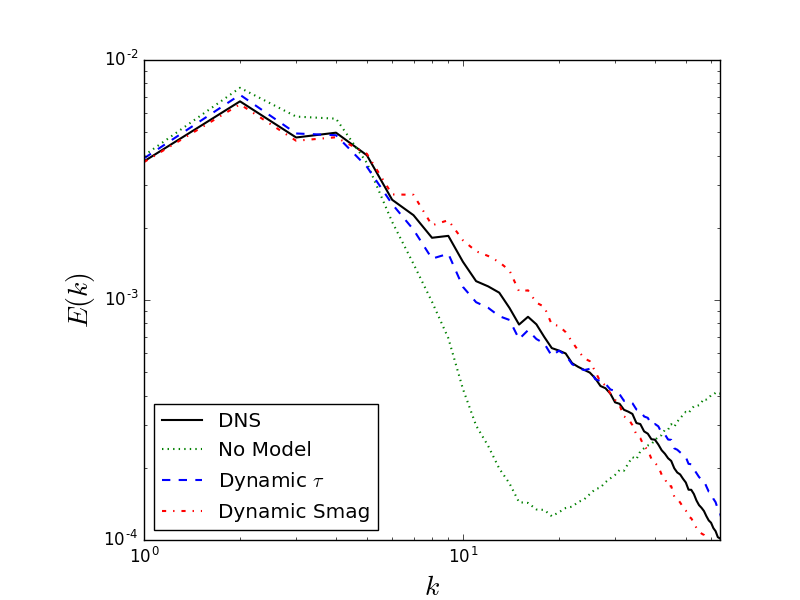}
	\caption{Energy spectrum at $t = 4.0$.}
	\end{subfigure}
	%\begin{subfigure}[t]{0.45\textwidth}
	%\includegraphics[clip,width=1.\textwidth]{HIT/spec3.pdf}
	%\caption{Filtered energy spectrum at $t = 3.0$.}
	%\end{subfigure}	
	\begin{subfigure}[t]{0.32\textwidth}
	\includegraphics[clip,width=1.\textwidth]{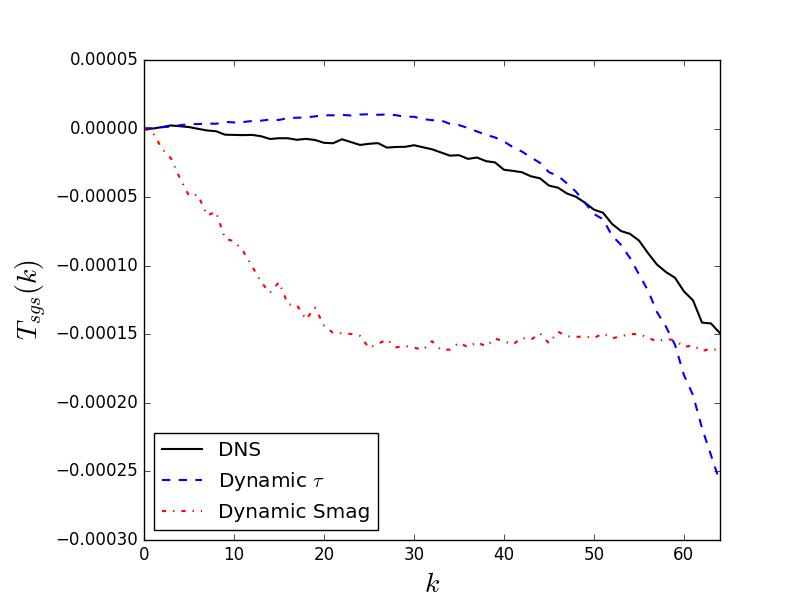}
	\caption{S-G energy transfer at $t = 4.0$..}
	\end{subfigure}		
	\caption{Summary of simulations of the high Reynolds number case using $128^3$ DOFs.}
	\label{fig:HIT_highRe2}
	\end{center}
\end{figure}

Homogeneous turbulence simulations at a high Reynolds number ($Re_{\lambda} \approx 164$) are now considered. At this Reynolds number, the DNS simulation requires $1024^3$ degrees of freedom to resolve up to $k\lambda_k = 0.5$. We consider large eddy simulations at both $64^3$ and $128^3$ degrees of freedom. These resolutions are $16$ and $8$ times coarser than the DNS, respectively. Figure~\ref{fig:HIT_highRe1} shows the evolution of resolved kinetic energy as well as the energy and subgrid transfer spectra at $t=4.0$ for the $64^3$ case. The same trends that were observed in the moderate Reynolds number case are again seen. The dynamic-$\tau$ model is slightly under dissipative. Not enough energy is removed from the low wave numbers, while too much energy is removed from the high wave numbers (as observed in Figure~\ref{fig:HIT_highRe1c}). The dynamic Smagorinsky model out performs the dynamic-$\tau$ model in this case. The dynamic-$\tau$ model still outperforms the case where no model is present. Figure~\ref{fig:HIT_highRe2} shows the same results for the $128^3$ case. At this increased resolution, the performance of the dynamic-$\tau$ model improves while the performance of the dynamic Smaogrinsky model is slightly worse than what it was in the $64^3$ case.

\subsection{Evolution of the coarse-graining time-scale}
\begin{figure}

	\begin{center}
	\begin{subfigure}[t]{0.45\textwidth}
	\includegraphics[clip,width=1.\textwidth]{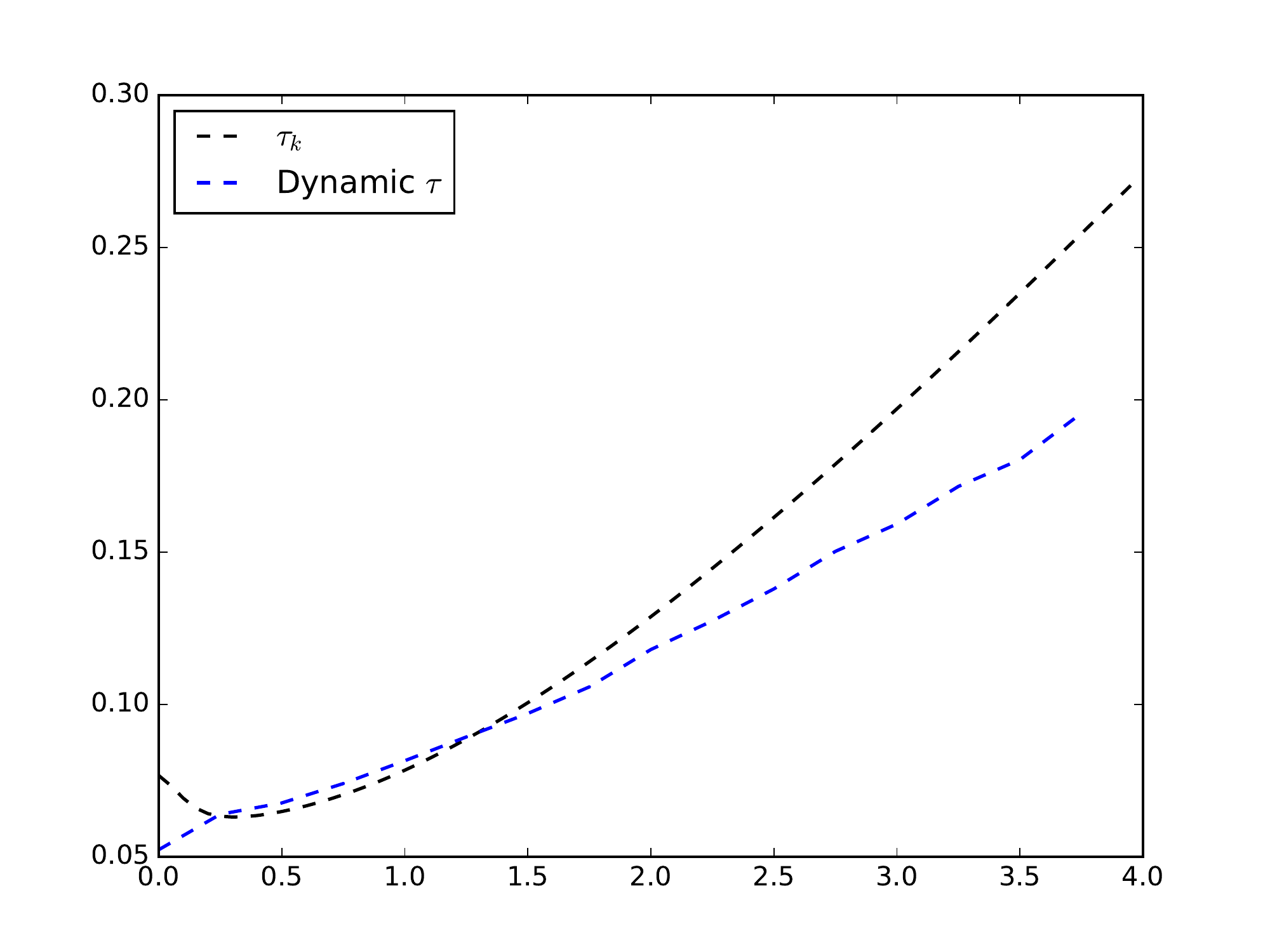}
	\caption{Low Reynolds number case.}
	\end{subfigure}
	\begin{subfigure}[t]{0.45\textwidth}
	\includegraphics[clip,width=1.\textwidth]{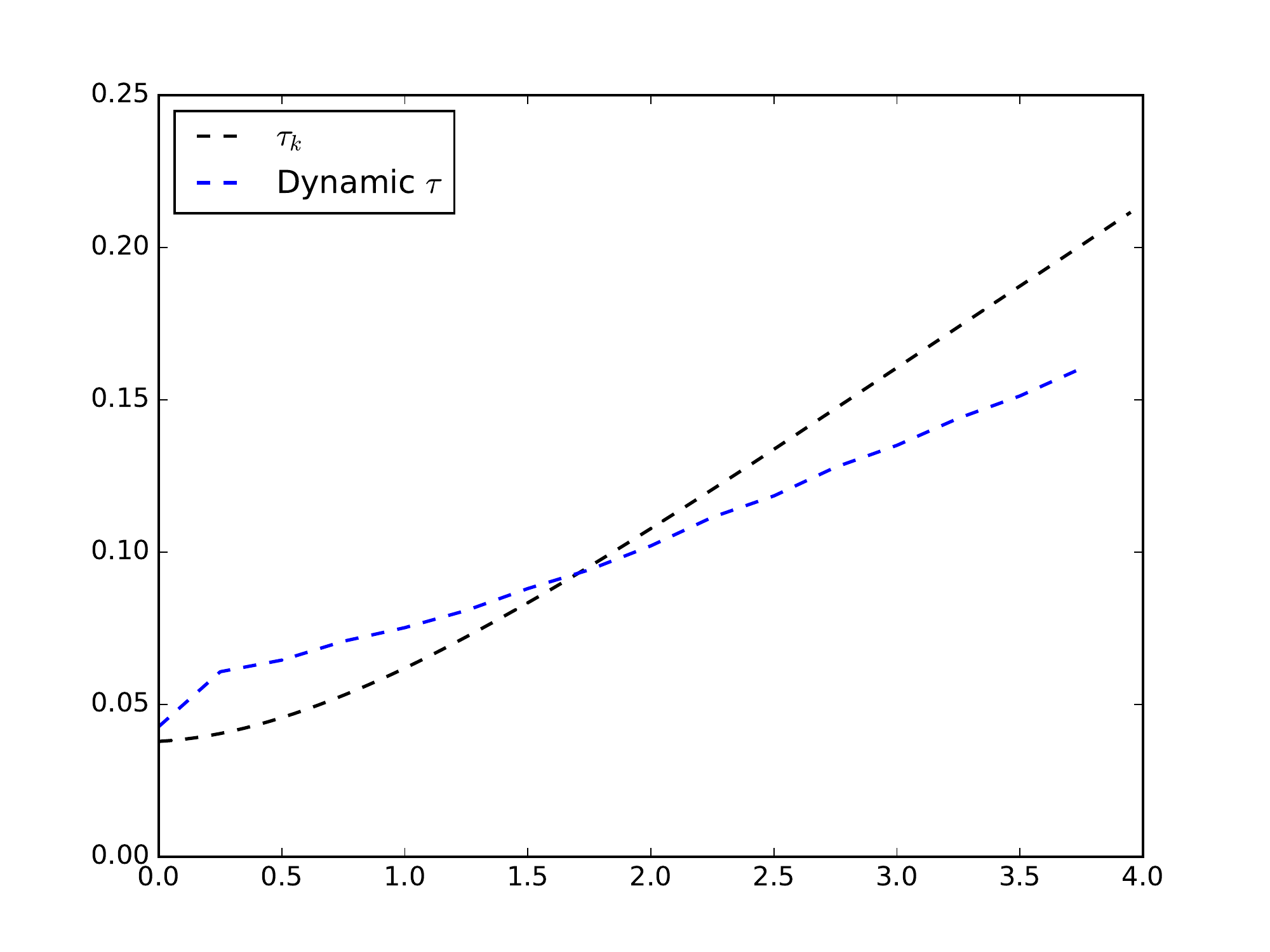}
	\caption{Moderate Reynolds number case.}
	\end{subfigure}
	\caption{Evolution of the coarse-graining time-scale as predicted by the dynamic procedure.}
	\label{fig:HIT_tau}
	\end{center}
\end{figure}

We now examine the evolution of the coarse-graining time-scale predicted by the dynamic procedure. Figure~\ref{fig:HIT_tau} shows the evolution of $\tau_{\MC{P}}$ as predicted by the dynamic procedure for both the low Reynolds number case and the moderate Reynolds number cases. As a reference, the predicted constant $\tau_{\MC{P}}$ is additionally compared to the Kolmogorov time-scale of the LES,  $\tau_k = \sqrt{\frac{\nu}{\epsilon} }.$ The Kolmogorov time-scale is the relevant physical quantity as it reflects the small-scale dynamics. The predicted time-scales are seen to grow approximately linearly with respect to time. The linear growth of $\tau_{\MC{P}}$ demonstrates that the dynamic-$\tau$ model behaves similarly to a renormalized $t$-model. The dynamic model, however, predicts the linear growth with respect to time while the renormalized $t$-model assumes it. It is further observed that $\tau_{\MC{P}}$ compares well with the Kolmogorov time-scale, demonstrating that there is a qualitative similarity between the time-scales of the memory kernel and that of the small-scale dynamics. 

\subsection{Validity of model assumptions}
\begin{figure}

	\begin{center}
	\begin{subfigure}[t]{0.45\textwidth}
	\includegraphics[clip,width=1.\textwidth]{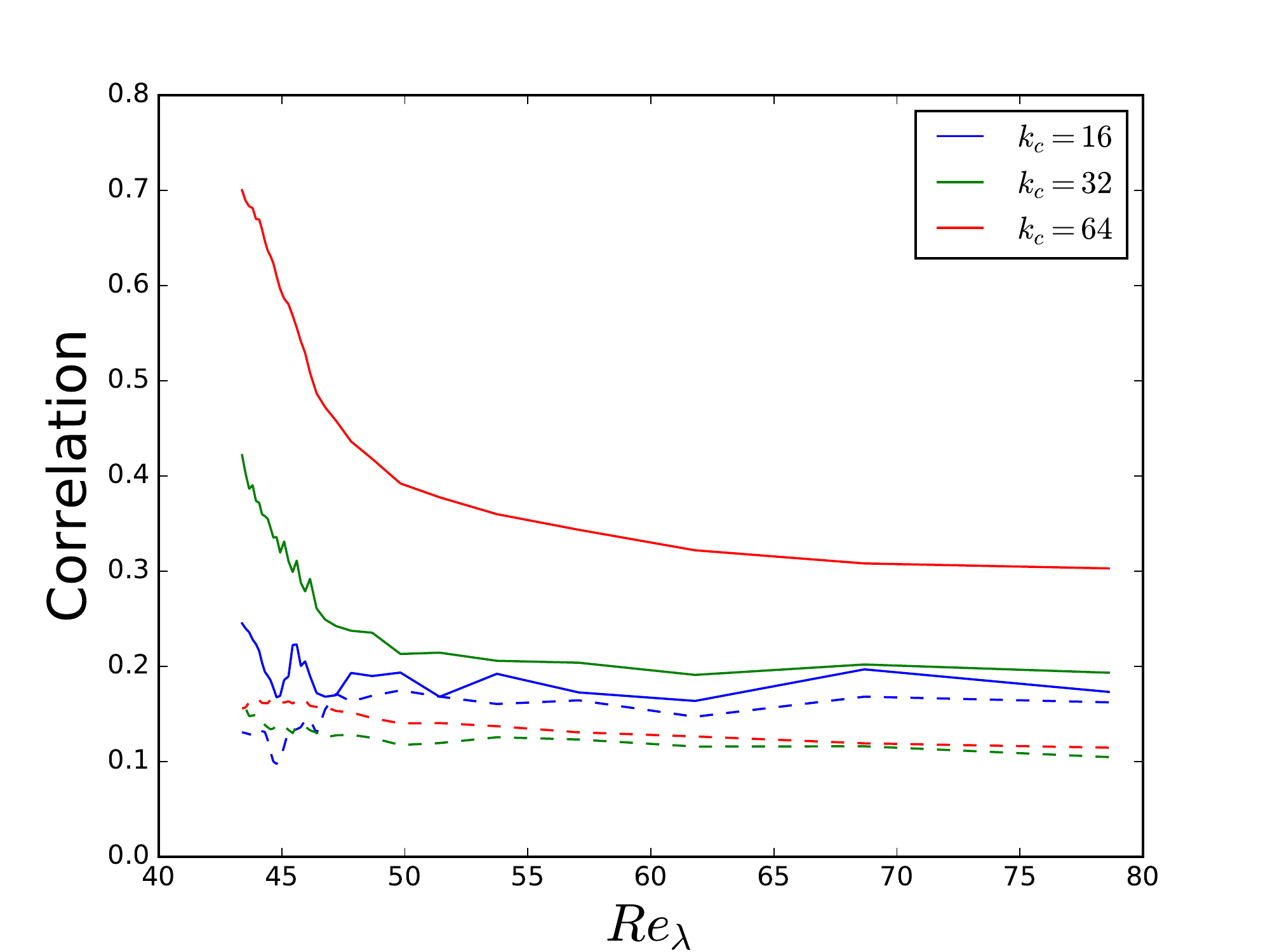}
	\end{subfigure}
	\caption{Correlation coefficient between the point-wise field of the approximated memory integral and the true memory integral. The solid lines are the dynamic-$\tau$ model, while the dashed lines are the dynamic Smagorinsky model. }
	\label{fig:corr_fig}
	\end{center}
\end{figure}

In the previous sections, it was seen that the performance of the dynamic-$\tau$ model degrades slightly as the LES becomes increasingly coarse. This degradation is due to the breakdown of the assumption that the memory integral is well correlated to its value at $s=0$. To demonstrate this, we compute the a priori correlation coefficient between the the exact memory integral and its value at $s=0$ as a function of the Reynolds number for several resolutions. The correlation coefficient between the memory integral and the term predicted by the Smagorinsky closure model is additionally computed for reference. Figure~\ref{fig:corr_fig} shows the resulting correlations. It is seen that as the Reynolds number increases, the correlation coefficient drops. It is further seen that the correlation coefficient drops as the model becomes increasingly coarse.  This drop in the correlation coefficient is due to the fact that the orthogonal dynamics becomes increasingly complex. It is noted, however, that the point-wise correlation between the field predicted by the dynamic-$\tau$ model is better than that predicted by the dynamic Smagorinsky model for all resolutions.  
\section{Application to Turbulent Channel flow}\label{sec:Channel}
Thus far, in this work as well as other applications of M-Z models in the literature, only decaying homogeneous problems have been discussed. These problems are characterized by the decay of kinetic energy and the growth in memory length. The simulation of fully developed turbulent channel flow with a Fourier-Chebyshev pseudo-spectral method is now considered. There are several significant differences between the channel and the triply periodic Navier-Stokes equations that are worth pointing out. First, the channel is only streamwise and spanwise periodic. The flow is inhomogeneous in the wall-normal direction with no-slip boundary conditions at the wall. Due to the inhomogeneity in the wall normal direction, no simple solution to the Poisson pressure equation exists for the channel (in contrast to Eq.~\ref{eq:NSspec}, where pressure simply appears as a projection). This complicates the derivation of M-Z-based models as pressure is determined by the solution of an elliptic equation. Pressure in incompressible flows is not a thermodynamic variable determined by an equation of state. In some sense, it can be viewed as a Lagrange multiplier that enforces the velocity field to be solenoidal at all times ~\cite{PressureIncompressible}. This viewpoint is adopted in this work. We choose to neglect the effects of pressure in the subgrid scale model, and instead ensure that the continuity equation is satisfied in the solution space of the resolved variables. 

For the channel flow, the Navier-Stokes equations are solved in skew-symmetric form. This form has been shown to minimize aliasing errors~\cite{SpectralMethodsInFluidDynamics}. 
The skew-symmetric form is, however, significantly more complicated than the primitive form of the Navier-Stokes equations and the analytic derivation of the M-Z models is considerably more challenging. 
%In previous work, the present authors have successfully used M-Z closures based on the primitive form of the N-S equations as closure models for the channel solved in skew-symmetric form. 
As such, the M-Z-based models are computed by numerically evaluating the Fr\'echet derivative, as discussed in Section~\ref{sec:numericalEvaluation}.
The dynamic model is constructed by coarse-graining in the periodic directions.  A no-slip boundary condition is used at the wall. Note that, since we only coarse-grain in the periodic directions, the effect of coarse-graining on the boundary conditions is not addressed in this work. The Navier-Stokes equations are solved using a coupled semi-implicit Adams-Bashforth  scheme for time integration, as in~\cite{MoinChannel1}. The continuity equation is directly enforced at each time-step, bypassing the need for pressure boundary conditions. The main solvers are written in Python and utilize mpi4py for parallelization. All FFT calculations (including the Chebyshev transforms) are de-aliased by the 3/2 rule. The solutions are compared to the dynamic Smagorinsky model. LES simulations at Reynolds numbers of $Re_{\tau} = \{180,395,590\}$ are considered. The relevant computational details are given in Table~\ref{table:2}. The domain sizes were selected to match those in Ref.~\cite{MoserChannel} and are designed to be long enough such that the periodicity constraint imposed by the Fourier ansatz is appropriate.
%The solutions are compared to the wall-damped Smagorinsky model, in which the filter width $\Delta$ is taken to include the Van Driest damping function
%$$\Delta  = \big(1 - e^{y^+/A} \big) \big(\Delta_1 \Delta_2 \Delta_3 \big)^3.$$ In this case, $\Delta_i$ is the filter width in the $i$-th direction and $A^+ = 25$. The %Smagorinsky constant was set to $C_s = 0.065$. 
%LES simulations at the relatively low Reynolds numbers of $Re_{\tau} = \{180,395\}$ are considered. The relevant computational details are given in Table~\ref{table:2}. The domain sizes were selected to match those in Ref.~\cite{MoserChannel} and are designed to be long enough such that the periodicity constraint imposed by the Fourier ansatz is appropriate.
\begin{table}%Table of simulation parameters
\begin{center}\scriptsize
\vskip -0.1in
\begin{tabular}{c c c c c c c c c } \hline
 $Re_{\tau}$ &  $L_x$ & $L_y$ & $L_z$ & $N_x$  & $N_y$ &$N_z$ & $\Delta t^+$ &   $\Delta t$ \\ \hline
180 & $4 \pi$ & 2 & $4/3 \pi$ &  32 & 64 & 32  & 0.24  & 0.02 \\ \hline 
395 & $2 \pi$ & 2 &  $\pi$ &  32 & 128 & 32  & 0.28 & 0.005\\ \hline
590 & $2 \pi$ & 2 &  $\pi$ &  32 & 192 & 32  & 0.31 & 0.0025\\ \hline
\end{tabular}
\caption{Physical and numerical details for large eddy simulations of the channel flow.}
\label{table:2}
\end{center}
\end{table}

\begin{figure}
	\begin{center}
	\includegraphics[width=0.99\textwidth]{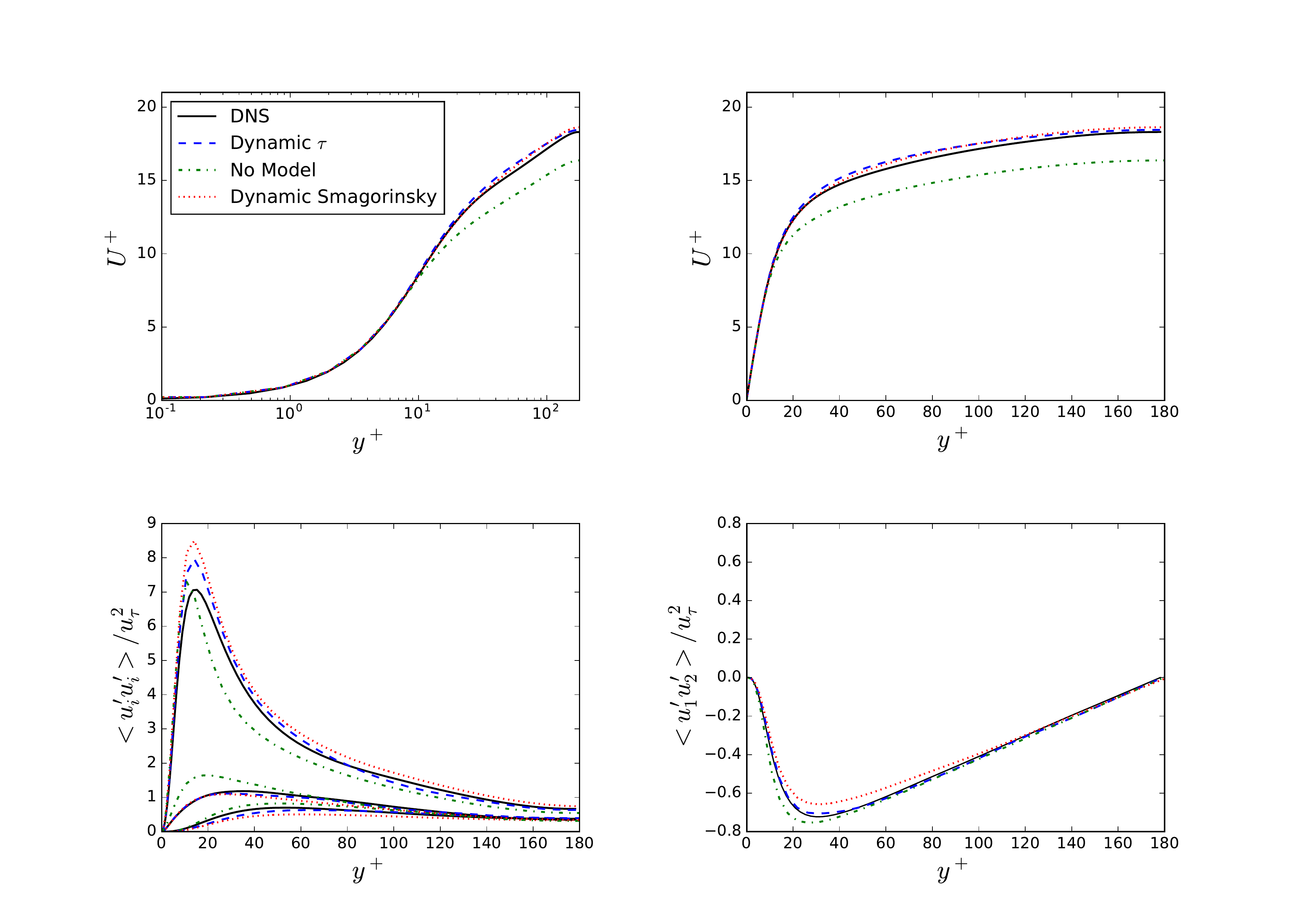}
	\caption{Statistical properties for fully-developed channel flow at $Re_{\tau} = 180$. }
	\label{fig:Channel1}
	\end{center}
\end{figure}

LES solutions at $Re_{\tau} = 180$ are shown in Figure~\ref{fig:Channel1} and are compared to unfiltered DNS data from~\cite{MoserChannel}.  The filtering process will not affect the mean velocity profiles, but the filtered Reynolds stress profiles are expected to be slightly different. The mean velocity profile predicted by the dynamic-$\tau$ model is slightly larger in magnitude than that of the DNS, but it is much improved from the simulation ran with no subgrid model. The model correctly damps the $R_{22}, R_{33},$ and $R_{12}$ Reynolds stresses. A slight amplification of the $R_{11}$ Reynolds stress is seen near the wall. The reason for this is not quite clear in the context of the dynamic-$\tau$ model. The mean non-dimenaional coarse-graining time-scale as predicted by the dynamic model is $\tau_{\MC{P}}^+ \approx 2.35$, where $\tau_{\MC{P}}^+ = \tau_{\MC{P}} u_{\tau}^2/\nu$. It is important to note that the dynamic procedure predicted a time-scale that did not increase with time. A growing time-scale, as was the case for the decaying problems, would be incorrect in this context.

\begin{figure}
	\begin{center}
	\includegraphics[width=0.45\textwidth]{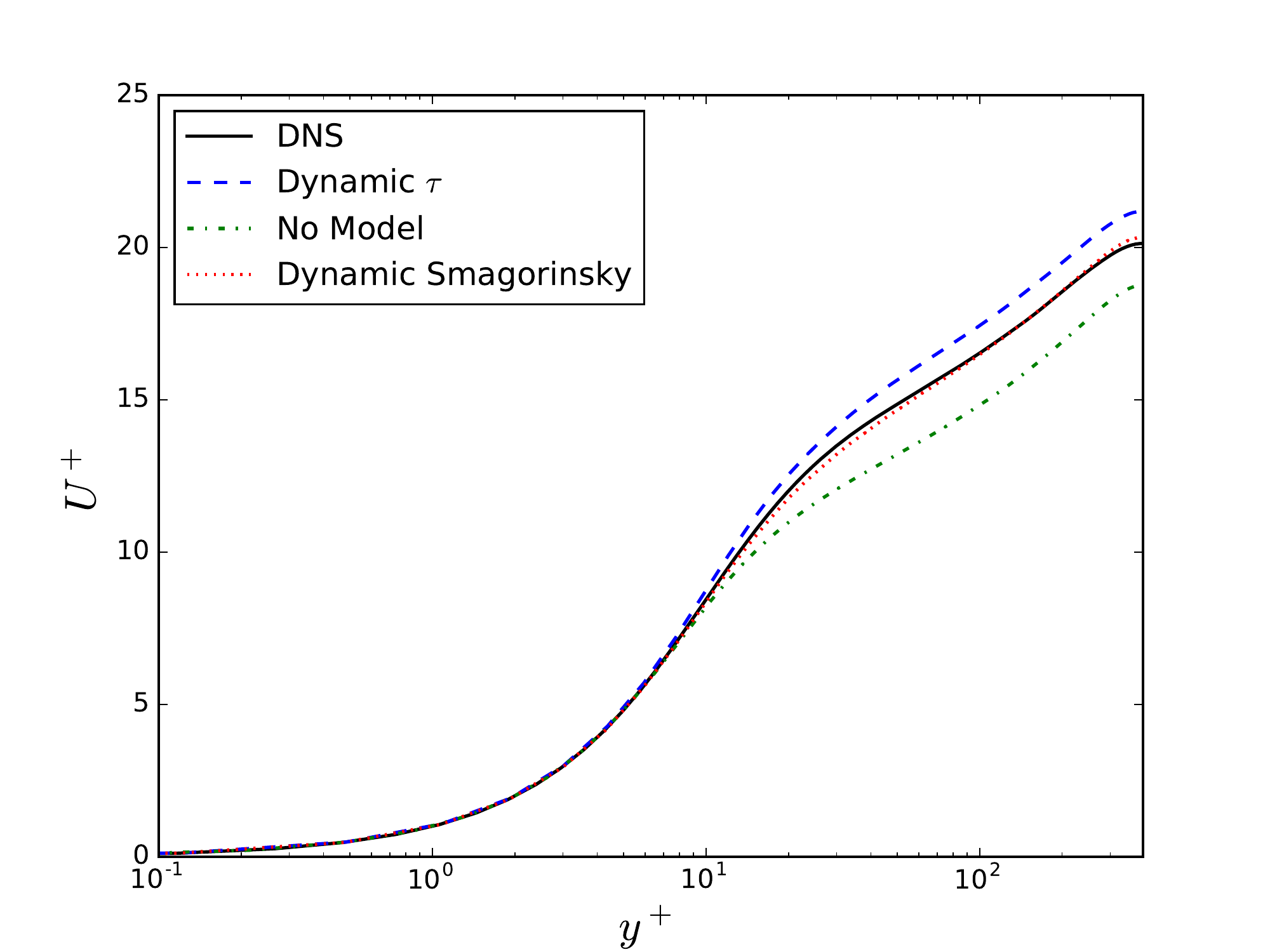}
	\includegraphics[width=0.45\textwidth]{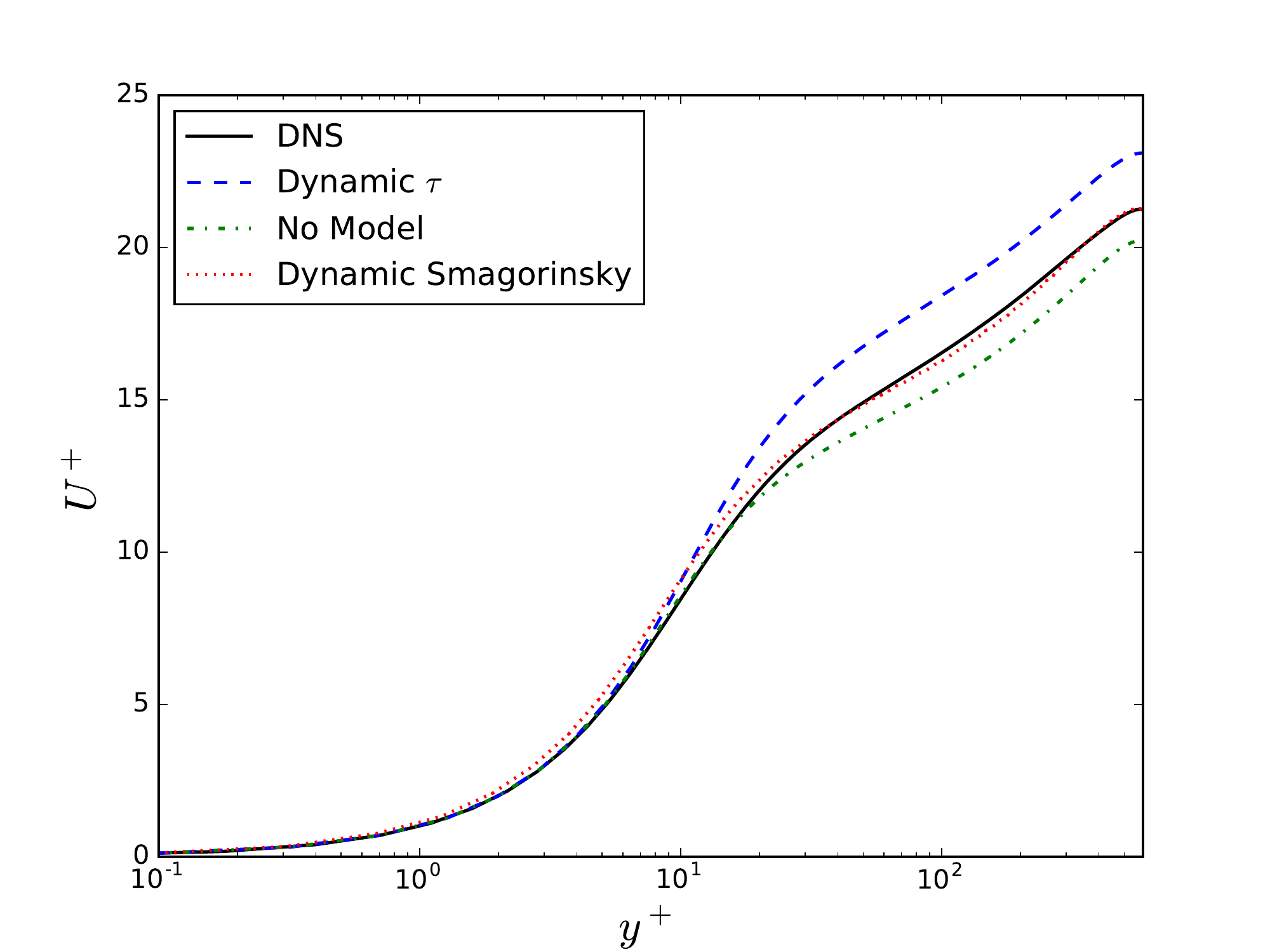}
	\caption{Velocity profiles for fully-developed channel flow at $Re_{\tau} = 395$ (left) and $Re_{\tau} = 590$ (right).}
	\label{fig:Channel2}
	\end{center}
\end{figure}

Figure~\ref{fig:Channel2} shows the velocity profiles for LES solutions at $Re_{\tau} = 395$ and $Re_{\tau} = 590$. As was seen in the case of homogeneous turbulence, increasing the Reynolds number (while holding the resolution constant) leads to a slight decrease in model performance. The mean velocity profiles for both cases are slightly over-predicted. The mean coarse-graining time-scales as predicted by the dynamic model are $\tau_{\MC{P}}^+ \approx \{2.2,2.5\}$ for the $395$ and $590$ case, respectively.  These non-dimensional time-scales are similar to those observed in the $Re_{\tau} = 180$ case. It is also worth noting that the predicted time-scale is roughly ten times the time step for all cases. This is due to the fact that the memory length scales with the spectral radius of the Jacobian~\cite{ParishAIAA2016}, as does the largest stable time step. This indicates that the time-step can be used as an alternative indicator to the memory length.

\section{Conclusions and Future Work}\label{sec:conclude}
The development of reduced models for systems that lack scale separation remains one of the prominent challenges in computational physics. Reduction of such systems to a reduced range of scales gives rise to a closure problem involving non-local memory effects. The optimal prediction framework developed by Chorin and co-workers uses the Mori-Zwanzig formalism to provide a mathematical framework to derive closure models for such systems. In this work, we presented a new M-Z-based model that approximates non-Markovian  effects using a Markovian closure. The model assumes that memory effects have a finite support in time. This time-scale is determined through a Germano-type procedure. The result of this process is a parameter-free closure model for coarse-grained systems, the structure of which is mathematically (rather than phenomenologically) imposed. 

The dynamic-MZ-$\tau$ model was shown to provide accurate results for Burgers equation and moderately resolved LES of fluid flows. For decaying problems (i.e. Burgers equation and decaying homogeneous turbulence) the dynamic procedure predicted a coarse-graining time-scale that grows linearly with time. The rates at which these time-scales increase are close to those obtained by Stinis' renormalized $t$-model. The  $t$-model, however, does not have a rigorous basis in problems that reach a statistically stationary state. In the case of turbulent channel flow, the dynamic procedure predicted a time-scale that did not grow with time - an attribute that the t-model is incapable of predicting. We believe that the flexibility of the model and consistency with renormalized M-Z models makes the dynamic-MZ-$\tau$ model a promising candidate. The dynamic model was shown to not perform as well for significantly under-resolved flows. This observation was also made regarding the $t$-model in ~\cite{ChandyFrankelLES}. 
%The present authors have also found other MZ-based models to be lacking for higher Reynolds number flows. 
The reason for this discrepancy is that the correlation between the integrated memory kernel and its value at $s=0$ degrades as the LES becomes increasingly coarse. 

Nevertheless, the results presented in this work highlight the potential of Mori-Zwanzig-based techniques in closure modeling as the structural form of the closure is imposed by the mathematics of the coarse-graining and not by the specific physics at play. Extensions can target higher order approximations of the memory integral. Further a dynamic procedure in the context of Stinis'~\cite{stinis_finitememory} hierarchical finite memory models  may also yield higher levels of accuracy.

The extension of the proposed model to the channel flow is one of the first applications of M-Z-based models to wall-bounded turbulence. However, the proposed model (and M-Z-based models in general) require further development to be of use in a general LES setting. We are currently pursuing the development of the dynamic model in a context similar to that of the variational multiscale method (VMS)~\cite{hughes}. The VMS formulation makes use of explicit scale-separation operators that are essential to the M-Z procedure. 

\section*{Acknowledgments}
%===================================================================
The authors would like to thank Prof. Peter Schmid for a number of insightful conversations which this work benefitted from. This research was funded by the AFOSR under the project {\em LES Modeling of Non-local effects using Statistical Coarse-graining} (Tech. Monitor: Jean-Luc Cambier). Computing resources were provided by the NSF via grant {\em MRI: Acquisition of Conflux, A Novel Platform for Data-Driven Computational Physics} (Tech. Monitor: Ed Walker).

\section*{Appendix A: Analytic evaluation of $\mathcal{PLQL}\phi_{0j}$ for Burgers Equation}
The analytic form of $\mathcal{PLQL}\phi_{0j}$ can be computed by evaluating the exact Fr\'echet derivative of $\MC{QL}\phi_{0j}$ and then projecting it. This corresponds to a linearization. For clarity, this process is illustrated on the inviscid Burgers equation (viscosity does not effect the final model form). The right hand side of Burgers equation at $t=0$ in Fourier space is
\begin{equation}\label{eq:VBE_freqApp}
\mathcal{L}u_{0k} = R_k(u_0) = - \frac{\imath  k}{2} \sum_{\substack{ p + q = k \\ p ,q \in F \cup G }}u_{p0} u_{q0} , \qquad k \in F \cup G.
\end{equation}
Split the convolution in Eq.~\ref{eq:VBE_freqApp} into resolved and unresolved terms,
\begin{equation}\label{eq:VBE_freqApp2}
\mathcal{L}u_{0k} =- \frac{\imath  k}{2} \sum_{\substack{ p + q = k \\ p ,q \in F }}u_{p0} u_{q0}  - \frac{\imath  k}{2} \sum_{\substack{ p + q = k \\ p \in F, q \in G }}u_{p0} u_{q0}  - \frac{\imath  k}{2} \sum_{\substack{ p + q = k \\ p \in G,q \in F  }}u_{p0} u_{q0} - \frac{\imath  k}{2} \sum_{\substack{ p + q = k \\ p ,q \in G  }}u_{p0} u_{q0}, \qquad k \in F \cup G,
\end{equation}
where modes in $F$ are resolved and in $G$ are unresolved. Application of the complimentary projector eliminates the first term on the RHS,
\begin{equation}\label{eq:VBE_freqApp3}
\mathcal{QL}u_{0k} =  - \frac{\imath  k}{2} \sum_{\substack{ p + q = k \\ p \in F, q \in G }}u_{p0} u_{q0}  - \frac{\imath  k}{2} \sum_{\substack{ p + q = k \\ p \in G,q \in F  }}u_{p0} u_{q0} - \frac{\imath  k}{2} \sum_{\substack{ p + q = k \\ p ,q \in G  }}u_{p0} u_{q0}, \qquad k \in F \cup G.
\end{equation}
Now linearize Eq.~\ref{eq:VBE_freqApp3} about $\mathbf{u_0}$,
\begin{equation}\label{eq:VBE_freqApp4}
\mathcal{QL}^{lin}u_{0k} =  - \frac{\imath  k}{2} \sum_{\substack{ p + q = k \\ p \in F, q \in G }}\big(u_{p0} u_{q0}' + u_{p0}' u_{q0}\big)  - \frac{\imath  k}{2} \sum_{\substack{ p + q = k \\ p \in G,q \in F  }}\big(u_{p0} u_{q0}' + u_{p0}' u_{q0}\big) - \frac{\imath  k}{2} \sum_{\substack{ p + q = k \\ p ,q \in G  }}\big(u_{p0} u_{q0}' + u_{p0}' u_{q0}\big) , \qquad k \in F \cup G.
\end{equation}
Application of the Liouville operator to Eq.~\ref{eq:VBE_freqApp3} is simply Eq.~\ref{eq:VBE_freqApp4} evaluated at $\mathbf{u_0}' = R(\mathbf{u}_0)$,
\begin{multline}\label{eq:VBE_freqApp5}
\mathcal{LQL}u_{0k} =  - \frac{\imath  k}{2} \sum_{\substack{ p + q = k \\ p \in F, q \in G }}\big(u_{p0} R_q(\mathbf{u_0}) + R_p(\mathbf{u_0}) u_{q0}\big)  -  \\ \frac{\imath  k}{2} \sum_{\substack{ p + q = k \\ p \in G,q \in F  }}\big(u_{p0} R_q(\mathbf{u_0}) + R_p(\mathbf{u_0}) u_{q0}\big) - \frac{\imath  k}{2} \sum_{\substack{ p + q = k \\ p ,q \in G  }}\big(u_{p0} R_q(\mathbf{u_0}) + R_p(\mathbf{u_0}) u_{q0}\big) , \qquad k \in F \cup G.
\end{multline}
Finally, project Eq.~\ref{eq:VBE_freqApp5},
\begin{equation}\label{eq:VBE_freqApp6}
\mathcal{PLQL}u_{0k} =  - \frac{\imath  k}{2} \sum_{\substack{ p + q = k \\ p \in F, q \in G }}u_{p0} R_q(\mathbf{\hat{u}_0})   -  \\ \frac{\imath  k}{2} \sum_{\substack{ p + q = k \\ p \in G,q \in F  }} R_p(\mathbf{\hat{u}_0})u_{q0}  , \qquad k \in F \cup G.
\end{equation}
Noting that the two separate terms on the RHS are identical and inserting in $R_q$, one recovers Eq.~\ref{eq:LESVBE_freqA},
\begin{equation}\label{eq:VBE_freqApp7}
\mathcal{PLQL}u_{0k} =  -\imath k \sum_{\substack{ p + q = k \\ p \in F,q \in G  }} u_{p0} \bigg[- \frac{\imath  q}{2} \sum_{\substack{ r + s = q \\ r ,s \in F }}u_{r0} u_{s0} \bigg], \qquad k \in F \cup G.
\end{equation}
The reader will note that the model form for the viscous Burgers equation is the same for modes in $F$.
\section*{References}
\bibliographystyle{aiaa}
\bibliography{refs}{}
\end{document}